\documentclass[12pt]{iopart}
\usepackage{epsfig}
\usepackage{graphicx}
\newcommand{\writingdate}{September 5th 2001}
\newcommand{\bra}{\langle}
\newcommand{\ket}{\rangle}

\newcommand{\minus}{{\!-\!}}
\newcommand{\bsigma}{\mbox{\boldmath $\sigma$}}
\newcommand{\bJ}{\mbox{\boldmath $J$}}
\newcommand{\be}{\begin{equation}}
\newcommand{\ee}{\end{equation}}
\newcommand{\bd}{\begin{displaymath}}
\newcommand{\ed}{\end{displaymath}}
\newcommand{\order}{{\cal O}}
\newcommand{\vsp}{\vspace*{3mm}}
\newcommand{\room}{\rule[-0.3cm]{0cm}{0.8cm}}
\begin{document}

\def\pin{}
\let\noin=\noindent
\def\Z{{\cal Z}_\beta}
\def\tZ{\tilde{\cal Z}_{\tilde \beta}}
\def\dS{\mbox{\boldmath $S$}}
\def\dT{\mbox{\boldmath $T$}}
\def\ds{\mbox{\boldmath $s$}}
\def\dt{\mbox{\boldmath $t$}}
\def\dm{\mbox{\boldmath $m$}}
\def\ddp{\mbox{\boldmath $p$}}
\def\dq{\mbox{\boldmath $q$}}
\def\bnul{\mbox{\boldmath $0$}}
\def\ptl#1{\frac{\partial}{\partial #1}}
\def\bfm#1{\mbox{\boldmath $#1$}}
\def\IK{{\rm I\!K}}
\def\ID{{\rm I\!D}}
\def\IB{{\rm I\!B}}
\def\IM{{\rm I\!M}}
\def\II{{\rm 1\!\!1}}
\def\IE{\mbox{I$\!$E}}
\def\IP{\mbox{I$\!$P}}
\def\IN{\mbox{I$\!$N}}
\def\IR{\mbox{I$\!$R}}

\title[Hierachical Self-Programming -- Version \writingdate]{Hierarchical Self-Programming\\ in Recurrent Neural Networks}
\author{
T Uezu\footnote{On leave from
            Graduate School of Human Culture, Nara Women's
            University,\\
            Kitauoyanishimachi, Nara City, 630-8506 Japan}
 and A C C Coolen}
\address{            Department of Mathematics, King's College London,\\
            The Strand, London WC2R 2LS, UK}

\begin{abstract}
We study self-programming in recurrent neural networks where both
neurons (the `processors') and synaptic interactions (`the
programme') evolve in time simultaneously, according to specific
coupled stochastic equations. The interactions are divided into a
hierarchy of $L$ groups with adiabatically separated and
monotonically increasing time-scales, representing sub-routines of
the system programme of decreasing volatility. We solve this model
in equilibrium, assuming ergodicity at every level, and find as
our replica-symmetric solution
a formalism with a structure similar but not identical to
Parisi's $L$-step replica symmetry breaking scheme. Apart from
differences in details of the equations (due to the fact that here
interactions, rather than spins, are grouped into clusters with
different time-scales), in the present model the block sizes $m_i$
of the emerging ultrametric solution are not restricted to the
interval $[0,1]$, but are independent control parameters, defined
in terms of the noise strengths of the various levels in the
hierarchy, which can take any value in $[0,\infty\ket$. This is
shown to lead to extremely rich phase diagrams, with an abundance
of first-order transitions especially when the level of
stochasticity in the interaction dynamics is chosen to be low.
\end{abstract}

 \ead{uezu@ki-rin.phys.nara-wu.ac.jp,
tcoolen@mth.kcl.ac.uk}


\section{Introduction}

In this paper we study  recurrent networks of binary neuronal
state variables, represented as Ising spins, with symmetric
couplings (or synaptic interactions) $J_{ij}$, taken to be of
infinite range. In contrast to most standard neural network
models, not only the neuron states but also the interactions are
 allowed to evolve in time (simultaneously), driven by
correlations in the states of the neurons (albeit slowly compared
to the dynamics of the latter), reflecting the effect of
`learning' or `long-term potentiation' in real nervous tissue.
Since the interactions represent the `programme' of the system,
and since the slow interaction dynamics are driven by the states
of the neurons (the `processors'), such models can be regarded as
describing self-programming information-processing systems, which
can be expected to  exhibit highly complex dynamical behaviour.

The first papers in which self-programming recurrent neural
networks were studied appear to be \cite{Sh,DH}. In the language
of self-programming systems one could say that these authors were
mostly concerned with the stability properties of embedded
`programmes' (usually taken to be those implementing
content-addressable or associative memories). In both \cite{Sh,DH}
the programme dynamics, i.e. that of the  $\{J_{ij}\}$, was
defined to be adiabatically slow compared to the neuronal
dynamics, and fully deterministic. However, the authors already
made the important observation that the natural type of
(deterministic) programme dynamics (from a biological point of
view), so-called Hebbian learning, could be written as a gradient
descent of the interactions $\{J_{ij}\}$ on the free energy
surface of a symmetric recurrent neural network equipped with
these interactions.

In order to study more generally the potential of such
self-programming systems, several authors (simultaneously and
independently) took the natural next step
\cite{CPS,PCS,DFM,PS,FD,C}: they generalized the interaction
dynamics by adding Gaussian white noise to the deterministic laws,
converting the process into one described by conservative Langevin
equations, and were thus able to set up an equilibrium statistical
mechanics of the self-programming process. This was (surprisingly)
found to take the form of a replica theory with finite replica
dimension, whose value was given by the ratio of the noise levels
in the neuronal dynamics and the interaction dynamics,
respectively. Furthermore, adding explicit quenched disorder to
the problem in the form of additional random (but frozen) forces
in  the interaction dynamics, led to theories with two nested
levels of replicas, one representing the disorder (with zero
replica dimension) and one representing the adiabatically slow
dynamics of the interactions \cite{PS,Jongenetal1,Jongenetal2}
(with nonzero replica dimension). The structure of these latter
theories was found to be more or less identical to those of
ordinary disordered spin systems such as the SK model \cite{SK},
with fixed interactions but quenched disorder, when described by
replica theories with one step replica symmetry breaking (RSB)
\cite{Parisi}. The only (yet crucial) difference was that in
ordinary disordered spin systems the size $m$ of the level-1 block
in the Parisi solution is determined by extremisation of the free
energy, which forces $m$ to lie in the interval $[0,1]$ see also
\cite{MC}, whereas in the self-programming neural networks of
\cite{PS,Jongenetal1,Jongenetal2} $m$ was an independent control
parameters, given by the ratio of two temperatures, which can take
any non-zero value. As a consequence one can observe in the
latter systems, for sufficiently large values of such dimensions,
much more complicated scenarios of (generally discontinuous) phase
transitions.

\begin{figure}[t]
\vspace*{-3mm}
\newcommand{\here}{\makebox(0,0)}
\newcommand{\frust}{\here{$\times$}}
\newcommand{\spinup}{\here{\circle*{20}}}
\newcommand{\spindown}{\here{\circle{20}}}
\setlength{\unitlength}{0.095mm}
\begin{center}\begin{picture}(850,850)(-80,30)
\multiput(100,200)(100,0){6}{\dashbox{10}(0,600)}
\multiput(50,750)(0,-100){6}{\dashbox{10}(600,0)}
\put(100,750){\line(1,0){300}} \put(050,650){\line(1,0){150}}
\put(400,650){\line(1,0){100}} \put(050,550){\line(1,0){50}}
\put(400,550){\line(1,0){100}} \put(100,450){\line(1,0){100}}
\put(300,450){\line(1,0){100}} \put(100,750){\line(0,-1){100}}
\put(200,750){\line(0,-1){100}} \put(200,550){\line(0,-1){200}}
\put(300,800){\line(0,-1){450}} \put(400,800){\line(0,-1){250}}
\put(400,450){\line(0,-1){100}} \put(500,750){\line(1,0){150}}
\put(050,350){\line(1,0){150}} \put(500,650){\line(1,0){100}}
\put(050,350){\line(1,0){50}} \put(200,250){\line(1,0){100}}
\put(300,350){\line(1,0){100}} \put(500,250){\line(1,0){150}}
\put(600,650){\line(0,-1){200}} \put(500,450){\line(0,-1){100}}
\put(300,250){\line(0,-1){50}} \put(600,250){\line(0,-1){50}}
\put(110,750){\spinup}  \put(110,650){\spinup}
\put(110,550){\spindown}\put(110,450){\spinup}
\put(110,350){\spinup}\put(110,250){\spinup}
\put(210,750){\spinup} \put(210,650){\spindown}
\put(210,550){\spinup} \put(210,450){\spinup}
\put(210,350){\spindown} \put(210,250){\spindown}
\put(310,750){\spinup} \put(310,650){\spinup}
\put(310,550){\spindown} \put(310,450){\spindown}
\put(310,350){\spinup} \put(310,250){\spindown}
\put(410,750){\spindown} \put(410,650){\spindown}
\put(410,550){\spinup}\put(410,450){\spindown}
\put(410,350){\spindown}\put(410,250){\spinup}
\put(510,750){\spindown} \put(510,650){\spindown}
\put(510,550){\spinup} \put(510,450){\spinup}
\put(510,350){\spindown} \put(510,250){\spinup}
\put(610,750){\spindown} \put(610,650){\spinup}
\put(610,550){\spinup}\put(610,450){\spindown}
\put(610,350){\spindown}\put(610,250){\spinup}
\end{picture}\end{center}
\vspace*{-13mm} \caption{Schematic illustration of the structure
and ingredients of our recurrent self-programming model for $L=2$.
The binary spin variables (spin up: $\circ$, spin down: $\bullet$)
evolve on time-scales of order 1. The level-1 interactions (solid
line segments) evolve on time-scales $\tau_1\gg 1$. The level-2
interactions (dashed line segments) evolve on time-scales
$\tau_2\gg \tau_1$. The interactions are randomly allocated to
levels. Our present model differs from this simple picture in two
ways: firstly, it is fully connected (i.e. infinite dimensional),
and secondly, we allow for an arbitrary number $L$ of interaction
types.}
 \label{fig:0}
\end{figure}
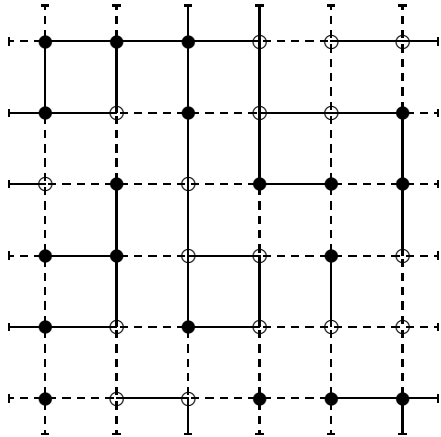

 In contrast to the previous models
involving coupled dynamics of fast neurons and slow interactions,
in this paper we study systems in which the interactions do not
evolve on a single time-scale, but where they are divided into a
hierarchy of $L$ different groups, each with their own
characteristic time-scale $\tau_\ell$ and noise level $T_\ell$ ($\ell=1,\ldots,L$),
describing a hierarchy of increasingly
non-volatile programming levels.  This appears to be a much more
realistic representation of self-programming systems; conventional
programmes generally take the form of hierarchies of routines,
sub-routines and so on, and it would appear appropriate to allow
low-level sub-routines to be more easily modifiable than
high-level ones. In order to retain analytical solvability we
choose the different groups of interactions randomly (prescibing
only their sizes), see figure \ref{fig:0}. We solve the model in
equilibrium, and find, upon making the replica-symmetric (i.e.
ergodic) ansatz within each level of our hierarchy, a theory which
resembles, but is not identical to Parisi's $L$-level replica
symmetry breaking solution for spin systems with frozen disorder.
Although Parisi's solution can also be traced back to the
existence of a hierarchy of adiabatically separated time-scales
\cite{MC}, in the present model the {\em interactions} are grouped
into clusters with different time-scales, rather than the spins.
Apart from quantitative differences in the details of the order
parameter equations, a major  consequence of this difference is
that in the present model the block sizes $m_i$ of the emerging
ultrametric solution are not restricted to the interval $[0,1]$,
but are independent control parameters, defined in terms of the
noise strengths of the various levels in the hierarchy. They can
consequently take any value in the interval $[0,\infty\ket$. We
show that this leads to extremely rich phase diagrams, with an
abundance of first-order transitions especially when the level of
stochasticity in the interaction dynamics is chosen to be low,
i.e. when the dimensions $\{m_i\}$ become large. We study our
model in full detail for the choices $L=2$ and $L=3$, including
phase diagrams, and we study the  asymptotic properties of our
model in the limits $m_1\to \infty$ for fixed $T$ (deterministic dynamics of the level-1
interactions) and $m_1\to 0$ for fixed $T_1$ (deterministic dynamics of the
neuronal processors), for arbitrary $L$.

\section{Model Definitions}

In this paper we will refer to our binary neurons as spins and to
the synaptic interactions as couplings. We will write the $N$-spin
state vector as $\bsigma=(\sigma_1,\ldots,\sigma_N)\in\{-1,1\}^N$,
and the matrix of interactions as $\bJ=\{J_{ij}\}$.  The spins are
taken to have a stochastic Glauber-type dynamics such that for
{\em stationary} choices of the couplings the microscopic spin
probability density would evolve towards a Boltzmann distribution
\begin{equation}
   p_\infty(\bsigma)=\frac{e^{-\beta H(\bsigma)}}{Z}
   ~~~~~~~~~~Z=\sum_{\bsigma}e^{-\beta H(\bsigma)}
   \label{eq:prob}
\end{equation}
with the conventional Hamiltonian
\begin{equation}
H(\bsigma)=-\sum_{i<j}J_{ij}\sigma_i\sigma_j
\label{eq:Hamiltonian}
\end{equation}
where $i,j\in\{1,\ldots,N\}$, and with the inverse temperature $\beta=T^{-1}$.

The couplings $J_{ij}$ also evolve in a stochastic manner, in
response to the states of the spins, but adiabatically slowly
compared to the spins, such that on the time-scales of the
couplings the spins are always in an equilibrium state described
by (\ref{eq:prob}). For the coupling dynamics the following
Langevin equations are proposed:
\begin{equation}
       \tau_{ij} \frac{d}{dt}J_{ij}=
                \frac{1}{N}\bra \sigma_i\sigma_j\ket_{\rm sp}
                -\mu_{ij} J_{ij}
                +\eta_{ij}(t)\sqrt{\frac{\tau_{ij}}{N}}
        \qquad ~~~~~~i<j=1\dots N \,
        \label{def:evcouplings}
\end{equation}
with $\tau_{ij}\gg 1$. In the adiabatic limit $\tau_{ij}\to
\infty$ the term $\bra \sigma_i\sigma_j\ket_{\rm sp}$,
representing spin correlations associated with the coupling
$J_{ij}$, becomes an average over the Boltzmann distribution
(\ref{eq:prob}) of the spins, given the instantaneous couplings
$\bJ$. The $\eta_{ij}(t)$ represent Gaussian white noise
contributions,  of zero mean and covariance $\bra
\eta_{ij}(t)\eta_{kl}(t')\ket =
         2 T_{ij}\delta _{ik}\delta _{jl}\delta (t-t')$,
with associated temperature $T_{ij} = \beta_{ij}^{-1}$.
Appropriate factors of $N$ have been introduced in order to ensure
non-trivial behaviour in the limit $N \rightarrow \infty$. We
classify the spin pairs $(i,j)$ according to the characteristic
time-scale $\tau_{ij}$ and the control parameters
$(T_{ij},\mu_{ij})$ associated with their interactions $J_{ij}$.
In contrast to papers such as
\cite{CPS,PCS,PS,Jongenetal1,Jongenetal2}, where $\tau_{ij}=\tau$,
$T_{ij}=\tilde{T}$ and $\mu_{ij}=\mu$ for all $(i,j)$, here the
various time-scales, temperatures and decay rates are no longer
assumed to be identical,  but to come in $L$ distinct
adiabatically separated groups $I_\ell$ (always with $i<j$):
\be
I_{\ell}=\{(i,j)|~\tau_{ij}=\tau_\ell,~T_{ij}=T_\ell,~\mu_{ij}=\mu_\ell\}
, l=1, \cdots, L,
\ee with $1 \ll \tau_1 \ll\tau_2\ll\ldots\ll\tau_L$. Thus
$\{(i,j)\}=\bigcup_{\ell\leq L}I_\ell$. We will write the set of
spin-interactions with time-scale $\tau_\ell$ as
$\bJ^\ell=\{J_{ij}|~(i,j)\in I_\ell\}$. The interactions in group
$I_2$ are adiabatically slow compared to those in group $I_1$, and
so on. The rationale of this set-up is that, in information
processing terms, this would represent a stochastic
self-programming neural information processing system equipped
with a program which consists of a hierarchy of increasingly less
volatile and less easily modifiable sub-routines.

Finally we have to define the detailed partitioning of the
$\frac{1}{2}N(N-1)$ interactions into the $L$ volatility groups.
We introduce $\epsilon_{ij}(\ell)\in\{0,1\}$ such that
$\epsilon_{ij}(\ell)=1$ if and only if $(i,j)\in I_\ell$, so
$\sum_{\ell=1}^L \epsilon_{ij}(\ell)=1$ for all $(i,j)$. In order
to arrive at a solvable mean-field problem, with full equivalence
of the $N$ sites, we will choose the $\epsilon_{ij}(\ell)$
independently at random for each pair $(i,j)$ with $i<j$, with
probabilities
\be
{\rm
Prob}[\epsilon_{ij}(\ell)=1]=\epsilon_\ell~~~~~~~~\sum_{\ell=1}^L\epsilon_\ell=1
\label{eq:epsilons} \ee These time-scale and temperature
allocation variables $\{\epsilon_{ij}(\ell)\}$ thus introduce
quenched disorder into our problem. Averaging over this disorder
will be denoted by $\overline{\cdots}$, as usual.

\section{Replica Analysis of the Stationary State}

\subsection{Equilibrium Statistical Mechanics of the Couplings}

We denote averages over the probability distribution of the
couplings at level $\ell$ in the hierarchy (for which
$\tau_{ij}=\tau_\ell$, $T_{ij}=T_\ell$ and $\mu_{ij}=\mu_\ell$) as
$\bra \ldots\ket_\ell$. At every level $\ell$, the stochastic
equation (\ref{def:evcouplings}) for those couplings which evolve
on that particular time-scale $\tau_\ell$ has now become
conservative:
\begin{eqnarray}
\tau_{\ell} \frac{d}{dt}J_{ij}&=&
                \frac{1}{N}\bra\ldots\bra~\bra
                \sigma_i\sigma_j\ket_{\rm sp}\ket_{1}\ldots\ket_{\ell-1}
                -\mu_\ell J_{ij}
                +\eta_{ij}(t)\sqrt{\frac{\tau_\ell}{N}}
\nonumber \\
&=&-\frac{1}{N}\ptl{J_{ij}} H_\ell(\bJ^\ell,\ldots,\bJ^L)
+\eta_{ij}(t)\sqrt{\frac{\tau_\ell}{N}}
\label{eq:conservative}
\end{eqnarray}
with the following effective Hamiltionian for the couplings at level $\ell$:
\begin{equation}
        H_1(\bJ^1,\ldots,\bJ^L)=-\beta^{-1}\log Z[\bJ^1,\ldots,\bJ^L]
\label{eq:couplinghamiltonian1}
\end{equation}
\begin{equation}
        H_{\ell+1}(\bJ^{\ell+1},\ldots,\bJ^L)=-\beta^{-1}_\ell\log
Z_\ell[\bJ^{\ell +1},\ldots,\bJ^L]
        ~~~~~~(1 \le \ell < L)
\label{eq:couplinghamiltonian2}
\end{equation}
and with the partition functions
\begin{equation}
        Z[\bJ^1,\ldots,\bJ^L]=\sum_{\bsigma}e^{-\beta H(\bsigma,\bJ)}
\label{eq:Z1}
\end{equation}
\begin{equation}
        Z_{\ell}[\bJ^{\ell+1},\ldots,\bJ^L]=
        \int\!d\bJ^\ell~e^{-\beta_\ell H_\ell(\bJ^l,\ldots,\bJ^L)}
        ~~~~~~(1 \le \ell < L) \label{eq:Z2}
\end{equation}
\begin{equation}
        Z_{L}=
        \int\!d\bJ^L~e^{-\beta_L H_L( \bJ^L)}
\label{eq:Z3}
\end{equation}
in which
\begin{equation}
H(\bsigma,\bJ)=-\sum_{i<j}J_{ij}\sigma_i\sigma_j +\frac12 N\sum_{i<j}\mu_{ij} J_{ij}^2
\label{eq:overallhamiltonian}
\end{equation}
This describes a hierarchy of nested equilibrations. At each
time-scale $\tau_\ell$ the interactions $\bJ^\ell$ equilibrate to
a Boltzmann distribution, with an effective Hamiltonian $H_\ell$
which is the free energy of the previous (faster)  level $\ell+1$
in the hierarchy, starting from the overall Hamiltonian
(\ref{eq:overallhamiltonian}) (for spins and couplings) at the
fastest (spin) level.
 As a result of having
different effective temperatures $T_\ell$ associated with each
level, the partition functions of subsequent levels are found to
generate replica theories with replica dimensions $m_\ell\geq 0$
which represent the ratios of the effective temperatures of the
two levels involved. This follows from substitution of
(\ref{eq:couplinghamiltonian2}) into (\ref{eq:Z2}):
\begin{equation}
        Z_{\ell}[\bJ^{l+1},\ldots,\bJ^L]=
        \int\!d\bJ^\ell~\left\{Z_{\ell-1}[\bJ^l,\ldots,\bJ^L]\right\}
^{m_{\ell}}
        \label{eq:iterateZ}
\end{equation}
\begin{equation}
        m_{\ell}=\beta_\ell/\beta_{\ell-1}
        ~~~~~~~~
        m_{1}=\beta_1/\beta
\label{eq:dimensions}
\end{equation}
The statics of the system, including the effect of the quenched
disorder,
 are governed by the disorder-averaged free energy ${\cal F}$
associated with the partition function $Z_L$ in (\ref{eq:Z3}),
where the slowest variables have finally been integrated out:
\begin{equation}
{\cal F}=-\frac{1}{\beta_L}\overline{\log Z_L}
=-\lim_{m_{L+1}\to 0}\frac{1}{m_{L+1}\beta_L}\log \overline{Z_L^{m_{L+1}}}
\label{eq:F}
\end{equation}
This function is found to act as the general generator of equilibrium values for
observables at any level in the hierarchy, since upon adding suitable
generating terms to the Hamiltonian (\ref{eq:overallhamiltonian}),
i.e.
$H(\bsigma,\bJ)\to H(\bsigma,\bJ)+\lambda \Phi(\bsigma,\bJ)$,
one finds:
\be
\overline{\bra\ldots\bra~\bra\Phi(\bsigma,\bJ)\ket_{\rm
sp}\ket_{1}\ldots\ket_{L}} =\lim_{\lambda\to
0}\frac{\partial}{\partial \lambda}{\cal F} \ee We can now combine
our previous results and write down an explicit expression for
$\cal F$, involving multiple replications due to
(\ref{eq:iterateZ},\ref{eq:F}). We find ${\cal F}=\lim_{m_{L+1}\to
0}F[m_1,\ldots,m_{L+1}]$, with
$F[m_1,\ldots,m_{L+1}]=-(m_{L+1}\beta_L)^{-1}\log
\overline{Z_1^{m_{L+1}}}$, and where $Z_L^{m_{L+1}}$ is written as
\begin{eqnarray}
 \hspace*{-0mm} Z_L^{m_{L+1}} &=&
 \int\!\left[\prod_{\alpha_{L+1}}\! d\bJ^{L,\alpha_{L+1}}\right]
 \prod_{\alpha_{L+1}}\!\left\{
 Z_{L-1}[\bJ^{L,\alpha_{L+1}}]\right\}^{m_{L}}
\nonumber\\
 &=&
 \int\!\left[\prod_{\alpha_{L+1}}\! d\bJ^{L,\alpha_{L+1}}\!\ldots
 \!\prod_{\alpha_2,\ldots,\alpha_{L+1}}\!
 d\bJ^{1,\alpha_2,\ldots,\alpha_{L+1}}\right]
 \nonumber
 \\
 && \hspace*{20mm}
 \times \prod_{\alpha_2,\ldots,\alpha_{L+1}}\!
 \left\{Z[\bJ^{1,\alpha_2,\ldots,\alpha_{L+1}},\!\ldots,\bJ^{L,\alpha_{L+1}}
 ]\right\}^{m_{1}}
 \nonumber \\
 &=&
 \sum_{\{\bsigma^{\alpha_1,\ldots,\alpha_{L+1}}\}}
 \int\!\left[\prod_{\alpha_{L+1}}\! d\bJ^{L,\alpha_{L+1}}\!\ldots
 \!\prod_{\alpha_2,\ldots,\alpha_{L+1}}\!d\bJ^{1,\alpha_2,\ldots,\alpha_{L+1}}
 \right]
 \nonumber
 \\
 &&
 \hspace*{20mm}
 \times \prod_{\alpha_1,\ldots,\alpha_{L+1}}\!
 e^{-\beta
 H(\bsigma^{\alpha_1,\ldots,\alpha_{L+1}},
 \bJ^{1,\alpha_2,\ldots,\alpha_{L+1}},\ldots,\bJ^{L,\alpha_{L+1}}
 )}
 \nonumber \\
 &=&
 \sum_{\{\bsigma^{\alpha_1,\ldots,\alpha_{L+1}}\}}\prod_{\ell=1}^L
 \prod_{(i<j)\in I_\ell}\prod_{\alpha_{\ell+1},\ldots,\alpha_{L+1}}
 \nonumber
 \\
 &&
 \hspace*{5mm}
 \times\int\!dz~
 e^{
  -\frac12\beta\mu_\ell N[m_{1}\ldots m_{\ell}z^2
 -2(\mu_\ell N)^{-1}z\sum_{\alpha_{1}\ldots\alpha_{\ell}}
 \sigma_i^{\alpha_1,\ldots,\alpha_{L+1}}\sigma_j^{\alpha_1,\ldots,
 \alpha_{L+1}}]
 }
 \nonumber\end{eqnarray}
\bd \hspace*{-7mm}
=
 \sum_{\{\bsigma^{\alpha_1,\ldots,\alpha_{L+1}}\}}
 e^{ \frac{\beta}{2 N}\sum_{\ell\leq L}\frac{1}{m_1 \cdots
 m_{\ell}\mu_\ell}
 \sum_{(i<j)\in I_\ell}\sum_{\alpha_{\ell+1},\ldots,\alpha_{L+1}}
  [\sum_{\alpha_{1}\ldots\alpha_{\ell}}
 \sigma_i^{\alpha_1,\ldots,\alpha_{L+1}}
 \sigma_j^{\alpha_1,\ldots,\alpha_{L+1}}] }
\ed
\be
 \label{eq:Finreplicas}
\ee
 (modulo irrelevant constants), in which always
$\alpha_\ell=1,\ldots,m_\ell$.

\subsection{Disorder Averaging}

In order to average (\ref{eq:Finreplicas}) over the disorder,
we note that the spin summation in the exponent can also be written in terms of the
allocation variables $\epsilon_{ij}(\ell)$:
\bd
\sum_{\ell\leq L}\frac{1}{m_1 \cdots m_{\ell}\mu_\ell}
\sum_{(i<j)\in
I_\ell}\sum_{\alpha_{l+1},\ldots,\alpha_{L+1}}\ldots= \ed \bd
\hspace*{20mm}
 \sum_{i<j}\sum_{\ell \leq L}
\frac{\beta}{\beta_\ell}\frac{\epsilon_{ij}(\ell)} {\mu_\ell}
\sum_{\alpha_{\ell+1},\ldots,\alpha_{L+1}}
 [\sum_{\alpha_{1}\ldots\alpha_{\ell}}
\sigma_i^{\alpha_1,\ldots,\alpha_{L+1}}
\sigma_j^{\alpha_1,\ldots,\alpha_{L+1}}]^2
\ed where we used $m_{1}\ldots m_{\ell}=\beta_\ell/\beta$. This
allows us to carry out the average over the (independent)
$\epsilon_{ij}(\ell)$ in (\ref{eq:Finreplicas}):
\begin{eqnarray}
F[\ldots]
&=&
-\frac{1}{m_{L+1} \beta_L }\log
\sum_{\{\bsigma^{\alpha_1,\ldots,\alpha_{L+1}}\}}
\nonumber\\
&&\times~\prod_{i<j}\overline{\left[e^{
 \frac{\beta^2}{2 N}
\sum_{\ell\leq L} \frac{\epsilon_{ij}(\ell)}{\beta_\ell \mu_\ell}
\sum_{\alpha_{\ell+1},\ldots,\alpha_{L+1}}
 [\sum_{\alpha_{1}\ldots\alpha_{\ell}}
\sigma_i^{\alpha_1,\ldots,\alpha_{L+1}}\sigma_j^{\alpha_1,\ldots,\alpha_{L+1}}]^2
}\right]}
\nonumber\\
&=&
-\frac{1}{m_{L+1} \beta_L}\log
\sum_{\{\bsigma^{\alpha_1,\ldots,\alpha_{L+1}}\}}
 e^{\order(N^0)}
\nonumber\\
&&\times~\prod_{i<j}\left[1+
 \frac{\beta^2}{2 N}
\sum_{\ell\leq L} \frac{\overline{\epsilon_{ij}(\ell)}}{\beta_\ell
\mu_\ell} \sum_{\alpha_{\ell+1},\ldots,\alpha_{L+1}}
 [\sum_{\alpha_{1}\ldots\alpha_{\ell}}
\sigma_i^{\alpha_1,\ldots,\alpha_{L+1}}\sigma_j^{\alpha_1,\ldots,\alpha_{L+1}}]^2
\right]
\nonumber
\\
&=&
 -\frac{1}{m_{L+1}\beta_L}\log
 \sum_{\{\bsigma^{\alpha_1,\ldots,\alpha_{L+1}}\}}
 e^{\order(N^0)}
\nonumber\\&&
 \times~\exp \left\{ \frac{N \beta^2}{4}\sum_{\alpha_1,\ldots,\alpha_{L+1}}
 \sum_{\beta_1,\ldots,\beta_{L+1}} \sum_{\ell\leq
 L}\frac{\epsilon_\ell}{\beta_\ell \mu_\ell}
 \delta_{(\alpha_{\ell+1},\ldots,\alpha_{L+1}),(\beta_{\ell+1},\ldots,\beta_{L+1})}
\right. \nonumber \\ && \hspace*{30mm}\left.\times~ [
\frac{1}{N}\sum_i \sigma_i^{\alpha_1,\ldots,\alpha_{L+1}}
\sigma_i^{\beta_1,\ldots,\beta_{L+1}}]^2 \right\} \nonumber
\end{eqnarray}
(again modulo irrelevant constants).
We abbreviate $a=(\alpha_1,\ldots,\alpha_{L+1})$,
and introduce the  spin-glass replica order parameters
$q_{ab}=
\frac{1}{N}\sum_i
\sigma_i^{\alpha_1,\ldots,\alpha_{L+1}}
\sigma_i^{\beta_1,\ldots,\beta_{L+1}}$ by inserting appropriate
integrals over $\delta$-distributions, and arrive at
\begin{eqnarray}
F[\ldots]
&=&
-\frac{1}{m_{L+1}\beta_L}\log\int\!\{dq_{ab}
d\hat{q}_{ab}\}e^{NG[\{q_{ab},\hat{q}_{ab}\}]+\order(N^0)}
\\
G[\{q_{ab},\hat{q}_{ab}\}] &=& i\sum_{ab}\hat{q}_{ab} q_{ab}
+\frac{\beta ^2}{4} \sum_{\ell\leq L}\frac{\epsilon_\ell}{\beta_
\ell \mu_\ell}\sum_{ab}
\delta_{(\alpha_{\ell+1},\ldots,\alpha_{L+1}),(\beta_{\ell+1},\ldots,\beta_{L+1})}q_{ab}^2
\nonumber
\\
&&+
\log \sum_{\{\sigma_{\alpha_1,\ldots,\alpha_{L+1}}\}}
e^{-i\sum_{ab}\hat{q}_{ab}
\sigma_{\alpha_1,\ldots,\alpha_{L+1}}\sigma_{\beta_1,\ldots,\beta_{L+1}}}
\end{eqnarray}
For $N\to\infty$ the above integral can be evaluated by steepest
descent, and upon elimination of the conjugate order parameters
$\{\hat{q}_{ab}\}$ by variation of $\{q_{ab}\}$, the
disorder-averaged free energy per spin $f=\lim_{N\to\infty}{\cal
F}/N$ is found to be
\begin{eqnarray}
f&=&\lim_{m_{L+1}\to 0} \frac{1}{m_{L+1}\beta_L}\left\{
\frac{\beta ^2}{4} \sum_{\ell\leq L}\frac{\epsilon_\ell}{\beta
_\ell\mu_\ell}\sum_{ab}
\delta_{(\alpha_{\ell+1},\ldots,\alpha_{L+1}),(\beta_{\ell+1},\ldots,\beta_{L+1})}
q_{ab}^2 \right. \nonumber \\ && \left.
-
\log \sum_{\{\sigma_{a}\}} e^{\frac{\beta ^2}{2}\sum_{\ell\leq
L}\frac{\epsilon_\ell} {\beta _\ell \mu_\ell}\sum_{ab}
\delta_{(\alpha_{\ell+1},\ldots,\alpha_{L+1}),
(\beta_{\ell+1},\ldots,\beta_{L+1})}q_{ab} \sigma_{a}\sigma_{b}}
\right\} \label{eq:RSBfreeenergy}
\end{eqnarray}
The saddle-point equations from which to solve the $\{q_{ab}\}$
are given by
\begin{eqnarray}
q_{cd} & = & \frac{ \sum_{ \{ \sigma_{a} \} } \sigma_c \sigma_d~
e^{ \frac{\beta^2}{2} \sum_{\ell \leq L} \epsilon_\ell / (\beta
_\ell \mu_\ell) \sum_{ab} \delta_{
(\alpha_{\ell+1},\ldots,\alpha_{L+1}
),(\beta_{\ell+1},\ldots,\beta_{L+1} )} q_{ab} \sigma_{a}
\sigma_{b} }} {  \sum_{ \{ \sigma_{a} \} } e^{ \frac{\beta^2}{2}
\sum_{\ell \leq L} \epsilon_\ell / (\beta _\ell \mu_\ell)
\sum_{ab}
\delta_{(\alpha_{\ell+1},\ldots,\alpha_{L+1}),(\beta_{\ell+1},\ldots,\beta_{L+1})}
q_{ab} \sigma_{a} \sigma_{b} }} \label{eq:RSBequations}
\end{eqnarray}
with our short-hand
$a=(\alpha_1,\ldots,\alpha_{L+1})$, and with
$\alpha_{\ell}\in \{1,\ldots,m_\ell \}$
for all $\ell$, where the dimensions $m_\ell$ are given by the
ratios of the temperatures of subsequent programming levels in the
hierarchy according to (\ref{eq:dimensions}).

\section{Replica Symmetric Solutions}

\subsection{Evaluation of the Replica-Symmetric Free Energy}

 Our full order parameters are $q_{ab}=
\frac{1}{N}\sum_i \sigma_i^{\alpha_1,\ldots,\alpha_{L+1}}
\sigma_i^{\beta_1,\ldots,\beta_{L+1}}$.
Note that $\alpha_{L+1}=\beta_{L+1}$.  Spin variables with
$\alpha_{L}=\beta_{L}$ have identical level-$L$ bonds. Those with
$\alpha_{L}=\beta_{L}$ and $\alpha_{L-1}=\beta_{L-1}$ have
identical level $L$ bonds  {\em and} identical level $L\!-\!1$
bonds, and so on. Hence, for the present model the
replica-symmetric (RS) ansatz (describing ergodicity at each level
of the hierarchy of time-scales) takes the following form:
\begin{eqnarray*}
 \alpha_L\!\neq \!\beta_L\!: &~~~& q_{ab}=q_L,\\
(\alpha_{\ell+1},\ldots,\alpha_{L+1})\!=\!(\beta_{\ell+1},\ldots,\beta_{L+1}),
~~\alpha_{\ell}\!\neq\! \beta_{\ell}\!: &~~~& q_{ab}=q_{\ell},\\
a=b: &~~~& q_{ab}=1
\end{eqnarray*}
or
\be
q_{ab}=\sum_{\ell=1}^{L}q_{\ell}~
\delta_{(\alpha_{\ell+1},\ldots,\alpha_{L+1})(\beta_{\ell+1},\ldots,
\beta_{L+1})} \overline{\delta}_{\alpha_{\ell}
\beta_{\ell}}+\delta_{ab}
\label{eq:RS}
\ee where
$\overline{\delta}_{ij}=1-\delta_{ij}$, and where
\be
0\le q_L \leq \ldots \leq q_{1} \leq 1
\ee
With a modest amount of foresight we introduce the abbreviation
\be
 \pi_\ell=\sum_{\ell^\prime=\ell}^{L}\frac{\beta
 \epsilon_{\ell^\prime}}{\beta_{\ell^\prime}\mu_{\ell^\prime}}
 =\sum_{\ell^\prime=\ell}^{L}\frac{1}{m_{1}\ldots m_{\ell^\prime}}
 ~\frac{\epsilon_{\ell^\prime}}{\mu_{\ell^\prime}}
 \label{eq:pis}
\ee
Insertion of the ansatz (\ref{eq:RS}) into
(\ref{eq:RSBfreeenergy}) gives (using the relation
$\beta_L=\beta\prod_{\ell=1}^{L}m_\ell$, and with the notational
convention $q_{L+1}=0$ to simplify summations):
\begin{eqnarray*}
\hspace*{-5mm}
 f
 &=& \frac{1}{4}\pi_1 + \lim_{m_{L+1}\to 0}
 \frac{1}{m_{L+1}\beta_L}\left\{\frac{\beta }{4} \sum_{\ell\leq
 L}\frac{\beta \epsilon_\ell}{\beta_\ell\mu_\ell} \sum_{ab}
 \sum_{\ell^\prime=1}^{L}q^2_{\ell^\prime}
 \delta_{(\alpha_{\ell^\prime +1},\ldots,\alpha_{L+1}),
 (\beta_{\ell^\prime +1},\ldots,\beta_{L+1})}
 \overline{\delta}_{\alpha_{\ell^\prime}\beta_{\ell^\prime}}
 \right. \nonumber
\\
&&
 \left. - \log \sum_{\{\sigma_{a}\}}
 e^{\frac{\beta}{2} \sum_{\ell\leq L}\frac{\beta \epsilon_\ell}
 {\beta_\ell \mu_\ell}\sum_{ab}
 \sum_{\ell^\prime=\ell}^{L}q_{\ell^\prime}
 \delta_{(\alpha_{\ell^\prime +1},\ldots,\alpha_{L+1}),
 (\beta_{\ell^\prime +1},\ldots,\beta_{L+1})}
 \overline{\delta}_{\alpha_{\ell^\prime},\beta_{\ell^\prime}}
 \sigma_{a}\sigma_{b}} \right\}
\nonumber\\
&=&
 \frac{1}{4}\pi_1 +\frac{1}{4}\sum_{\ell=2}^{L}q^2 _{\ell} \pi_\ell
 m_1\cdots m_{\ell-1}(m_\ell \minus 1)+\frac{1}{4}q_1^2 \pi_1  (m_1 \minus 1)
\nonumber\\
&&
 -\lim_{m_{L+1}\to 0} \frac{1}{ m_{L+1}\beta _L}
 \log\left\{ \sum_{\{\sigma_{a}\}}
e^{ \frac{\beta \pi _1}{2} m_1
 \cdots m_{L+1} +\frac{\beta}{2} \sum_{\ell=1}^{L}
 q_{\ell}\pi_{\ell} \sum_{\alpha _{\ell+1},\cdots, \alpha _{L+1}}
 (\sum_{\alpha _{1},\cdots, \alpha_{\ell}} \sigma _a)^2}
 \right.
\nonumber \\ &&
 \left.\hspace*{10mm}
 \times ~e^{-\frac{\beta}{2} \sum_{\ell=1}^{L} q_{\ell+1}\pi_{\ell+1}
 \sum_{\alpha _{\ell+1},\cdots, \alpha _{L+1}}
 (\sum_{\alpha_{1},\cdots, \alpha_{\ell}} \sigma _a)^2 -\frac{\beta}{2}q_1 \pi_1
 \sum_{\alpha _{1},\cdots, \alpha_{L+1}} (\sigma _a)^2 }
 \right\}
\nonumber \\
&=&
 -\frac{1}{4}\pi_1
 +\frac{1}{4}\sum_{\ell=2}^{L}q^2 _{\ell} \pi_\ell m_1\cdots
 m_{\ell-1}(m_\ell \minus 1)+\frac{1}{4}q_1^2 \pi_1  (m_1 \minus 1)
 +\frac{1}{2}q_{1}\pi_1
\nonumber \\ &&
 -\lim_{m_{L+1}\to 0}
 \frac{1}{ m_{L+1}\beta _L} \log \sum_{\{\sigma_{a}\}} e^{
 \frac{\beta}{2} \sum_{\ell=1}^{L} (q_{\ell}\pi_{\ell}
 -q_{\ell+1}\pi_{\ell+1})
  \sum_{\alpha_{\ell+1},\cdots, \alpha _{L+1}}
 (\sum_{\alpha_{1},\cdots, \alpha _{\ell}} \sigma _a)^2 }
\nonumber\\
 &=&
  -\frac{1}{4}\pi_1
 +\frac{1}{4}\sum_{\ell=2}^{L}q^2 _{\ell} \pi_\ell
 m_1\cdots m_{\ell-1}(m_\ell \minus 1)+\frac{1}{4}q_1^2 \pi_1  (m_1 \minus 1)
 +\frac{1}{2}q_{1}\pi_1
 \nonumber \\
 &&
 -\lim_{m_{L+1}\to 0}\frac{1}{ m_{L+1}\beta _L}
 \log\left\{
 \int\!
 \left[\prod_{\ell=1}^{L}\prod_{\alpha_{\ell+1}\ldots\alpha_{L+1}}
 \!\!\!Dz_{\alpha_{\ell+1}\ldots\alpha_{L+1}}\right]
\right.\nonumber \\ && \left.
 \hspace*{10mm}\times~
  \prod_{a}\left[ 2\cosh[
 \sum_{\ell=1}^{L}z_{\alpha_{\ell+1}\ldots\alpha_{L+1}}
 \sqrt{\beta(q_{\ell}\pi_\ell-q_{\ell+1}\pi_{\ell+1})}]
 \right]
 \right\}
 \end{eqnarray*}
Alternatively, we can reduce the number of Gaussian integrations
in this expression to a smaller number of nested ones:
\begin{eqnarray*}
\hspace*{-5mm} f &=&
 -\frac{1}{4}\pi_L-\frac{1}{\beta} \log 2
 +\frac{1}{4}\sum_{\ell=2}^{L}q^2 _{\ell} \pi_\ell
 m_1\cdots m_{\ell-1}(m_\ell \minus 1)+\frac{1}{4}q_1^2 \pi_1  (m_1 \minus 1)
 +\frac{1}{2}q_{1}\pi_1
 \nonumber \\
 &&
 -\lim_{m_{L+1}\to 0}\frac{1}{ m_{L+1}\beta _L}
 \log\left\{
 \int\!
 \left[\prod_{\ell=2}^{L}\prod_{\alpha_{\ell+1}\ldots\alpha_{L+1}}
 \!\!\!Dz_{\alpha_{\ell+1}\ldots\alpha_{L+1}}
 \right]\right.
 \nonumber \\
 && \left.\times\!
 \prod_{\alpha_2,\ldots,\alpha_{L}}\!\!
 Dz_1 \cosh^{m_{1}}
 \left[
 z_1 \sqrt{\beta(q_1 \pi_1 \minus q_{2}\pi_{2})}
 \!+
 \sum_{\ell=2}^{L}z_{\alpha_{\ell+1}\ldots\alpha_{L+1}}
 \sqrt{\beta(q_{\ell}\pi_\ell\minus q_{\ell+1}\pi_{\ell+1})}
 \right]\right\}\\
& =&
-\frac{1}{4}\pi_L-\frac{1}{\beta} \log 2 +
 \frac{1}{4}\sum_{\ell=2}^{L}q^2 _{\ell} \pi_\ell m_1\cdots
 m_{\ell-1}(m_\ell \minus 1)+\frac{1}{4}q_1^2 \pi_1  (m_1 \minus 1)
 +\frac{1}{2}q_{1}\pi_1
\nonumber \\ &&
  -\frac{1}{
 \beta m_1 \cdots m_L} \log  \int\! Dz_L
\nonumber\\ &&
 \hspace*{10mm}\times \left\{
 \ldots
  \left\{
 \int\! Dz_1 \cosh^{m_{1}}
 \left[
 \sum_{\ell=1}^{L}z_\ell \sqrt{\beta(q_{\ell}\pi_\ell- q_{\ell+1}\pi_{\ell+1})}
 \right]
 \right \} ^{m_2}
 \ldots
 \right \} ^{m_L}
\end{eqnarray*}
For $L=1$ this  reduces to (modulo the irrelevant constant):
\begin{eqnarray}
\hspace*{-3mm} f&=& \frac{1}{2}q_1 \pi_1 +\frac{1}{4}q_1^2 \pi_1
(m_{1}\!-\!1) -\frac{1}{\beta} \log 2 -\frac{1}{\beta
m_1}\log\int\!Dz_1 ~ \cosh^{m_{1}}\left[z_1 \sqrt{\beta q_{1}
\pi_1}\right] \nonumber
\\ \label{eq:f_L1}
\end{eqnarray}
whereas for $L>1$ we can simplify our result to (modulo the
irrelevant constant):
\begin{eqnarray}
\hspace*{-3mm} f&=& \frac{1}{2}\pi_1 q_1-\frac{1}{\beta} \log 2
 +\frac{1}{4}\sum_{\ell=2}^{L}q^2_{\ell}\pi_\ell
 \left[\prod_{k=1}^{\ell-1}m_{k}\right](m_\ell-1) +\frac{1}{4}q_1^2
 \pi_1(m_1\!-\!1)
\nonumber \\ &&
 -\frac{1}{\beta\prod_{\ell=1}^{L}m_{\ell}}\log \int\!Dz_L \left\{
 \int\!Dz_{L-1} \left\{ \ldots \left\{ \int\!Dz_1
 \right.\right.\right.
\nonumber \\ &&
 \left.\left.\left.
 \left\{ \cosh\left[ z_L\sqrt{\beta q_{L}\pi_L}
 +\!\sum_{\ell=1}^{L-1}z_\ell \sqrt{\beta(q_{\ell}\pi_{\ell}\minus
 q_{\ell+1}\pi_{\ell+1})} \right] \right\}^{m_1} \right\}^{m_{2}}
 \!\!\!\! \ldots \right\}^{m_{L-1}} \right\}^{m_L}
\label{eq:FRS2}
\end{eqnarray}
For systems with a single coupling time-scale (\ref{eq:f_L1}) we
again observe the similarity with the free energy of Parisi's
RSB-1 ansatz \cite{Parisi} (see also e.g. \cite{PS,MC}). For
systems with multiple coupling time-scales (\ref{eq:FRS2}) this
similarity is lost: Parisi's RSB-$L$ solution \cite{Parisi,MC}
emerges only when $\pi_\ell=\pi_L$ for all $\ell$, i.e. when we
return to a single coupling time-scale.

\subsection{The Single-Level Benchmark}

A simple and convenient benchmark test of the above equations is
obtained upon putting $\epsilon_\ell\to \delta_{r\ell}$ for some
$r\in\{1,\ldots,L\}$. Here we retain only a single bond
time-scale, and our theory should effectively reduce to that of
\cite{CPS}. For $L=1$ this is indeed true, according to equation
(\ref{eq:f_L1}); here we will address the generic case $L >1$ and
$1 <r < L$. The $\epsilon_\ell$ occur only in the quantities
$\pi_\ell$ of (\ref{eq:pis}), which now simplify to
\be
\pi_{\ell\leq r}=\frac{\epsilon_r }{\mu_r m_1 \cdots
m_r}~~~~~~~~\pi_{\ell
> r}=0 \ee Insertion into the replica-symmetric disorder-averaged
free energy per spin (\ref{eq:FRS2}) gives
\begin{eqnarray*}
 f&=& \frac{1}{2}\pi_r q_1-\frac{1}{\beta} \log 2 +\frac{1}{4}\pi_r
 \sum_{\ell=2}^{r}q^2_{\ell}
 \left[\prod_{k=1}^{\ell-1}m_{k}\right](m_\ell-1) +\frac{1}{4}q_1^2
 \pi_r(m_1\!-\!1) \nonumber \\ &&
 -\frac{1}{\beta\prod_{\ell=1}^{r}m_{\ell}}\log \int\!Dz_r \left\{
 \int\!Dz_{r-1} \left\{ \ldots \left\{ \int\!Dz_1
 \right.\right.\right.
\nonumber \\ &&
 \left.\left.\left. \times
 \left\{ \cosh\left[ z_r\sqrt{\beta q_{r}\pi_r}
 +\sum_{\ell=1}^{r-1}z_\ell \sqrt{\beta(q_{\ell}\;\minus\;
 q_{\ell+1})\pi_{r}} \right] \right\}^{m_1} \right\}^{m_{2}} \!\!\!
 \ldots \right\}^{m_{r-1}} \right\}^{m_r}
\end{eqnarray*}
Since $q_L$ measures the overlap between spin states with
different level $L$ bonds, $q_{L-1}$ measures the overlap between
spin states with identical level $L$ bonds but different level
$(L- 1)$ bonds, and so on, the relevant state for
$\epsilon_\ell=\delta_{r\ell}$ must be one where $q_\ell=q_r$ for
all $\ell\leq r$. Insertion into the above expression gives
\begin{eqnarray*}
f &=& \frac{1}{2}\pi_r q_r-\frac{1}{\beta} \log 2
+\frac{1}{4}\pi_r q^2_{r} (\prod_{k=1}^{r}m_{k} -1) \nonumber \\
&& -\frac{1}{\beta\prod_{\ell=1}^{r}m_{\ell}}\log \int\!Dz_r
\left\{ \cosh\left[z_r\sqrt{\beta q_{r}\pi_r} \right]
\right\}^{m_1 m_2 \cdots m_r}\\
\end{eqnarray*}
This expression is indeed equivalent to (\ref{eq:f_L1}), from
which it can be obtained by making the replacements $\pi_1\to
\pi_r$ and $m_1 \to m_1 \cdots m_r$.

\subsection{Nature of the Physical Saddle-Point}

Due to our previous elimination of (imaginary) conjugate order
parameters and the possible curvature sign changes occurring in
replica theories, we can no longer be sure that the relevant
saddle-point (\ref{eq:f_L1},\ref{eq:FRS2}) gives the minimum of
$f$. In order to allow us to select the physical saddle-point in
situations where multiple saddle-points exist, we now first
determine the nature of the physical saddle-point by inspection of
the high-temperature state. Since there are multiple temperatures
in this problem (one for the spins, and $L$ for the various bond
levels), there are also different ways to send all temperatures to
infinity, with potentially different outcomes. Here we consider
the limit which appears most natural, where $\beta\to 0$ for fixed
$\{m_1,\ldots,m_L\}$. We then obtain from expression
(\ref{eq:FRS2}), with $\tilde{f}=f+\frac{1}{\beta}\log 2$:
\begin{eqnarray}
 \hspace*{-5mm} \lim_{\beta\to 0}\tilde{f} &=& \frac{1}{2}\pi_1 q_1
 +\frac{1}{4}\sum_{\ell=2}^{L}q^2_{\ell}\pi_\ell
 \left[\prod_{k=1}^{l-1}m_{k}\right](m_\ell-1) +\frac{1}{4}q_1^2
 \pi_1(m_1\!-\!1) \nonumber\\ && -\lim_{\beta\to
 0}\frac{1}{\beta\prod_{\ell=1}^{L}m_{\ell}}\log \int\!Dz_L \left\{
 \int\!Dz_{L-1} \left\{ \ldots \left\{1+ \frac{1}{2}\beta m_1
 \room
 \right.\right.\right.
\nonumber \\ && \left.\left.\left. \hspace*{-4mm}
 \times\int\!Dz_{1}\left[
 z_L\sqrt{q_{L}\pi_L}
 +\sum_{\ell=1}^{L-1}z_\ell\sqrt{q_{\ell}\pi_{\ell}\minus
 q_{\ell+1}\pi_{\ell+1}} \right]^2\!\!\!+\order(\beta^2) \right\}^{m_{2}}
 \!\!\! \ldots \right\}^{m_{L-1}} \right\}^{m_L}
 \nonumber \\
 &=&
 \frac{1}{2}\pi_1 q_1
 +\frac{1}{4}\sum_{\ell=2}^{L}q^2_{\ell}\pi_\ell
 \left[\prod_{k=1}^{\ell-1}m_{k}\right](m_\ell-1)
 +\frac{1}{4}q_1^2 \pi_1(m_1\!-\!1)
 +\order(\beta^2)
 \nonumber \\
 &&
 -\lim_{\beta\to
 0}\frac{1}{\beta\prod_{\ell=1}^{L}m_{\ell}}\log\left\{
 1+\frac{1}{2}\beta [\prod_{\ell=1}^L m_\ell]\left[
  q_{L}\pi_L+\sum_{\ell=1}^{L-1}(q_{\ell}\pi_{\ell}
 \minus q_{\ell+1}\pi_{\ell+1})
 \right]
 \right\}
 \nonumber
\\
&=& \frac{1}{4}\sum_{\ell=2}^{L}q^2_{\ell}\pi_\ell
\left[\prod_{k=1}^{\ell-1}m_{k}\right](m_\ell-1) +\frac{1}{4}q_1^2
\pi_1(m_1\!-\!1)+\order(\beta^2)
\label{eq:highT}
\end{eqnarray}
This shows explicitly that the nature of the physical state, for
$\beta\to 0$ given by  $q_\ell=0$ for all $\ell$, depends on the
specific choice made for the various coupling temperatures, which
determine the replica dimensions $\{m_\ell\}$ via
(\ref{eq:dimensions}). This is in sharp contrast with the standard
Parisi ansatz \cite{Parisi} where one always has $m_\ell\leq 1$.
Here we find:
\be
\begin{array} {lll}
m_\ell > 1:&& {\rm minimisation~of}~f~{\rm with~respect~to}~q_\ell\\
m_\ell < 1:&& {\rm maximisation~of}~f~{\rm with~respect~to}~q_\ell
\end{array}
\label{eq:extremisation}
\ee

\subsection{The Full Saddle-Point Equations}

In this section  we derive the full saddle point equations of
(\ref{eq:FRS2}) for the replica-symmetric order parameters $0\leq
q_L\leq q_{L-1}\leq \cdots q_1 \leq 1$, in the general $L$-level
hierarchy. This is most efficiently done by first writing the free
energy (\ref{eq:FRS2}) as
\begin{eqnarray}
f& = &
 \frac{1}{2}\pi_1 q_1-\frac{1}{\beta} \log 2
 +\frac{1}{4}\sum_{\ell=2}^L q^2_{\ell}\pi_\ell
 \left[\prod_{k=1}^{\ell -1}m_{k}\right](m_\ell -1)
 +\frac{1}{4}q_1^2 \pi_1(m_1\!-\!1)
 \nonumber\\
 &&\hspace*{50mm}
 -[\beta\prod_{\ell=1}^{L}m_{\ell}]^{-1}\log K
\label{eq:f_iterative}
\\
K&=& \int\!Dz_L \left\{
 \int\!Dz_{L-1} \left\{ \ldots \left\{ \int\!Dz_1
\left\{ \cosh\Xi \right\}^{m_1} \right\}^{m_{2}}
 \!\!\!\! \ldots \right\}^{m_{L-1}} \right\}^{m_L}
\label{eq:K_L}
\\
\Xi &=& z_L\sqrt{\beta q_{L}\pi_L}
 +\!\sum_{\ell=1}^{L-1}z_\ell \sqrt{\beta(q_{\ell}\pi_{\ell}\minus
 q_{\ell+1}\pi_{\ell+1})}=\!\sum_{\ell=1}^{L}a_lz_l,
\label{eq:Xi}\\
a_{\ell(<L)} &=& \sqrt{\beta(q_{\ell}\pi_{\ell}\minus
 q_{\ell+1}\pi_{\ell+1})},\;\; a_{L} = \sqrt{\beta q_L \pi_L}
\label{eq:aell}
\end{eqnarray}
From this it follows, upon taking derivatives of $f$ with respect
to $q_1$ and $q_{\ell>1}$, respectively, that the saddle-point
equations can be written as
\begin{eqnarray}
 0& = & \frac{1}{2}\pi_1
 +\frac{1}{2}q_1 \pi_1(m_1\!-\!1) -\frac{1}{\beta m_1 \cdots
 m_L}\frac{1}{K} \frac{\partial K }{\partial q_1}
\label{eq:spRS1}\\ 0& = &
 \frac{1}{2}q_\ell \pi_\ell [\prod _{k=1}^{\ell-1} m_k](m_\ell\!-\!1)
 -\frac{1}{\beta m_1 \cdots m_L}\frac{1}{K} \frac{\partial K
 }{\partial q_\ell},~~~~~ {\rm for}~~ \ell >1.
\label{eq:spRS2}
\end{eqnarray}
The remaining problem is to calculate the various derivatives of
$K$, which is complicated by the nesting of integrals. Note that
$K$ can be written as $K=K_L$, where $K_L$ is defined iteratively:
\be
K_0 = \cosh \Xi,~~~~~~~~ K_L= \int\!Dz_{L} K_{L-1} ^{m_L}
\ee
 To suppress notation we also define
\be
M_0=1,~~~~~~ M_{\ell>0} = \prod_{\ell^\prime=1}^\ell
m_{\ell^\prime},~~~~~~
 \bra f\ket_\ell=
K_\ell^{-1}\int\!Dz_\ell~K_{\ell-1}^{m_\ell} f(z_\ell) \ee In
\ref{app:saddle} we show that, with these short-hands, the
relevant derivatives of $K$ are given by the following
expressions:
\begin{eqnarray}
  \frac{\partial K_L}{\partial q_1} & = &
\frac{1}{2}\beta \pi_1  M_L
K \left\{ 1 + (m_1 -1)  \bra \cdots \bra\tanh^2 \Xi\ket_1 \cdots \ket_L \right\}\\
  \frac{\partial K_L}{\partial q_{\ell>1}}
&=&  \frac{1}{2}\beta\pi_\ell M_L M_{\ell-1}(m_\ell-1)K \bra
\cdots \bra
 [\bra \cdots \bra  \tanh \Xi\ket_1  \cdots \ket_{\ell-1}]^2
 \ket_{\ell} \cdots  \ket_{L}
\end{eqnarray}
Hence the saddle-point equations (\ref{eq:spRS2},\ref{eq:spRS2})
reduce to the, in view of the definition (\ref{eq:RS}), appealing
and transparent   relations
\begin{eqnarray}
 q_1 &=&\bra\ldots \bra \tanh^2\Xi\ket_1\ldots \ket_L
\label{eq:spRSfinal1}
\\
q_{\ell>1} &=&\bra \ldots \bra ~[\bra\ldots\bra
\tanh\Xi\ket_1\ldots\ket_{\ell-1}]^2~
 \ket_{\ell} \ldots  \ket_{L}
\label{eq:spRSfinal2}
\end{eqnarray}

\section{The Two-Level Hierarchy: $L=2$}

\subsection{General Properties}

We next apply our results to the case where we have just two
time-scales in the bond dynamics, and calculate the phase diagram.
We found in studying the single level limit that our theory will
simply self-consistently `forget' about non-existent intermediate
levels (as it should), and substitute the right temperature
definitions in the expressions that would have been obtained if we
had considered subsequent levels. Hence we may choose $L=2$,
without loss of generality, and simply put $L=2$ in expression
(\ref{eq:FRS2}):
\begin{eqnarray}
\hspace*{-15mm} f_{L=2}(q_1, q_2)& = & \frac{1}{2}\pi_1
 q_1-\frac{1}{\beta} \log 2 +\frac{1}{4} q^2_2\pi_2 m_1( m_2-1)
 +\frac{1}{4}q_1^2 \pi_1(m_1\!-\!1)
\nonumber \\ && \hspace*{-20mm}
 -\frac{1}{\beta
 m_1 m_2 } \log
  \int\!Dz_{2} \left\{  \int\!Dz_{1}
 \left\{ \cosh\left[
 z_2\sqrt{\beta q_2\pi_2}
 +z_1\sqrt{\beta(q_1\pi_1\minus q_{2}\pi_2)}\right]
 \right\}^{m_1}
 \right\}^{m_{2}}
\label{eq:FRS_L2}
\end{eqnarray}
For $L=2$ the saddle-point equations
(\ref{eq:spRSfinal1},\ref{eq:spRSfinal2}) reduce in explicit form
to
\begin{eqnarray}
 q_1 &=&
 \frac{\int\!Dz_2[\int\!Dz_1~\cosh^{m_1}\Xi]^{m_2}
    \left[\frac{\int\!Dz_1~\cosh^{m_1}\Xi\tanh^2\Xi}{\int\!Dz_1~\cosh^{m_1}\Xi}\right]
    }
 {\int\!Dz_2[\int\!Dz_1~\cosh^{m_1}\Xi]^{m_2}}
\label{eq:spRS_L2a}
\\
 q_2 &=&
 \frac{\int\!Dz_2[\int\!Dz_1~\cosh^{m_1}\Xi]^{m_2}
    \left[\frac{\int\!Dz_1~\cosh^{m_1}\Xi\tanh\Xi}{\int\!Dz_1~\cosh^{m_1}\Xi}\right]^2
    }
 {\int\!Dz_2[\int\!Dz_1~\cosh^{m_1}\Xi]^{m_2}}
\label{eq:spRS_L2b}
\end{eqnarray}
(note: $q_1\geq q_2$), with \bd
 \Xi= z_2\sqrt{\beta q_{2}\pi_2}
 +z_1 \sqrt{\beta(q_{1}\pi_{1}-
 q_{2}\pi_{2})}
\ed We can distinguish between three phases. The first is a
paramagnetic phase  ($P$), corresponding to the solution
$q_1=q_2=0$ of (\ref{eq:spRS_L2a},\ref{eq:spRS_L2b});
according to (\ref{eq:highT})
it is the sole saddle-point of (\ref{eq:FRS_L2}) in the high
temperature regime, as it should. The second is a spin-glass phase
($SG_1$), corresponding to a solution of the form $q_1>0,~q_2=0$,
describing a (spin-glass type) state with freezing of spins and
couplings at time-scale $\tau_1$ but not on the time-scale
$\tau_2\gg\tau_1$. Spins and level-1 couplings `freeze' into a
state determined by the level-2 couplings, but the latter slowly
but continually evolve, given sufficient time. Insertion of
$q_2=0$ into (\ref{eq:spRS_L2a},\ref{eq:spRS_L2b}) shows that in
this $SG_1$ phase $q_1$ is the solution of
\begin{eqnarray}
 q_1 &=&
    \frac{
    \int\!Dz~\cosh^{m_1}(z\sqrt{\beta q_{1}\pi_{1}})\tanh^2(z\sqrt{\beta
    q_{1}\pi_{1}})}
    {\int\!Dz~\cosh^{m_1}(z\sqrt{\beta q_{1}\pi_{1}})}
\label{eq:spSG1}
\end{eqnarray}
In the third phase ($SG_2$) one has $q_1\neq 0$ and $q_2\neq 0$,
and the system freezes on all time-scales. Now one has to solve
(\ref{eq:spRS_L2a},\ref{eq:spRS_L2b}) in full. We indicate the
various potential transition temperatures separating these phases
as follows:
\begin{eqnarray}
 T^{\rm 2nd}_{p,1}: &~~~ P\to SG_1    &~~~ {\rm 2nd~order}\nonumber \\
 T^{\rm 2nd}_{1,2}: &~~~ SG_1\to SG_2 &~~~ {\rm 2nd~order}\label{eq:L2_define_lines}\\
 T^{\rm 1st}_{p,1}: &~~~ P\to SG_1 &~~~ {\rm 1st~order}\nonumber\\
 T^{\rm 1st}_{p,2}: &~~~ P\to SG_2 &~~~ {\rm 1st~order}\nonumber
\end{eqnarray}
We will find that this list contains all transitions occurring.

\subsection{$P\to SG_1$ Transitions}

The locations and properties of the two types of $P\to SG_1$
transitions follow upon expanding $f(q_1,0)$ in powers of $q_1$:
\bd
 -\beta m_1 f(q_1,0)= F_0 + a(T) q_1 ^2 + b(T)q_1 ^3+ c(T) q_1^4
 +\cdots
\ed
The coefficients are found to be
\begin{eqnarray}
 a(T)&=&\frac{1}{4} m_1 (m_1-1)\beta  \pi_1(\beta \pi_1-1)
\nonumber\\
 b(T)&=&\frac{1}{6} m_1 (m_1-1)(m_1-2)(\beta \pi_1)^3
\nonumber \\
 c(T)&=& \frac{1}{96}m_1(\beta  \pi_1)^4  (3m_1^3-72
m_1 ^2 +128 m_1-68).
\end{eqnarray}
Note that $c(T)<0$ for $0 < m_1 \le m_+$, where $m_+ \approx 15$.
A continuous bifurcation occurs when $a(T)=0$, i.e. when
$T=\pi_1$, so
\be
T_{p,1}^{\rm 2nd}=\epsilon_1/m_1\mu_1+\epsilon_2/m_1m_2\mu_2
\label{eq:trans_L2_PSG1_2nd}
\ee
This bifurcation can, however, dependent on the values of
$\{m_1,m_2\}$, be preceded by a discontinuous one.  Upon varying
$m_1$ one finds the following scenario:
\begin{tabbing}
zzzzzzzzzzzzzz\= zzzzzzzzzzzzzz\= zzzzzzzzzzzz\=\kill
\vspace*{1mm}
 \> $0<m_1<1$: \> $b(T)>0$,
     \> $f(q_1,0)$  maximal at $q_1=0$ for $T>T_{p,1}^{\rm 2nd}$\\
 \>\>\> $f(q_1,0)$  maximal at $q_1>0$ for $T<T_{p,1}^{\rm 2nd}$\\[2mm]
 \> $1<m_1<2$: \> $b(T)<0$,
     \> $f(q_1,0)$  minimal at $q_1=0$ for $T>T_{p,1}^{\rm 2nd}$\\
 \>\>\> $f(q_1,0)$  minimal at $q_1>0$ for $T<T_{p,1}^{\rm 2nd}$\\[2mm]
 \> $2<m_1$:   \> $b(T)>0$,
     \> $f(q_1,0)$  minimal at $q_1=0$ for $T>T_{p,1}^{\rm 2nd}$\\
 \>\>\> $f(q_1,0)$  minimal at $q_1>0$ for $T<T_{p,1}^{\rm 1st}$
 \vspace*{1mm}
\end{tabbing}
where $T_{p,1}^{\rm 1st}$ signals a
discontinuous (i.e. first order) transition, with $T_{p,1}^{\rm 1st}>T_{p,1}^{\rm 2nd}$. The condition for
this latter transition follows from putting the derivative of the
right-hand side of (\ref{eq:spSG1}) with respect to $q_1$
equal to unity, giving
\begin{eqnarray}
1&=&
 \frac{1}{2}\beta \pi_1 \left[ \room m_1(1-m_1)q_1^2
 + 4(m_1-2) q_1 +2
 \right.
 \nonumber\\
&& \left. +~(m_1-2) (m_1-3)
 \frac{
    \int\!Dz~\cosh^{m_1}(z\sqrt{\beta q_{1}\pi_{1}})\tanh^4(z\sqrt{\beta
    q_{1}\pi_{1}})}
    {\int\!Dz~\cosh^{m_1}(z\sqrt{\beta q_{1}\pi_{1}})}
  \right]
\label{eq:trans_L2_PSG1_1st}
\end{eqnarray}
Together with (\ref{eq:extremisation}) we may now conclude that
the $P\to SG_1$ transition is second order for $m_1<2$, in which
case it is given by (\ref{eq:trans_L2_PSG1_2nd}), and first order
for $m_1>2$, in which case it is given by the solution of
(\ref{eq:trans_L2_PSG1_1st}). In a subsequent section we will show
that for $m_1\to\infty$ the first order transition temperature
$T_{p,1}^{\rm 1st}$ tends to a finite and nonzero value.

\subsection{Transitions to $SG_2$}

There are found to be two such transition temperatures:
$T_{p,2}^{\rm 1st}$ (where one goes from a paramagnetic state to
one with $q_1\geq q_2>0$) and $T_{1,2}^{\rm 2nd}$ (where one goes
from $q_1>q_2=0$ to $q_1\geq q_2>0$). The transition condition
defining $T_{1,2}^{\rm 2nd}$  follows from putting the derivative
of the right-hand side of (\ref{eq:spRS_L2b}) with respect to $q_2$
equal to unity,
followed by putting $q_2=0$, which gives
\be
 1= \sqrt{\beta \pi_2}[1+(m_1-1)q_1]
\label{eq:trans_L2_SG1SG2_2nd} \ee We will show in \ref{app:asym} that
$T_{1,2}^{\rm 2nd} \propto  \sqrt{T/m_1}=\sqrt{T_1}$ as $m_1\to \infty$.

To determine $T_{p,2}^{\rm 1st}$, in contrast, one must demand an
instability of (\ref{eq:spRS_L2a},\ref{eq:spRS_L2b}) with
$q_{1,2}>0$. This implies meeting the more general condition
\be
(\frac{\partial \varphi_1}{\partial q_1}-1) (\frac{\partial
\varphi _2}{\partial q_2}-1) = \frac{\partial \varphi_1}{\partial
q_2} \frac{\partial \varphi_2}{\partial q_1}
\label{eq:trans_L2_PSG2_1st} \ee where $\varphi_1(q_1,q_2)$ and
$\varphi_2(q_1,q_2)$ denote the right-hand sides of
(\ref{eq:spRS_L2a}) and (\ref{eq:spRS_L2b}), respectively. For
$m_1\to \infty$ we will show, as with $T_{p,1}^{\rm 1st}$,  that
also $T_{p,2}^{\rm 1st}$ tends to a finite and nonzero value, and
that $T_{p,2}^{{\rm (1st)}} >T_{p,1}^{{\rm (1st)}}$. Hence,
$T_{p,2}^{\rm 1st}$ will ultimately become the true physical transition as  $m_1\to \infty$
(or, equivalently, $T_1\to 0$ for fixed $T$).

\subsection{The $L=2$ Phase Diagram}

For $L=2$ there are still five independent control parameters in
our model, viz. $T$, $m_1$, $m_2$, $\pi_1$ and $\pi_2$ (where
$\pi_2=\epsilon_{2}/m_1 m_2\mu_2$ and
$\pi_1=\pi_2+\epsilon_{1}/m_1\mu_{1}$) or any combination of these.
Hence we can only present representative cross-sections of the
full phase diagram. In this section we will restrict ourselves to
showing the transition lines in the $(T_1,T)$ plane, solved
numerically from the various appropriate conditions as derived
above, for the choice $\epsilon_1/\mu_1=1$, $\epsilon_2/\mu_2=2$ with
$m_2\in\{0.5, 1.5\}$ (which was found to cover more or less the generic
scenarios).
According to (\ref{eq:trans_L2_PSG1_2nd}),
for this choice of parameters
the second order $P\to SG_1$ transition temperature $T_{p,1}^{\rm 2nd}$ will be at
\be
T_{p,1}^{\rm 2nd}=\frac{1}{m_1}(1+\frac{2}{m_2})
\ee
and it is replaced by a first order $P\to SG_1$ transition at
$m_1=2$, which, together with the relation $m_1=T/T_1$, tells us that the
first and second order $P\to SG_1$ lines meet at
\begin{eqnarray}
T_{p,1}^{\rm 1st}=T_{p,2}^{\rm 2nd}: &~~~~& T_1=\frac{1}{4}(1+\frac{2}{m_2})
\label{eq:change_point}
\end{eqnarray}
In section \ref{sec:asymptotics} we will prove that, for any $L$,
the discontinuous transition temperatures $T_{p,1}^{{\rm 1st}}$ and $T_{p,2}^{{\rm 1st}}$ always tend to
a finite non-zero value as $m_1\to \infty$, whereas $\lim_{m_1\to
\infty} T_{p,1}^{{\rm 2nd}}=\lim_{m_1\to\infty}T_{1,2}^{{\rm
 2nd}}=0$.
Thus, for sufficiently small $T_1$ (i.e. large $m_1$) the first order transitions are always the ones
which will actually take place.
Secondly, we will find that there is always a critical value $T_c^{(0)}$ for
$T_1$ such that $T_{1,2}^{\rm 2nd}=0$ at $T_1=T_c^{(0)}$, with $T_{1,2}^{\rm 2nd}$
simply absent for $T_1>T_c^{(0)}$;
as a consequence, the phase $SG_2$ no longer exists for $T_1 > T_c^{(0)}$.
These two properties will be shown to be specific cases of more
general results concerning spin-glass phases of arbitrary order.
\vsp

\begin{figure}[t]
\vspace*{1mm} \setlength{\unitlength}{1.04mm}
\begin{picture}(150,85)
 \put(7,20){\epsfxsize=91\unitlength\epsfbox{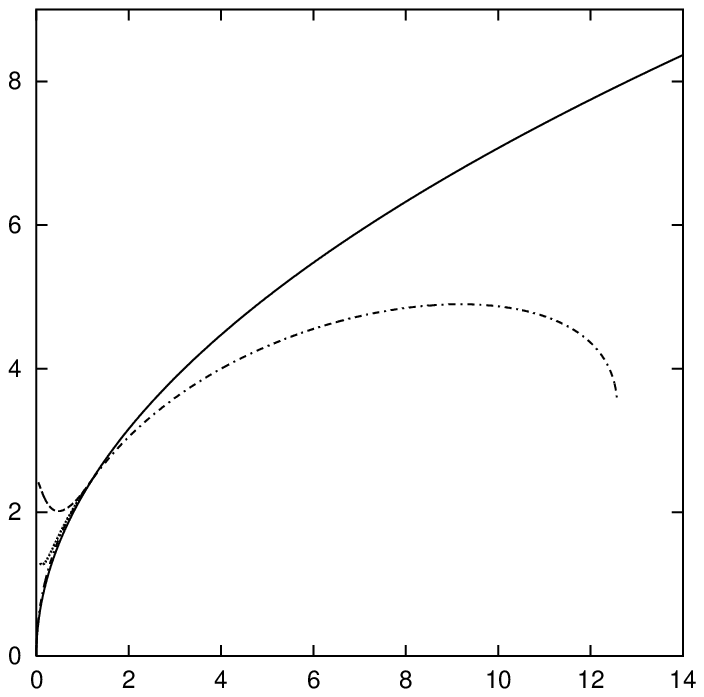}}
 \put(84,20){\epsfxsize=91\unitlength\epsfbox{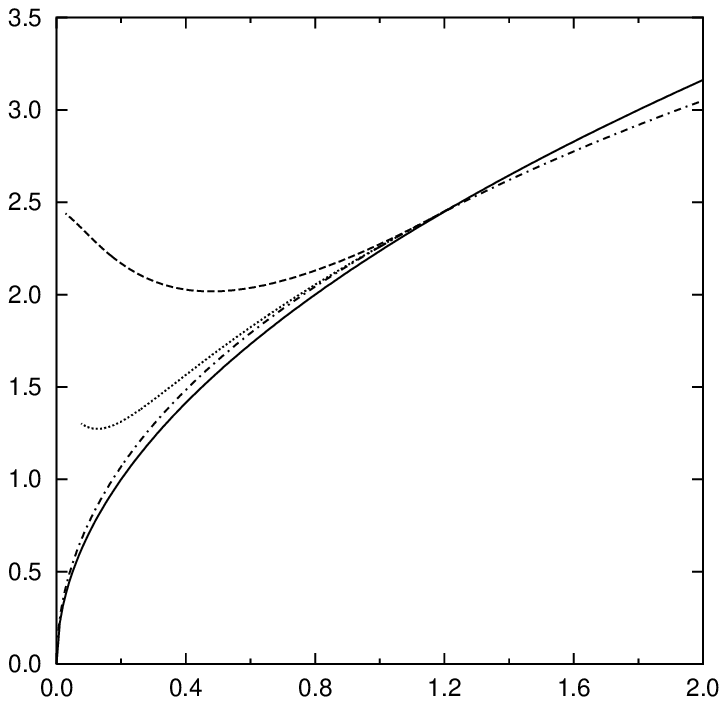}}
 \put(0,52){ $T$}   \put(39,13){$T_1$}
 \put(80,52){$T$} \put(119,13){$T_1$}
 \put(25,70){$P$} \put(55,60){$SG_1$} \put(45,35){$SG_2$}
 \put(105,70){$P$} \put(94,50){$SG_1$} \put(122,35){$SG_2$}
\end{picture}
\vspace*{-16mm}
\caption{Phase diagram in the $(T_1,T)$ plane for
$L=2$, $\epsilon_1/\mu_1=1$, $\epsilon_2/\mu_2=2$ and $m_2=0.5$,
obtained
by solving numerically the equations defining the various
bifurcation lines (note: $T_1=T/m_1$). Left panel: the large-scale
structure. Right panel: enlargement of the low $T_1$ (or high
$m_1$) region, where the first order transitions are found.
The relevant phases are $P$ (paramagnetic phase, $q_1=q_2=0$), $SG_1$
(first spin-glass phase, $q_1>0,~q_2=0$), and $SG_2$ (second
spin-glass phase, $q_1>0,~q_2>0$). The transition lines shown, as
defined in (\ref{eq:L2_define_lines}), are $T_{p,1}^{\rm 2nd}$
(solid curve), $T_{1,2}^{\rm 2nd}$ (dot-dash curve), $T_{p,1}^{\rm
1st}$ (dashed curve), and $T_{p,2}^{1st}$ (dotted curve).
Note: the curve $T_{1,2}^{\rm 2nd}$ jumps to zero discontinuously
at $T_1=T_c^{(0)}\approx 12.575$ (see main text).
} \label{fig:1}
\end{figure}

\begin{figure}[t]
\vspace*{1mm} \setlength{\unitlength}{1.04mm}
\begin{picture}(150,85)
 \put(37,20){\epsfxsize=91\unitlength\epsfbox{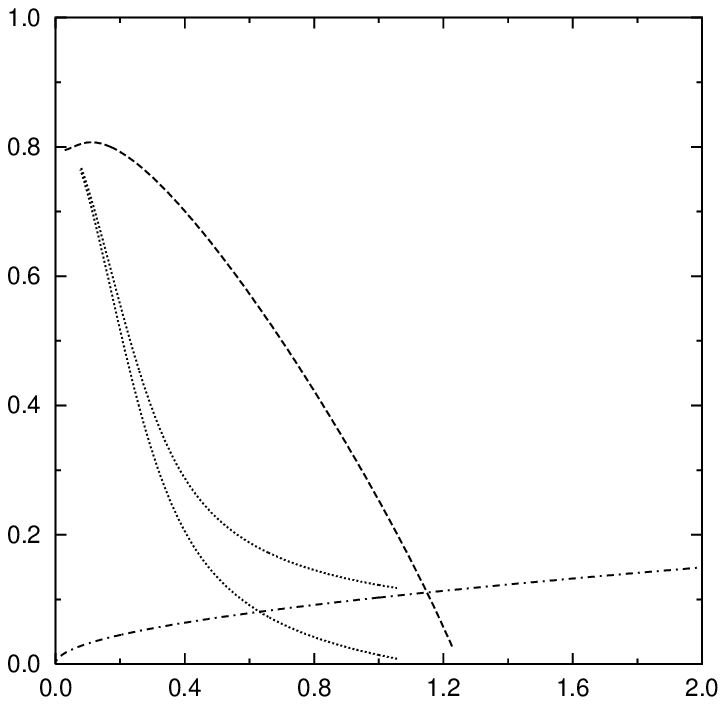}}
 \put(31,52){ $q_{1,2}$}   \put(71,13){$T_1$}
\end{picture}
\vspace*{-16mm}
\caption{Values of the order parameters $q_1$ and $q_2$ as functions of $T_1$,
along some of the transition lines of the previous figure, for
$L=2$, $\epsilon_1/\mu_1=1$, $\epsilon_2/\mu_2=2$ and $m_2=0.5$.
 The values shown are $q_1$ along $T_{1,2}^{\rm 2nd}$ (dot-dash curve), $q_1$ along $T_{p,1}^{\rm
1st}$ (dashed curve), and both $q_1$ and $q_2$ for $T_{p,2}^{1st}$ (dotted curves).
Note: the order parameters which are not shown in this figure are zero by
definition.
} \label{fig:2}
\end{figure}

Let us first turn to $m_2=0.5$, i.e. relatively high noise levels
in the level-2 couplings compared to that of the spins.
In figure \ref{fig:1} we show the phase diagram and the various bifurcation lines
of the $L=2$
system for $m_2=0.5$, as obtained by numerical solution of the relevant
equations. The actual physical transitions are the ones which
occur first as the temperature $T$ is lowered from within the
paramagnetic region. It will be clear from the figure that upon lowering $T$ one can
find different types of transition sequences, dependent on the
value of $T_1$, which prompts us to define the following critical
values for $T_1$ (these depend on the values chosen for the control parameters):
\begin{eqnarray*}
 T_c^{(0)}\approx 12.575&:~~~~~~& {\rm creation~point~of~phase}~SG_2
\nonumber\\
 T_c^{(1)}=1.25         &:& {\rm creation~point~of~}T_{p,1}^{\rm 1st},~~~ T_{p,1}^{\rm 2nd}=T_{p,1}^{\rm 1st}
\nonumber\\
 T_c^{(2)} \approx 1.15 &:& T_{p,1}^{\rm 1st}~ {\rm and}~ T_{1,2}^{\rm 2nd}~ {\rm
 touch~tangentially}
\nonumber \\
 T_c^{(3)} \approx 1.06 &:& {\rm creation~point~of~} T_{p,2}^{\rm 1st},~~~ T_{1,2}^{\rm 2nd}=T_{p,2}^{\rm 1st}
\end{eqnarray*}
Note: $T_c^{(1)}$ is the point given by expression
(\ref{eq:change_point}). In terms of these critical values we can
classify the different transition scenarios encountered upon
reducing $T$ down to zero, starting in the paramagnetic region, as
follows:
\begin{eqnarray*}
 T_c^{(0)} < T_1:            &~~~~ P \to SG_1~{\rm (2nd~order)}
\\
 T_c^{(1)} < T_1 < T_c^{(0)}:  &~~~~ P \to SG_1~{\rm (2nd~order)}, ~~~&~~ SG_1 \to SG_2 ~{\rm (2nd~order)}
\\
 T_c^{(2)} < T_1 < T_c^{(1)}:&~~~~ P\to SG_1~{\rm (1st~order)}, ~~~&~~ SG_1 \to SG_2 ~{\rm (2nd~order)}
\\
 T_c^{(3)} < T_1<T_c^{(2)}:&~~~~ P\to SG_1~{\rm (1st~order)}, ~~~&~~ SG_1\to SG_2~{\rm (2nd~order)}
\\
 T_1 < T_c^{(3)}:           &~~~~ P\to SG_1~{\rm (1st~ order)}, ~~~&~~ SG_1\to  SG_2 ~{\rm (1st~ order)}
\end{eqnarray*}
In figure \ref{fig:2} we show the corresponding values of the
order parameters along the various transition lines, as functions
of $T_1$ (with line types identical to those used in figure
\ref{fig:1} for the same transitions), except for those order
parameters  which are zero by definition (such as the values of
$q_2$ when bifurcating continuously from zero). Curves in figures
\ref{fig:1} and \ref{fig:2} which terminate for small but nonzero
values of $T_1$ do so due to computational limitations; they can
be shown to extend down to $T_1=0$. We will calculate the values
of the order parameters in the $m_1\to \infty$ (i.e. $T_1\to 0$)
limit analytically in a subsequent section.

The general tendency in the phase diagram, as expected, is for the
system to have more phases, increasingly discontinuous transitions
separating them, and hence the most interesting physics, for small
values of $T_1$ (i.e. large $m_1$). We should emphasize once more
in this context that in our model we have the freedom to choose
$m_1>1$ (or $T_1<T$), in contrast to the standard Parisi solution
for ordinary complex systems \cite{Parisi} (where one always has
$m_1\leq 1$). \vsp

\begin{figure}[t]
\vspace*{1mm} \setlength{\unitlength}{1.04mm}
\begin{picture}(150,85)
 \put(7,20){\epsfxsize=91\unitlength\epsfbox{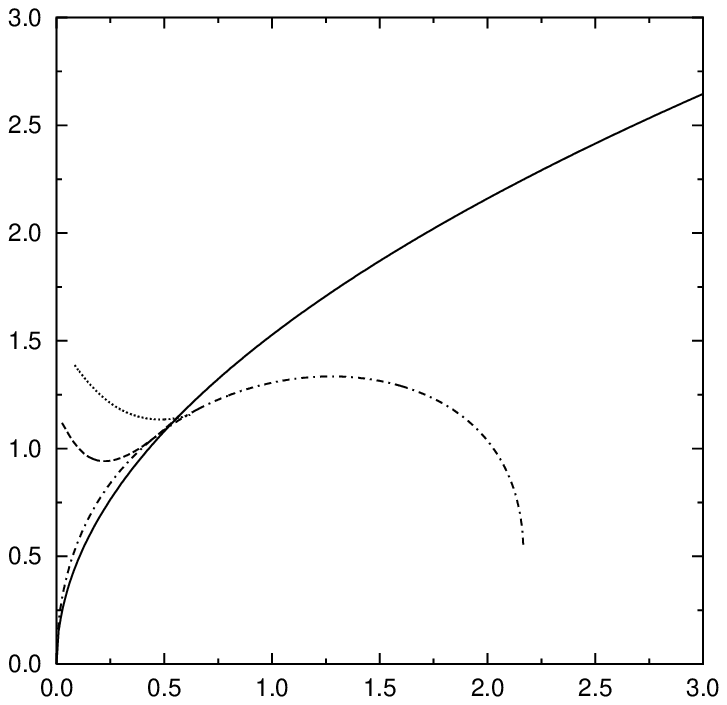}}
 \put(84,20){\epsfxsize=91\unitlength\epsfbox{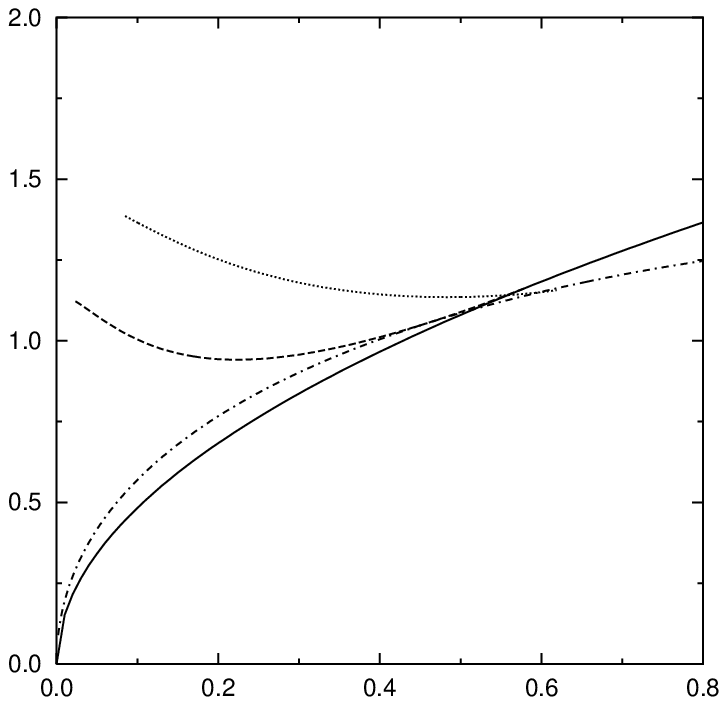}}
 \put(0,52){ $T$}   \put(39,13){$T_1$}
 \put(80,52){$T$} \put(119,13){$T_1$}
 \put(25,70){$P$} \put(55,55){$SG_1$} \put(35,31){$SG_2$}
 \put(105,70){$P$} \put(99,53){$SG_2$} \put(122,35){$SG_2$}
\end{picture}
\vspace*{-16mm} \caption{Phase diagram in the $(T_1,T)$ plane for
$L=2$, $\epsilon_1/\mu_1$, $\epsilon_2/\mu_2=2$ and $m_2=1.5$,
obtained by solving numerically the equations defining the various
bifurcation lines (note: $T_1=T/m_1$). Left panel: the large-scale
structure. Right panel: enlargement of the low $T_1$ (or high
$m_1$) region, where the first order transitions are found. The
relevant phases are $P$ (paramagnetic phase, $q_1=q_2=0$), $SG_1$
(first spin-glass phase, $q_1>0,~q_2=0$), and $SG_2$ (second
spin-glass phase, $q_1>0,~q_2>0$). The transition lines shown, as
defined in (\ref{eq:L2_define_lines}), are $T_{p,1}^{\rm 2nd}$
(solid curve), $T_{1,2}^{\rm 2nd}$ (dot-dash curve), $T_{p,1}^{\rm
1st}$ (dashed curve), and $T_{p,2}^{1st}$ (dotted curve). Note:
the curve $T_{1,2}^{\rm 2nd}$ jumps to zero discontinuously at
$T_1=T_c^{(0)}\approx 2.17$ (see main text). } \label{fig:3}
\end{figure}

\begin{figure}[t]
\vspace*{1mm} \setlength{\unitlength}{1.04mm}
\begin{picture}(150,85)
 \put(37,20){\epsfxsize=91\unitlength\epsfbox{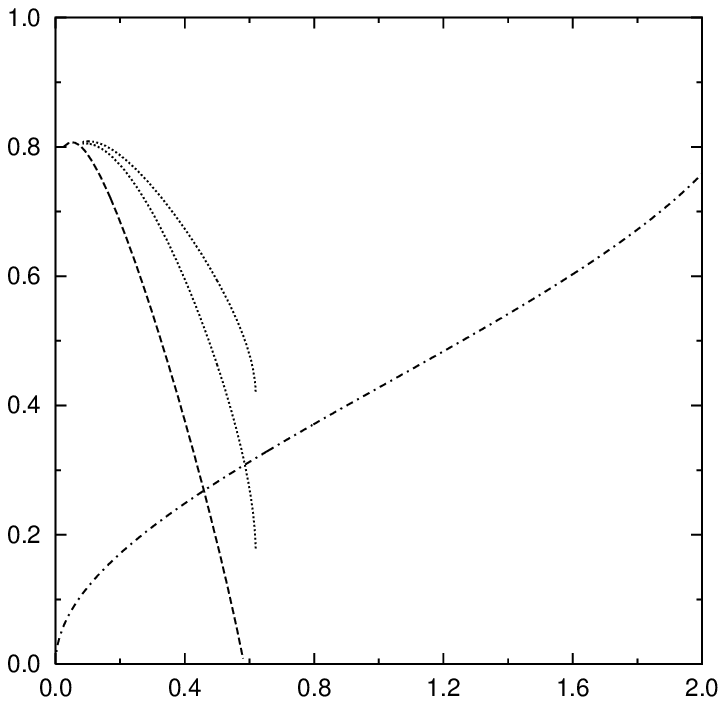}}
 \put(31,52){ $q_{1,2}$}   \put(71,13){$T_1$}
\end{picture}
\vspace*{-16mm}
\caption{Values of the order parameters $q_1$ and $q_2$ as functions of $T_1$,
along some of the transition lines of the previous figure, for
$L=2$, $\epsilon_1/\mu_1=1$, $\epsilon_2/\mu_2=2$ and $m_2=1.5$.
 The values shown are $q_1$ along $T_{1,2}^{\rm 2nd}$ (dot-dash curve), $q_1$ along $T_{p,1}^{\rm
1st}$ (dashed curve), and both $q_1$ and $q_2$ for $T_{p,2}^{1st}$
(dotted curves). Note: the order parameters which are not shown in
this figure are zero by definition. } \label{fig:4}
\end{figure}

Let us next turn to $m_2=1.5$, i.e. relatively low noise levels in
the level-2 couplings compared to that of the spins. In figure
\ref{fig:3} we show the phase diagram and the various bifurcation
lines of the $L=2$ system for $m_2=1.5$, as obtained by numerical
solution of the relevant equations.  It is again clear from the
figure that upon lowering $T$ one can find different types of
transition sequences, dependent on the value of $T_1$. Now the
relevant critical values for $T_1$ are as follows:
\begin{eqnarray*}
 T_c^{(0)}\approx 2.17&:~~~~~~& {\rm creation~point~of~phase}~SG_2
\nonumber\\
 T_c^{(1)} \approx 0.597 &:& {\rm creation~point~of~}
             T_{p,2}^{\rm 1st},~~~ T_{1,2}^{\rm 2nd}=T_{p,2}^{\rm 1st}
\nonumber\\
 T_c^{(2)}= 7/12\approx 0.583 &:& T_{p,1}^{\rm 2nd}=T_{p,1}^{\rm 1st}
\nonumber\\
 T_c^{(3)} \approx 0.5572 &:& T_{p,1}^{\rm 1st}=T_{p,2}^{\rm 1st}
\end{eqnarray*}
Note: $T_c^{(2)}$ is the point given by expression
(\ref{eq:change_point}). In terms of these critical values we can
classify the different transition scenarios encountered upon
reducing $T$, as before, as follows:
\begin{eqnarray*}
 T_c^{(0)} < T_1:            &~~~~ P \to SG_1~{\rm (2nd~order)}
\\
 T_c^{(1)} < T_1 < T_c^{(0)}:  &~~~~ P \to SG_1~{\rm (2nd~order)}, ~~~&~~ SG_1 \to SG_2 ~{\rm (2nd~order)}
\\
 T_c^{(2)} < T_1 < T_c^{(1)}:&~~~~ P\to SG_1~{\rm (2nd~order)}, ~~~&~~ SG_1 \to SG_2 ~{\rm (1st~order)}
\\
 T_c^{(3)} < T_1 < T_c^{(2)}:&~~~~ P\to SG_1~{\rm (1st~order)}, ~~~&~~
SG_1 \to SG_2 ~{\rm (1st~order)}
\\
  T_1<T_c^{(3)}:&~~~~ P\to SG_2~{\rm (1st~order)}
\end{eqnarray*}
In figure \ref{fig:4} we show the corresponding values of the
order parameters along the various transition lines, as functions
of $T_1$ (with line types identical to those used in figure
\ref{fig:3} for the same transitions), except for those which are
zero by definition. Again, curves in figures \ref{fig:3} and
\ref{fig:4} which terminate for small but nonzero values of $T_1$
do so due to computational limitations. The values of the order
parameters in the $m_1\to \infty$ (i.e. $T_1\to 0$) limit will be
calculated  analytically in a subsequent section.

The main difference between $m_2=1.5$ and $m_2=0.5$ is that in the
former case, for sufficiently low $T_1$, when lowering $T$ the
system goes straight from the paramagnetic state $q_1=q_2=0$ into
the $SG_2$ state $q_{1,2}\neq 0$, without an intermediate $SG_1$
phase. This is in agreement with the physical picture sketched so
far: high values of the replica dimension $m_2$, corresponding to
low coupling temperatures (here $T_2$), are found to induce more
pronounced discontinuities, similar to the effect of choosing
large values for $m_1$.

\section{The Three-Level Hierarchy: $L=3$}

\subsection{General Properties}

We now apply our results to the case $L=3$, where we have three
time-scales in the bond dynamics. Putting $L=3$ in expression
(\ref{eq:FRS2}) for the free energy per spin gives:
\begin{eqnarray}
\hspace*{-10mm} f_{L=3}(q_1, q_2,q_3)& =&
 \frac{1}{2}\pi_1 q_1-\frac{1}{\beta} \log 2
  +\frac{1}{4}q_1^2 \pi_1(m_1\!-\!1)
\nonumber\\&&\hspace*{-10mm}
  +\frac{1}{4}q^2_{2}\pi_2 m_{1}(m_2-1)
  +\frac{1}{4}q^2_{3}\pi_3 m_1 m_2(m_3-1)
\nonumber \\ &&\hspace*{-10mm}
 -\frac{1}{\beta m_1 m_2 m_3}\log \int\!Dz_3 \left\{
 \int\!Dz_{2} \left\{ \int\!Dz_1
 \left\{\room \cosh[\Xi] \right\}^{m_1} \right\}^{m_{2}}
\right\}^{m_3} \label{eq:FRS_L3}
\end{eqnarray}
 with
\bd
 \Xi=  z_3\sqrt{\beta q_{3}\pi_3}
 +z_2 \sqrt{\beta(q_{2}\pi_{2}- q_{3}\pi_{3})}
  +z_1 \sqrt{\beta(q_{1}\pi_{1}-
 q_{2}\pi_{2})}
\ed Due to the increased complexity of formulae induced by the
nesting of integrals, it is for $L= 3$ no longer helpful to write
the saddle-point equations
 for the three order
parameters $q_1\geq q_2\geq q_3$ in explicit form. However, in
order to suppress notation in deriving the conditions for the
various phase transitions, we will introduce the following
abbreviations for the right-hand sides of the saddle-point
equations (\ref{eq:spRSfinal1},\ref{eq:spRSfinal2}):
\begin{eqnarray*}
 q_1 = &  \bra\bra \bra \tanh^2 \Xi\ket_1\ket_2
 \ket_3 &= \varphi_1(q_1, q_2, q_3)
\\
 q_2 = & \bra \bra \left\{ \bra \tanh \Xi\ket_1 \right \} ^2 \ket_2\ket_{3}
&= \varphi_2(q_1,q_2, q_3)\\
 q_3 =& \bra  \left\{\bra  \bra  \tanh \Xi\ket_1
\ket_{2} \right \} ^2\ket_{3} & = \varphi_3(q_1, q_1, q_3)
\end{eqnarray*}
For $L=3$ we can distinguish between four phases. The first is a
paramagnetic phase $P$, where $q_1=q_2=q_3=0$; this always solves
(\ref{eq:spRSfinal1},\ref{eq:spRSfinal2}), and according to
(\ref{eq:highT}), is the sole saddle-point of (\ref{eq:FRS_L3}) in
the high temperature regime. In addition we now have three
spin-glass phases: $SG_1$, corresponding to a solution of the form
$q_1>q_2=q_3=0$ (i.e. freezing of spins and couplings on
time-scale $\tau_1$ but not on the time-scale $\tau_2\gg\tau_1$),
 $SG_2$, corresponding to a solution
$q_1\geq q_2>q_3=0$ (i.e. freezing of spins and couplings on
time-scales $\tau_1$ and $\tau_2$ but not on the time-scale
$\tau_3\gg\tau_2$), and
 $SG_3$, corresponding to a solution
$q_1\geq q_2\geq q_3>0$ (i.e. freezing of spins and couplings on
all time-scales). It immediately follows from the simple relation
$f_{L=3}(q_1,q_2,0)=f_{L=2}(q_1,q_2)$ that the phases $SG_1$ and
$SG_2$ are fully identical to those of the $L=2$ model, including
the values of the order parameters and the locations of transition
lines. However, there will now be new transition lines which are
related to transitions into phase $SG_3$. \vsp

All partial derivatives of the type $\partial
\varphi_\ell/\partial q_{\ell^\prime}$ which occur in the
transition conditions discussed below are calculated in detail in
\ref{app:L3}. In the various derivations we will save ink and
paper by using the short-hands $\psi_\ell=\bra\ldots\bra\tanh
\Xi\ket_1\ldots\ket_\ell$, e.g. \bd
 \psi_1=\bra\tanh \Xi\ket_1,~~~~~~
 \psi_2=\bra\bra\tanh \Xi\ket_1\ket_2,~~~~~~{\rm etc.}
\ed
The result is
\begin{eqnarray}
\frac{\partial \varphi _1}{\partial q_2} &=&
 \frac{\beta \pi_2}{2} m_1(m_2 -1) \left\{
 -m_1 m_2 m_3\varphi_1 \varphi_2
  + m_1 (m_2 -2) \bra \bra \psi_1^2 \bra\tanh^2 \Xi
\ket_1\ket_2 \ket_3 \right.\nonumber
\\ && \left. + m_1 m_2(m_3-1)
\bra \bra\bra\tanh^2\Xi\ket_1\ket_2
  \bra \psi_1^2\ket_2\ket_3
  \right.\nonumber\\
&& \left. + 2 (m_1-2)\bra \bra \psi_1\bra\tanh^3 \Xi\ket_1 \ket_2
\ket_3 + 4 \varphi_2  \right\} \label{eq:d1d2}
\end{eqnarray}
\begin{eqnarray}
\frac{\partial \varphi_2 }{ \partial q_1} &=& \frac{\beta
\pi_1}{2} (m_1 -1) \left\{  -m_1 m_2 m_3 \varphi_1 \varphi_2 + m_1
(m_2 -2) \bra \bra \psi_1 ^2 \bra\tanh^2 \Xi \ket_1\ket_2 \ket_3
\right.\nonumber
\\
&& \left.
 + m_1 m_2(m_3-1) \bra
\bra\bra\tanh^2\Xi\ket_1\ket_2
  \bra \psi_1^2\ket_2\ket_3
 \right.\nonumber \\
&& \left. + 2 (m_1-2)\bra \bra \psi_1\bra\tanh^3 \Xi\ket_1 \ket_2
\ket_3+4 \varphi_2 \right\} \label{eq:d2d1}
\end{eqnarray}
\begin{eqnarray}
\frac{\partial \varphi _1}{\partial q_1}& =&  \frac{\beta
\pi_1}{2} \left\{ - (m_1 \minus 1)m_1 m_2 m_3 (\varphi_1)^2
 + m_1(m_1\minus 1)(m_2 \minus 1)\bra\bra\bra\tanh^2 \Xi\ket_1^2\ket_2 \ket_3
 \right.\nonumber\\
&&
+ (m_1 -1)m_1 m_2(m_3 -1)
 \bra \bra\bra\tanh^2\Xi\ket_1\ket^2_2 \ket_3
 \nonumber \\
&& \left. +(m_1 -2) (m_1 -3)\bra \bra\bra \tanh^4 \Xi \ket_1\ket_2
\ket_3
 + 4(m_1 -2)\varphi_1 +2 \right\}
 \label{eq:d1d1}
\end{eqnarray}
\begin{eqnarray}
\frac{\partial \varphi _2}{\partial q_2}& =& \frac{\beta \pi_2}{2}
\left\{ -  m^2_1 m_2 m_3(m_2 -1)(\varphi_2 )^2
+m_{1}^2 (m_2-2)(m_2 -3)\bra \bra \psi_{1}^4 \ket_2 \ket_3
\right.\nonumber\\
&&
+m_1^2 m_2 (m_2-1)(m_3-1)
\bra
  \bra\psi_1^2\ket^2_2
\ket_3
\nonumber
\\ && +4m_{1}(m_2-2)\varphi _2 +4m_{1}(m_1 -1)(m_2-2)
\bra \bra \psi_{1}^2 \bra\tanh^2 \Xi \ket_1 \ket_2 \ket_3
\nonumber \\ && \left. +2 + 4(m_1 -1)\varphi _1  +2(m_1 -1)^2 \bra
\bra \bra\tanh^2 \Xi \ket_1^2 \ket_2 \ket_3 \right\}
\label{eq:d2d2}
\end{eqnarray}
\begin{eqnarray}
 \frac{\partial \varphi_3}{\partial q_1} & =&
 \frac{\beta
 \pi_1}{2}(m_1-1) \left\{
 2 m_1 (m_2-1) \bra\psi_{2}\bra\bra\tanh^2 \Xi\ket_1 \psi_{1}\ket_2 \ket_3
 \right. \nonumber
\\
&& \left.
 + (m_3-2)m_1 m_2
\bra \bra\bra\tanh^2\Xi\ket_1\ket_2 \psi_2^2\ket_3 \right.
\nonumber
 \\
&& \left. +  2(m_1 -2) \bra\psi_{2} \bra   \bra \tanh^3 \Xi\ket_1
\ket_2 \ket_3 +4 \varphi _3 -m_1 m_2 m_3\varphi_1 \varphi_3
 \right\} \label{eq:d3d1}
\end{eqnarray}
\begin{eqnarray}
  \frac{\partial
 \varphi _3}{\partial q_2} & =& \frac{\beta \pi_2}{2}(m_2-1)
 \left\{
  m^2_1 m_2(m_3-2) \bra \bra\psi_1^2\ket_2
 \psi_2^2
 \ket_3
\right.\nonumber
 \\ &&
+4 m_1(m_1-1)  \bra\psi_{2} \bra  \psi _1 \bra\tanh^2 \Xi\ket_1
\ket_2  \ket_3
 \nonumber
\\ &&\left.
 +2m_1^2(m_2-2)  \bra\psi_{2} \bra\psi_1^3 \ket_2 \ket_3
 + 4 m_1 \varphi_3  - m^2_1 m_2
m_3\varphi_2 \varphi _3 \right\} \label{eq:d3d2}
\end{eqnarray}
\begin{eqnarray}
 \frac{\partial \varphi _3}{\partial q_3} &=& \frac{\beta
 \pi_3}{2} \left\{2 - (m_3-1)m_1^2 m_2^2 m_3 \varphi^2_3
 + m^2_1 m^2_2  (m_3-2) (m_3 -3)\bra \psi_{2}^4\ket_3 \right.
\nonumber\\ && \left.
 +4(m_3-2)m_1 m_2\varphi_3
+ 4(m_1-1)\varphi_1 + 4m_1(m_2-1)\varphi_2
 \right.
\nonumber\\&&\left.
 + 4(m_1-1)(m_3-2)m_1 m_2
 \bra \bra\bra\tanh^2\Xi\ket_1\ket_2
  \psi_2^2
 \ket_3
\right.
 \nonumber
\\ &&
\left.
 + 4 m^2_1m_2(m_2-1)(m_3-2) \bra
  \bra\psi^2_1\ket_2
\psi_2^2
 \ket_3 \right.
\nonumber
 \\ &&
 \left.
 + 4m_1 (m_1-1)(m_2-1)\bra
  \bra\bra\tanh^2\Xi\ket_1\ket_2
  \bra\psi^2_1\ket_2
 \ket_3
\nonumber \right.
\\ &&
\left. + 2(m_1\minus 1)^2
 \bra\bra\bra\tanh^2\Xi\ket_1\ket^2_2
\ket_3  + 2m_1^2(m_2\minus 1)^2 \bra \bra\psi^2_1\ket^2_2 \ket_3
\right\} \label{eq:d3d3}
\end{eqnarray}
\begin{eqnarray}
 \frac{\partial \varphi_1}{\partial q_3}& =&
\frac{\beta \pi_3}{2} m_1 m_2 (m_3 -1)
 \left\{
4 \varphi _3 -m_1 m_2 m_3\varphi_1 \varphi_3 \right.\nonumber
\\
&&\left. +2 m_1 (m_2\minus 1) \bra\psi_{2}\bra\bra\tanh^2
\Xi\ket_1 \psi_{1}\ket_2 \ket_3 +  2(m_1 \minus 2) \bra\psi_{2}
\bra \bra \tanh^3 \Xi\ket_1 \ket_2 \ket_3
 \right.\nonumber
 \\
&& \left.
 + (m_3-2)m_1 m_2
 \bra
  \bra\bra\tanh^2\Xi\ket_1\ket_2
 \psi_2^2
 \ket_3
\right\} \label{eq:d1d3}
\end{eqnarray}
\begin{eqnarray}
 \frac{\partial
\varphi _2}{\partial q_3}&=& \frac{\beta \pi_3}{2}m_2(m_3-1)
\left\{
 4 m_1 \varphi_3- m^2_1 m_2 m_3  \varphi_2 \varphi _3
\right.
\nonumber
 \\ &&
+2m_1^2(m_2-2)  \bra\psi_{2} \bra\psi_1 ^3 \ket_2  \ket_3
 +4 m_1(m_1 -1)  \bra\psi_{2} \bra  \psi _1 \bra\tanh^2
\Xi\ket_1 \ket_2  \ket_3
 \nonumber
 \\
&&\left.
+ m^2_1(m_3-2)m_2 \bra
 \bra\psi_1^2\ket_2
 \psi_2^2
\ket_3
\right\} \label{eq:d2d3}
\end{eqnarray}

\subsection{Transitions Identical to Those of $L=2$}

All transitions relating only to the three phases $P$, $SG_1$ and
$SG_2$, as well as the second order transition $SG_1\to SG_2$,
where $q_3=0$, must be identical to those derived in the previous
section for $L=2$. The general condition for such transitions is
\begin{eqnarray}
0&=& (\frac{\partial \varphi_1}{\partial q_1}-1 ) (\frac{\partial
\varphi_2}{\partial q_2}-1 )- \frac{\partial \varphi_1}{\partial
q_2}
 \frac{\partial \varphi_2}{\partial q_1}
 \label{eq:trans_L2}
\end{eqnarray}
This involves only the  partial derivatives
(\ref{eq:d1d2},\ref{eq:d2d1},\ref{eq:d1d1},\ref{eq:d2d2}). Due to
$q_3=0$ the averages $\bra\ldots\ket_3$ drop out, and together
with the identities \be
 \bra \bra \tanh^2 \Xi\ket_1\ket_2=q_1,~~~~~
  \bra \bra \tanh \Xi\ket^2_1  \ket_2=q_2,~~~~~
  \bra \psi^2_1\ket_2=q_2,~~~~~
\psi_2=0
\label{eq:q3=0simplifications}
 \ee
 the four relevant
partial derivatives simplify considerably to
\begin{eqnarray}
\frac{\partial \varphi _1}{\partial q_2} &=&
 \frac{\beta \pi_2}{2} m_1(m_2-1) \left\{
  (4- m_1 m_2 q_1) q_2
  + m_1 (m_2 -2)  \bra \psi_1^2 \bra\tanh^2 \Xi
\ket_1\ket_2  \right.\nonumber
\\
&& \left. + 2 (m_1-2) \bra \psi_1\bra\tanh^3 \Xi\ket_1 \ket_2
\right\} \label{eq:d1d2simple}
\end{eqnarray}
\begin{eqnarray}
\frac{\partial \varphi_2 }{ \partial q_1} &=& \frac{\beta
\pi_1}{2} (m_1- 1) \left\{(4- m_1 m_2q_1) q_2  + m_1 (m_2- 2) \bra
\psi_1^2 \bra\tanh^2 \Xi \ket_1\ket_2 \right.\nonumber
\\
&& \left. + 2 (m_1-2)\bra \psi_1\bra\tanh^3 \Xi\ket_1 \ket_2
\right\} \label{eq:d2d1simple}
\end{eqnarray}
\begin{eqnarray}
\frac{\partial \varphi _1}{\partial q_1}& =&  \frac{\beta
\pi_1}{2} \left\{
2 -(m_1 -1)m_1 m_2 q_1^2+ m_1(m_1\minus 1)(m_2 \minus 1)\bra\bra\tanh^2 \Xi\ket_1^2\ket_2
 \right.\nonumber\\
&& \left. +(m_1 -2) (m_1 -3) \bra\bra \tanh^4 \Xi \ket_1\ket_2
 + 4(m_1 -2)q_1 \right\}
 \label{eq:d1d1simple}
\end{eqnarray}
\begin{eqnarray}
\frac{\partial \varphi _2}{\partial q_2}& =& \frac{\beta \pi_2}{2}
\left\{ -  m^2_1 m_2 m_3(m_2 -1)q_2^2 +m_{1}^2 (m_2-2)(m_2 -3)
\bra \psi_{1}^4 \ket_2 \right.\nonumber\\ && +m_1^2 m_2
(m_2-1)(m_3-1)
  \bra\psi_1^2\ket^2_2
\nonumber
\\ && +4m_{1}(m_2-2)q_2 +4m_{1}(m_1 -1)(m_2-2)
 \bra \psi_{1}^2 \bra\tanh^2 \Xi \ket_1 \ket_2
\nonumber \\ && \left. +2 + 4(m_1 -1)q_1  +2(m_1 -1)^2 \bra
\bra\tanh^2 \Xi \ket_1^2 \ket_2  \right\} \label{eq:d2d2simple}
\end{eqnarray}
For the $P\to SG_1$ transitions and the continuous $SG_1\to SG_2$
transition we have to put $q_2=0$, in
(\ref{eq:d1d2simple},\ref{eq:d2d1simple},\ref{eq:d1d1simple},\ref{eq:d2d2simple}).
We then arrive, together with the dropping out of $\bra
\ldots\ket_2$ and  the relations $\psi_1=\psi_2=0$,
$\bra\tanh^2\Xi\ket_1=q_1$, at the simple expressions $\partial
\varphi _1/\partial q_2=
\partial \varphi_2 / \partial q_1= 0$ and
\bd
 \frac{\partial \varphi _1}{\partial q_1}= \frac{\beta
\pi_1}{2} \left\{
 2 - m_1(m_1\minus 1)q_1^2
 +(m_1 \minus 2) (m_1 \minus 3) \bra \tanh^4 \Xi
\ket_1
 + 4(m_1 \minus 2)q_1 \right\}
\ed
\bd
 \frac{\partial \varphi _2}{\partial q_2}=\beta \pi_2
[1+(m_1-1)q_1]^2 \ed
Thus
 (\ref{eq:trans_L2}) gives us the transition conditions
\be
 \frac{\beta
 \pi_1}{2} \left\{
 2 - m_1(m_1\minus 1)q_1^2
 +(m_1 \minus 2) (m_1 \minus 3) \bra \tanh^4 \Xi
 \ket_1
 + 4(m_1 \minus 2)q_1 \right\}=1
\label{eq:trans1_L2}
\ee
\be
 \beta \pi_2 [
 1+(m_1-1)q_1]^2 =1
 \label{eq:trans2_L2}
\ee For $q_1=0$ (where also $\Xi=0$) we recover from
(\ref{eq:trans1_L2}) the condition (\ref{eq:trans_L2_PSG1_2nd})
for the line $T_{p,1}^{\rm 2nd}$, for $q_1>0$ equation
(\ref{eq:trans1_L2}) is seen to be identical to condition
$T=\pi_1$ (which for $L=2$ led to (\ref{eq:trans_L2_PSG1_1st}))
for the line $T_{p,1}^{\rm 1st}$, and (\ref{eq:trans2_L2}) is
identical to condition (\ref{eq:trans_L2_SG1SG2_2nd}) for the line
$T_{1,2}^{\rm 2nd}$. Finally, the condition for the first order
$SG_1\to SG_2$ transition line $T_{1,2}^{\rm 1st}$ of $L=2$ is
recovered by combining the full expressions
(\ref{eq:d1d2simple},\ref{eq:d2d1simple},\ref{eq:d1d1simple},\ref{eq:d2d2simple})
with the bifurcation condition
 (\ref{eq:trans_L2}), as it should.
Thus, from our $L=3$ saddle-point equations we do indeed extract
fully those transitions encountered earlier for $L=2$, which do
not involve the $SG_3$ phase.

\subsection{Transitions to $SG_3$}

The novel transitions induced by going from $L=2$ to $L=3$ are
those where the new phase $SG_3$ is concerned. The condition for
second order $SG_2\to SG_3$ transitions is simply given by
 $\partial \varphi_3/\partial q_3 |_{q_3=0}=1$.
Inserting $q_3=0$ into (\ref{eq:d3d3}), followed by usage of the
 simplifying relations
(\ref{eq:q3=0simplifications}) which follow from $q_3=0$, leads to
\begin{eqnarray*}
 \frac{\partial \varphi _3}{\partial q_3}|_{q_3=0} &=& \beta
 \pi_3[1
+ (m_1- 1)
 q_1
 + m_1(m_2- 1)  q_2]^2
\end{eqnarray*}
And the condition defining the critical temperature $T_{2,3}^{\rm
2nd}$ for the second order transition $SG_2\to SG_3$ is seen to be
simply
\begin{eqnarray}
\beta
 \pi_3[1
+ (m_1- 1)
 q_1
 + m_1(m_2- 1)  q_2]^2=1
\label{eq:trans_L3_SG2SG3_2nd}
\end{eqnarray}
Finally, the most general bifurcation condition defining first
order $SG_2\to SG_3$ transitions is given by
\be
\left|
\begin{array}{lll}
\partial \varphi_1/\partial q_1-1 &~ \partial
\varphi_1/\partial q_2 &~ \partial \varphi_1/\partial q_3
\\[2mm] \partial \varphi_2/\partial q_1&~ \partial
\varphi_2/\partial q_2-1 &~ \partial \varphi_2/\partial q_3\\[2mm]
\partial \varphi_3/\partial q_1&~ \partial
\varphi_3/\partial q_2 &~ \partial \varphi_3/\partial q_3-1\\
\end{array} \right|=0
\label{eq:trans_L3_SG2SG3_1st} \ee where $ q_1 \ge q_2 \ge q_3 >0
$. Here there are no further simplifying properties to be
exploited, and hence (\ref{eq:trans_L3_SG2SG3_1st}) must be solved
numerically with the full expressions
(\ref{eq:d2d1}-\ref{eq:d2d3}) for the nine partial derivatives.

\subsection{The $L=3$ Phase Diagram}

For $L=3$ we have seven independent control parameters, viz. $T$,
$m_1$, $m_2$, $m_3$, $\pi_1$, $\pi_2$ and $\pi_3$ (where
$\pi_3=\epsilon_{3}/m_1 m_2 m_3\mu_3$,
$\pi_2=\pi_3+\epsilon_{2}/m_1m_2\mu_{2}$ and
$\pi_1=\pi_2+\epsilon_1/m_1\mu_1$), so we again have to resort to
phase diagram cross-sections. As in the previous case $L=2$ we
will show the transition lines in the $(T_1,T)$ plane, solved
numerically from the various bifurcation conditions, now for the
choice $\epsilon_1/\mu_1=\epsilon_2/\mu_2=1$, $\epsilon_3/\mu_3=2$
with $m_2\in\{0.5, 1.5\}$ and with $m_3=2$  (i.e. relatively low
noise levels in the level-3 bonds). According to
(\ref{eq:trans_L2_PSG1_2nd}), for this choice of parameters the
second order $P\to SG_1$ transition condition $T_{p,1}^{\rm
2nd}=\pi_1$ will again reduce to
\be
T_{p,1}^{\rm 2nd}=\frac{1}{m_1}(1+\frac{2}{m_2}) \ee and is
replaced by a first order $P\to SG_1$ transition at $m_1=2$, or
\begin{eqnarray}
T_{p,1}^{\rm 1st}=T_{p,2}^{\rm 2nd}: &~~~~&
T_1=\frac{1}{4}(1+\frac{2}{m_2}) \label{eq:change_point2}
\end{eqnarray}
As in the case $L=2$,
 we will generally find (see section \ref{sec:asymptotics}) that for any $L$ and for any $\ell\geq 1$
 there are always non-zero critical
values for $T_1$ above which $T_{\ell,\ell+1}^{\rm 2nd}=0$. Hence
all spin-glass phases $SG_{\ell>1}$ will at some finite point
cease to exist as $T_1$ is increased. The values of the order
parameters in the $m_1\to \infty$ (i.e. $T_1\to 0$) limit will be
also be calculated  analytically in section \ref{sec:asymptotics}.
\vsp

\begin{figure}[t]
\vspace*{1mm} \setlength{\unitlength}{1.04mm}
\begin{picture}(150,85)
 \put(7,20){\epsfxsize=91\unitlength\epsfbox{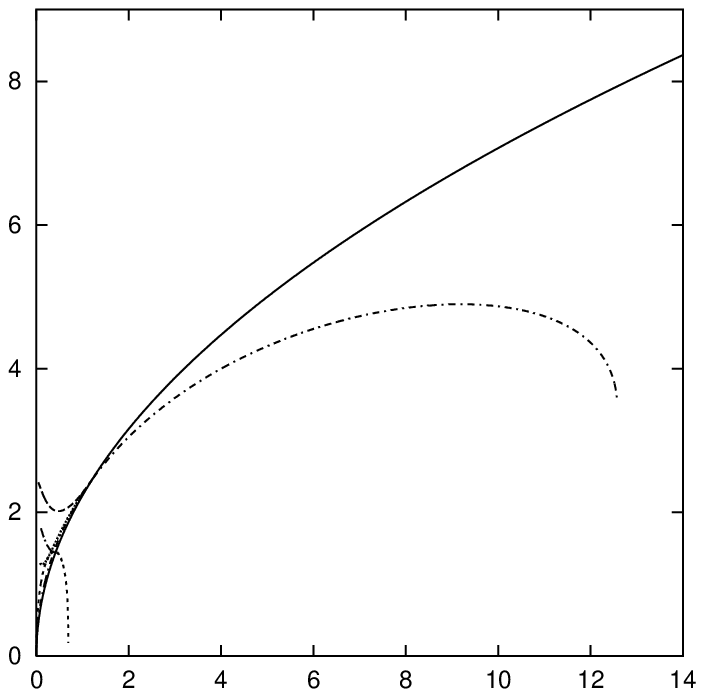}}
 \put(84,20){\epsfxsize=91\unitlength\epsfbox{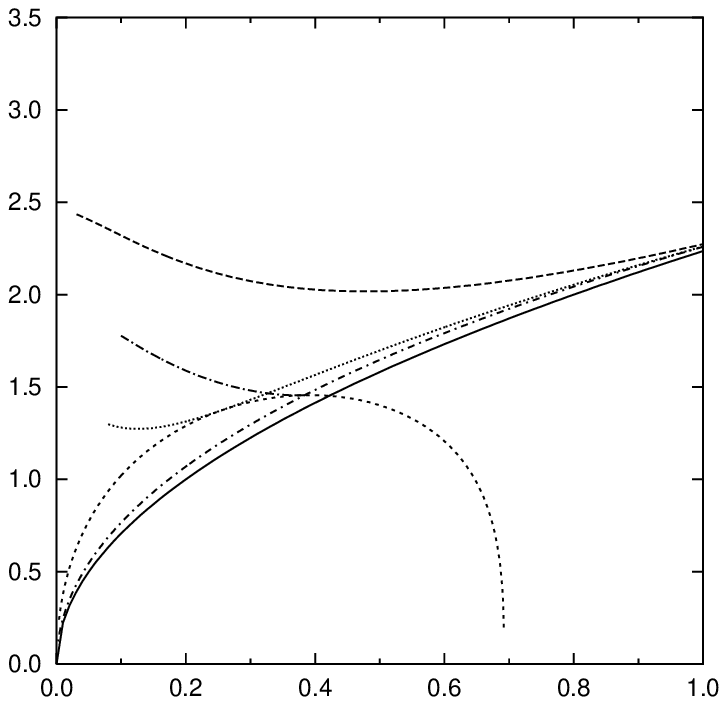}}
 \put(0,52){ $T$}   \put(39,13){$T_1$}
 \put(80,52){$T$} \put(119,13){$T_1$}
 \put(25,70){$P$} \put(55,60){$SG_1$} \put(45,35){$SG_2$}
 \put(14,26){$SG_3$}
 \put(105,70){$P$} \put(105,52){$SG_1$} \put(136,37){$SG_2$}
 \put(93,47){$SG_3$}\put(108,29){$SG_3$}
\end{picture}
\vspace*{-16mm} \caption{Phase diagram in the $(T_1,T)$ plane for
$L=3$, $\epsilon_1/\mu_1=\epsilon_2/\mu_2=1$,
$\epsilon_3/\mu_3=2$, $m_3=2$ and $m_2=0.5$, obtained by solving
numerically the equations defining the various bifurcation lines
(note: $T_1=T/m_1$). Left panel: the large-scale structure. Right
panel: enlargement of the low $T_1$ (or high $m_1$) region, where
the first order transitions are found. The relevant phases are $P$
(paramagnetic phase, $q_1=q_2=0$), $SG_1$ (first spin-glass phase,
$q_1>q_2=q_3=0$), $SG_2$ (second spin-glass phase, $q_{1}\geq q_2
> q_3=0$), and $SG_3$ (third spin-glass phase,
$q_{1}\geq q_2\geq q_3>0$). The transition lines shown are
$T_{p,1}^{\rm 2nd}$ (solid curve), $T_{1,2}^{\rm 2nd}$ (dot-dash
curve), $T_{p,1}^{\rm 1st}$ (dashed curve), $T_{p,2}^{1st}$
(dotted curve), $T_{2,3}^{\rm 2nd}$ (short dashes), and
$T_{2,3}^{\rm 1st}$ (dot-long dash curve). Note: the curves
$T_{1,2}^{\rm 2nd}$ and $T_{2,3}^{\rm 2nd}$ jump to zero
discontinuously at $T_1=T_c^{(0)}\approx 12.575$ and
$T_1=T_c^{(4)}\approx 0.692$, respectively (see main text). }
\label{fig:5}
\end{figure}

\begin{figure}[t]
\vspace*{1mm} \setlength{\unitlength}{1.04mm}
\begin{picture}(150,85)
 \put(37,20){\epsfxsize=91\unitlength\epsfbox{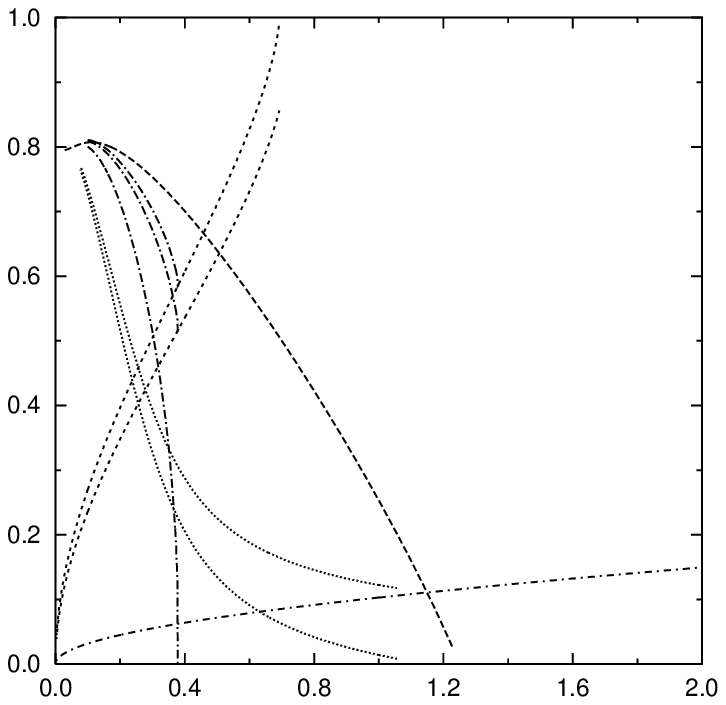}}
 \put(31,52){ $q_{1,2}$}   \put(71,13){$T_1$}
\end{picture}
\vspace*{-16mm} \caption{Values of the order parameters $q_1$ and
$q_2$ as functions of $T_1$, along some of the transition lines of
the previous figure, for $L=3$,
$\epsilon_1/\mu_1=\epsilon_2/\mu_2=1$, $\epsilon_3/\mu_3=2$,
$m_3=2$ and $m_2=0.5$.
 The values shown are $q_1$ along $T_{1,2}^{\rm 2nd}$ (dot-dash curve), $q_1$ along $T_{p,1}^{\rm
1st}$ (dashed curve), $q_1$ and $q_2$ along $T_{p,2}^{1st}$
(dotted curves), $q_1$ and $q_2$ along
 $T_{2,3}^{\rm 2nd}$ (short dashes), and $q_1$, $q_2$ and $q_3$
 for $T_{2,3}^{\rm 1st}$ (dot-long dash curves)
 Note: the order parameters which are not shown in this
figure are zero by definition. } \label{fig:6}
\end{figure}

In figure \ref{fig:5} we show the phase diagram and the various
bifurcation lines of the $L=3$ system for $m_2=0.5$ (relatively
high noise levels in the level-2 couplings), obtained by numerical
solution of the relevant equations. The physical transitions are
the ones which occur first as the temperature $T$ is lowered from
within the paramagnetic region. In the present case the different
$T_1$-dependent  types of transition sequences are separated by
the following critical values for $T_1$:
\begin{eqnarray*}
 T_c^{(0)}\approx 12.575&:~~~~~~& {\rm creation~point~of~phase}~SG_2
\nonumber\\
 T_c^{(1)}=1.25         &:& {\rm creation~point~of~}T_{p,1}^{\rm 1st},~~~ T_{p,1}^{\rm 2nd}=T_{p,1}^{\rm 1st}
\nonumber\\
 T_c^{(2)} \approx 1.15 &:& T_{p,1}^{\rm 1st}~ {\rm and}~ T_{1,2}^{\rm 2nd}~ {\rm
 touch~tangentially}
\nonumber \\
 T_c^{(3)} \approx 1.06 &:& {\rm creation~point~of~} T_{p,2}^{\rm 1st},~~~ T_{1,2}^{\rm 2nd}=T_{p,2}^{\rm 1st}
\nonumber\\
 T_c^{(4)} \approx 0.692&:& {\rm creation~point~of~phase}~SG_3
\nonumber \\
 T_c^{(5)} \approx 0.38 &:& {\rm creation~point~of~} T_{p,3}^{\rm
 1st},~~~T_{2,3}^{\rm 2nd}=T_{p,3}^{\rm 1st}
\nonumber \\
 T_c^{(6)} \approx 0.325 &:& T_{p,2}^{\rm 1st}=T_{p,3}^{\rm 1st}
\end{eqnarray*}
Note: $T_c^{(1)}$ is the point given by expression
(\ref{eq:change_point2}). In terms of these critical values we can
classify the different transition scenarios encountered upon
reducing $T$ down to zero (and the orders of the transitions),
starting in the paramagnetic region, as follows:
\begin{eqnarray*}
\hspace*{-15mm}
 T_c^{(0)} < T_1:            &~~~ P \to SG_1~{\rm (2nd)}
\\ \hspace*{-15mm}
 T_c^{(1)} < T_1 < T_c^{(0)}:  &~~~ P \to SG_1~{\rm (2nd)},
 ~~~&~ SG_1 \to SG_2 ~{\rm (2nd)}
\\ \hspace*{-15mm}
 T_c^{(2)} < T_1 < T_c^{(1)}:&~~~ P\to SG_1~{\rm (1st)},
 ~~~&~ SG_1 \to SG_2 ~{\rm (2nd)}
\\ \hspace*{-15mm}
 T_c^{(3)} < T_1<T_c^{(2)}:&~~~ P\to SG_1~{\rm (1st)},
 ~~~&~ SG_1\to SG_2~{\rm (2nd)}
\\ \hspace*{-15mm}
 T_c^{(4)}< T_1 < T_c^{(3)}: &~~~ P\to SG_1~{\rm (1st)},
 ~~~&~ SG_1\to  SG_2 ~{\rm (1st)}
\\ \hspace*{-15mm}
 T_c^{(5)}< T_1 < T_c^{(4)}: &~~~ P\to SG_1~{\rm (1st)},
 ~~~&~ SG_1\to SG_2 ~{\rm (1st)},
~~~~~ SG_2\to  SG_3~ {\rm (2nd)}
\\ \hspace*{-15mm}
 T_c^{(6)}< T_1 < T_c^{(5)}: &~~~P\to  SG_1~{\rm (1st)},
 ~~~&~SG_1\to SG_2 ~{\rm (1st)},
~~~~~ SG_2\to SG_3~{\rm  (1st)}
\\ \hspace*{-15mm}
 T_1 < T_c^{(6)}: &~~~ P\to SG_1~ {\rm (1st)},
 ~~~&~SG_1\to  SG_3~{\rm (1st)}
\end{eqnarray*}
In figure \ref{fig:6} we show the corresponding values of the
order parameters along the various transition lines, as functions
of $T_1$ (with line types identical to those used in figure
\ref{fig:5} for the same transitions), except for those order
parameters  which are zero by definition (such as the values at
the transition of order parameters which bifurcate continuously
from zero). Curves in figures \ref{fig:5} and \ref{fig:6} which
terminate for small but nonzero values of $T_1$ do so due to
computational limitations.

\begin{figure}[t]
\vspace*{1mm} \setlength{\unitlength}{1.04mm}
\begin{picture}(150,85)
 \put(7,20){\epsfxsize=91\unitlength\epsfbox{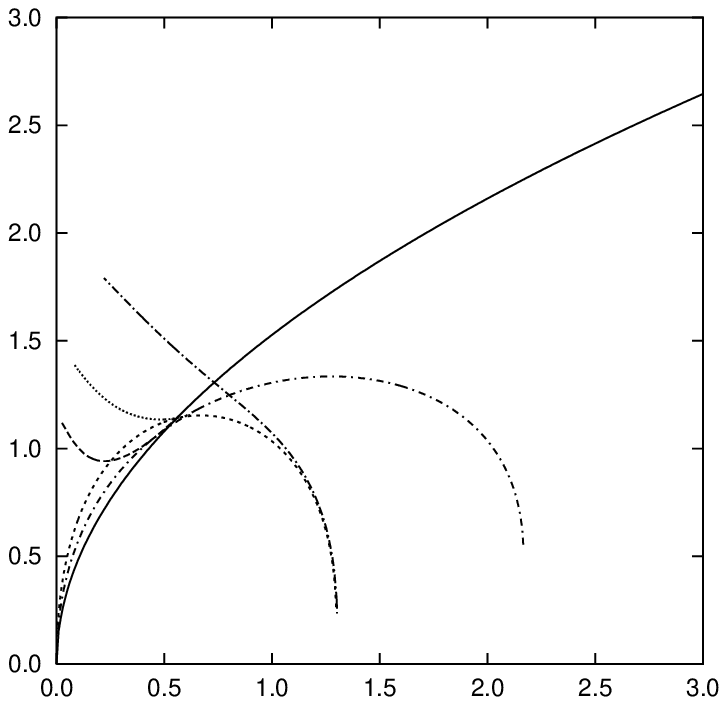}}
 \put(84,20){\epsfxsize=91\unitlength\epsfbox{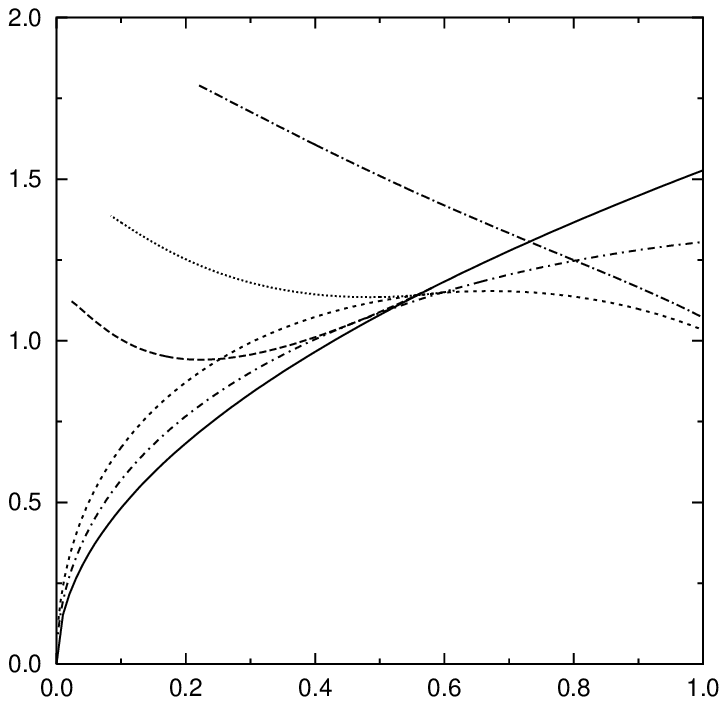}}
 \put(0,52){ $T$}   \put(39,13){$T_1$}
 \put(80,52){$T$} \put(119,13){$T_1$}
 \put(25,70){$P$} \put(58,50){$SG_1$} \put(44,33){$SG_2$}
 \put(23,29){$SG_3$}
 \put(129,72){$P$} \put(105,63){$SG_3$} \put(122,35){$SG_3$}
\end{picture}
\vspace*{-16mm} \caption{Phase diagram in the $(T_1,T)$ plane for
$L=3$, $\epsilon_1/\mu_1=\epsilon_2/\mu_2=1$,
$\epsilon_3/\mu_3=2$, $m_3=2$ and $m_2=1.5$, obtained by solving
numerically the equations defining the various bifurcation lines
(note: $T_1=T/m_1$). Left panel: the large-scale structure. Right
panel: enlargement of the low $T_1$ (or high $m_1$) region, where
the first order transitions are found. The relevant phases are $P$
(paramagnetic phase, $q_1=q_2=0$), $SG_1$ (first spin-glass phase,
$q_1>q_2=q_3=0$), $SG_2$ (second spin-glass phase, $q_{1}\geq q_2
> q_3=0$), and $SG_3$ (third spin-glass phase,
$q_{1}\geq q_2\geq q_3>0$). The transition lines shown are
$T_{p,1}^{\rm 2nd}$ (solid curve), $T_{1,2}^{\rm 2nd}$ (dot-dash
curve), $T_{p,1}^{\rm 1st}$ (dashed curve), $T_{p,2}^{1st}$
(dotted curve), $T_{2,3}^{\rm 2nd}$ (short dashes), and
$T_{2,3}^{\rm 1st}$ (dot-long dash curve). Note: the curves
$T_{1,2}^{\rm 2nd}$ and $T_{2,3}^{\rm 2nd}$ jump to zero
discontinuously at $T_1=T_c^{(0)}\approx 2.17$ and
$T_1=T_c^{(1)}\approx 1.3$, respectively (see main text). }
\label{fig:7}
\end{figure}

\begin{figure}[t]
\vspace*{1mm} \setlength{\unitlength}{1.04mm}
\begin{picture}(150,85)
 \put(37,20){\epsfxsize=91\unitlength\epsfbox{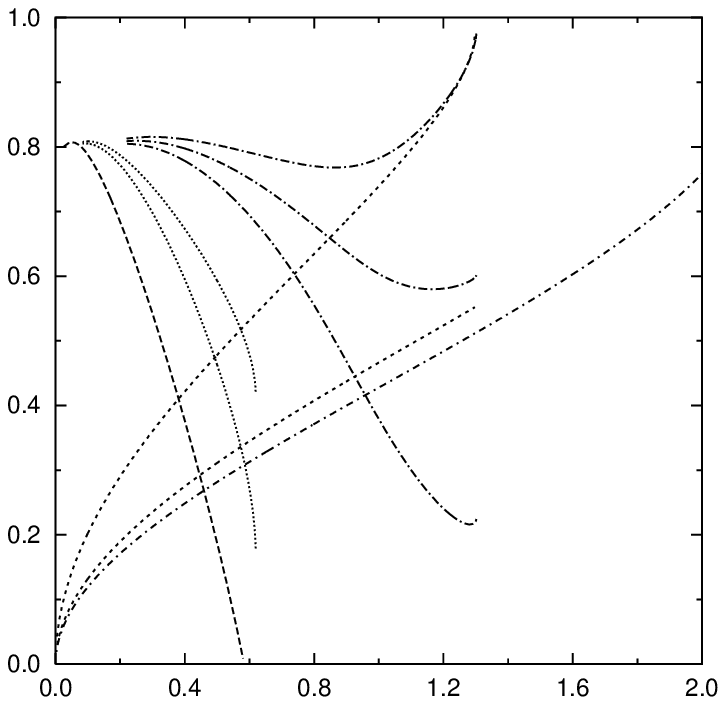}}
 \put(31,52){ $q_{1,2}$}   \put(71,13){$T_1$}
\end{picture}
\vspace*{-16mm} \caption{Values of the order parameters $q_1$ and
$q_2$ as functions of $T_1$, along some of the transition lines of
the previous figure, for $L=3$,
$\epsilon_1/\mu_1=\epsilon_2/\mu_2=1$, $\epsilon_3/\mu_3=2$,
$m_3=2$ and $m_2=1.5$.
 The values shown are $q_1$ along $T_{1,2}^{\rm 2nd}$ (dot-dash curve), $q_1$ along $T_{p,1}^{\rm
1st}$ (dashed curve), $q_1$ and $q_2$ along $T_{p,2}^{1st}$
(dotted curves), $q_1$ and $q_2$ along
 $T_{2,3}^{\rm 2nd}$ (short dashes), and $q_1$, $q_2$ and $q_3$
 for $T_{2,3}^{\rm 1st}$ (dot-long dash curves)
 Note: the order parameters which are not shown in this
figure are zero by definition. } \label{fig:8}
\end{figure}

 In figure
\ref{fig:7} we show the phase diagram and the various bifurcation
lines of the $L=3$ system for $m_2=1.5$ (i.e. relatively low noise
levels in the level-2 couplings), as obtained by numerical
solution of the relevant equations. Now the relevant critical
values for $T_1$ are as follows:
\begin{eqnarray*}
 T_c^{(0)}\approx 2.17&:~~~~~~& {\rm creation~point~of~phase}~SG_2
 \nonumber\\
 T_c^{(1)} \approx 1.3 &:& {\rm creation~point~of~phase}~SG_3,~{\rm via}~ T_{2,3}^{\rm 1st}
\nonumber\\
 T_c^{(2)}\approx 1.3 &:& {\rm creation~point~of~} T_{2,3}^{\rm 2nd}
\nonumber \\
 T_c^{(3)}\approx 0.8 &:& T_{1,2}^{\rm 2nd}=T_{2,3}^{\rm 1st}
\nonumber \\
 T_c^{(4)} \approx 0.73 &:& T_{p,1}^{\rm 2nd}= T_{2,3}^{\rm 1st}
\nonumber\\
 T_c^{(5)} \approx 0.597 &:& {\rm creation~point~of~}
             T_{p,2}^{\rm 1st},~~~ T_{1,2}^{\rm 2nd}=T_{p,2}^{\rm 1st}
\nonumber\\
 T_c^{(6)} = 7/12\approx 0.583 &:& T_{p,1}^{\rm 2nd}= T_{p,1}^{\rm 1st}
\nonumber\\
 T_c^{(7)} \approx 0.5572 &:& T_{p,1}^{\rm 1st}= T_{p,2}^{\rm 1st}
\end{eqnarray*}
Note:  $T_c^{(2)}<T_c^{(1)}$. In terms of these critical values we
can classify the different transition scenarios encountered upon
reducing $T$, as before, as follows:
\begin{eqnarray*}
\hspace*{-15mm}
 T_c^{(0)} < T_1:            &~~~ P \to SG_1~{\rm (2nd)}
\\ \hspace*{-15mm}
 T_c^{(1)} < T_1 < T_c^{(0)}:  &~~~ P \to SG_1~{\rm (2nd)},
 ~~~&~ SG_1 \to SG_2 ~{\rm (2nd)}
\\ \hspace*{-15mm}
 T_c^{(2)} < T_1 < T_c^{(1)}:&~~~ P\to SG_1 {\rm (2nd)},
 ~~~&~SG_1\to SG_2 ~{\rm (2nd)},
 ~~~~~SG_2\to SG_3 ~{\rm (1st)}
\\ \hspace*{-15mm}
 T_c^{(3)}< T_1 < T_c^{(2)}:&~~~ P\to SG_1 {\rm (2nd)},
 ~~~&~SG_1\to SG_3 ~{\rm (1st)}
\\ \hspace*{-15mm}
 T_1 < T_c^{(3)}:&~~~ P\to SG_3 ~{\rm (1st)}
\end{eqnarray*}
In figure \ref{fig:8} we show the corresponding values of the
order parameters along the various transition lines, as functions
of $T_1$ (with line types identical to those used in figure
\ref{fig:7} for the same transitions), except for those which are
zero by definition. Again, curves in figures \ref{fig:7} and
\ref{fig:8} which terminate for small but nonzero values of $T_1$
do so due to computational limitations.

The general picture for $L=3$ is obviously
 more complicated but
qualitatively similar to that which we arrived at for $L=2$, with
increasingly discontinuous and non-trivial  phase behaviour of the
system for decreasing relative noise levels of the couplings,
whether measured by $m_1$, $m_2$ or $m_3$. For instance, for $L=2$
we observed that for sufficiently large values of $m_1$ and $m_2$,
upon lowering $T$  the system goes straight from the paramagnetic
state $q_1=q_2=0$ into the $SG_2$ state $q_{1}\geq q_2> 0$,
without an intermediate $SG_1$ phase. Here we see, similarly, that
for sufficiently large values of $m_1$, $m_2$ and $m_2$, upon
lowering $T$  the system goes straight from the paramagnetic state
$q_1=q_2=q_3=0$ into the $SG_3$ state $q_{1}\geq q_2\geq q_3> 0$,
without intermediate $SG_1$ or $SG_2$  phases.

Finally, in the next section we will show that the jump of the
$SG_\ell\to SG_{\ell+1}$ transitions, as observed in the
$L=2$ and $L=3$ phase diagrams for small $T$, reflect non-physical re-entrance
phenomena, which can be regarded as indicators that replica
symmetry will be broken for values of $m_1=T/T_1$ close to zero.

\section{Asymptotic behaviour in $L$-level Hierarchy}
\label{sec:asymptotics}

In this section we study the saddle-point equations of the general
$L$-level hierarchy in the two limits $T_1\to 0$ for fixed $T$
(i.e. $m_1\to \infty$, fully deterministic evolution of level-1
couplings) and $T\to 0$ for fixed $T_1$ (i.e. $m_1\to 0$, fully
deterministic evolution of spins). For the particular choices
$L=2$ and $L=3$ this implies solving our model along the left and
bottom boundaries in the phase diagrams of figures \ref{fig:1},
\ref{fig:3}, \ref{fig:5} and \ref{fig:7}.

\subsection{The Limit $m_1\to\infty$ in the $L$-level Hierarchy}

In this subsection we study the order parameter equations in the $\ell$-th spin-glass
phase $SG_\ell$, characterized by $q_1\geq \ldots \geq q_\ell>0$ and $q_{\ell+1}=\ldots=q_L=0$.
We calculate the order parameters in the limit $m_1\to\infty$
(deterministic evolution of level-1 couplings), for fixed $\beta$, and we derive
in this limit an expression for the first order $P\to SG_\ell$ transition temperature
$T_{p,\ell}^{\rm 1st}$.
Insertion of $q_{\ell+1}= q_{\ell+2}=\ldots= q_L=0$ into
expression (\ref{eq:f_iterative}) for the free energy per spin
gives
\begin{eqnarray}
f& = &
\frac{1}{2}\pi_1 q_1-\frac{1}{\beta} \log 2
+\frac{1}{4}\sum_{k=2}^\ell q^2_k\pi_k
m_1 m_2 \cdots m_{k-1}(m_k-1)
\nonumber \\
&& +\frac{1}{4}q_1^2 \pi_1(m_1\!-\!1)
-[\beta\prod_{k=1}^{\ell}m_{k}]^{-1}\log K_\ell
\label{eq:f_iterative_L}\\
K_\ell & = & \int\!Dz_\ell
\left\{
\int\!Dz_{\ell-1}
\left\{
\ldots
\int\!Dz_2
\left\{
\int\!Dz_{1}
\left\{
\cosh \Xi
\right\}^{m_1}
\right\}^{m_{2}}
\!\!\!
\ldots
\right\}^{m_{\ell-1}}
\right\}^{m_\ell}
\nonumber
\end{eqnarray}
with
\bd
\Xi =\sum_{k=1}^{\ell}z_k a_k,~~~~~~~
a_\ell=\sqrt{\beta q_\ell \pi_\ell},~~~~~~~
a_k=\sqrt{\beta(q_k\pi_k\minus q_{k+1}\pi_{k+1})}~~~~(k<\ell)
\ed
It is not a priori obvious how the various order parameters will
scale with $m_1$ for $m_1\to \infty$. Here we will make a scaling ansatz
which we then show to lead selfconsistently to solutions of our
saddle-point equations, which satisfy all physical and
mathematical requirements. We assume that $\Xi=\order(m_1^0)$ as
$m_1\to\infty$, and we consequently put
\bd
a_k=\tilde{a}_k \zeta_k,~~~~~~
z_k= \tilde{z}_k \zeta^{-1}_k ~~~~~~(k=1, \ldots, \ell)
\ed
such that $\Xi =  \sum_{k=1}^{\ell}\tilde {z}_k \tilde{a}_k$, with
$\tilde{a}_k=\order(m_1^0)$ and $\tilde{z}_k=\order(m_1^0)$, and
with scaling functions $\zeta_k=\zeta_k(m_1)$ which will be determined in
due course.
As a consequence, we can work out the term $K_\ell$ in
(\ref{eq:f_iterative_L}) for large $m_1$ as follows:
\begin{eqnarray*}
K_\ell & = &\int\!\!Dz_\ell
\left\{
\int\!\!Dz_{\ell-1}
\left\{
\ldots
\int\!\!Dz_2
\left\{
\int\!\! \frac{d \tilde{z}_{1}}{\sqrt{2 \pi}  \zeta_1}~
e^{(
\zeta _1 ^2 m_1 \ln \cosh \Xi - \frac{1}{2} \tilde{z}_1 ^2)/\zeta _1 ^2}
\right\} ^{m_{2}}
\!\!\!
\ldots
\right\}^{m_{\ell-1}}
\right\}^{m_\ell}
\end{eqnarray*}
This expression immediately suggests that the appropriate $m_1\to\infty$ scaling
is obtained upon choosing  $\zeta _1 = m_1 ^{-1/2}$. As a result
of this choice, the $\tilde{z}_1$ integration can for $m_1\to\infty$ in leading order be carried out by steepest
descent, and we arrive at:
\bd
K_\ell  \simeq  \int\!Dz_\ell
\left\{
\int\!Dz_{\ell-1}
\left\{
\ldots
\left\{
\int\!Dz_2~
e^{ m_1 m_2  k_1( \tilde{z}_1 ^0,\tilde{z}_2,\cdots,\tilde{z}_\ell) }
\right\}^{m_3}
\!\!\!
\ldots
\right\}^{m_{\ell-1}}
\right\}^{m_\ell}
\ed
\bd
k_1( \tilde{z}_1,\tilde{z}_2,\cdots,\tilde{z}_\ell)
= \ln \cosh \Xi - \frac{1}{2}\tilde{z}_1 ^2
\ed
where $\tilde{z}_1^0$ denotes the saddle-point of $k_1$ with respect to variation of
$\tilde{z}_1$, i.e.
\begin{eqnarray*}
\frac{\partial k_1}{\partial \tilde{z}_1}|
_{( \tilde{z}_1 ^0 ,\tilde{z}_2,\cdots,\tilde{z}_\ell)}=0&&:~~~~~~
\tilde{z}_1 ^0  =  \tilde{a}_1 \tanh
\Xi(\tilde{z}_1 ^0, \tilde{z}_2,\cdots,\tilde{z}_\ell)
\end{eqnarray*}
Hence
$ \tilde{z}_1 ^0$ is a function of $\tilde{z}_2, \cdots,\tilde{z}_\ell$.
We now see that one can repeat this procedure and carry out all the Gaussian
integrations iteratively by steepest descent, upon setting $\zeta_k=(m_1 m_2 \cdots m_k)^{-1/2}$, which gives
in leading order
\begin{eqnarray}
K_\ell & \simeq &
e^{ m_1 m_2  \ldots m_{\ell-1} m_\ell
k_\ell( \tilde{z}_1 ^0,\tilde{z}_2^0,\ldots,\tilde{z}_{\ell-1}^0,\tilde{z}_\ell^0 )}
\label{eq:leading_K}
\end{eqnarray}
in which the functions $k_j(\ldots)$ are defined recursively via:
\begin{eqnarray*}
 k_j ( \tilde{z}_1^0,\ldots,
\tilde{z}_{j-1}^0, \tilde{z}_j,\ldots,\tilde{z}_\ell)
 &=&
 k_{j-1} ( \tilde{z}_1^0,\ldots,
\tilde{z}_{j-1}^0, \tilde{z}_j,\ldots,\tilde{z}_\ell) - \frac{1}{2}\tilde{z}_j^2
~~~~~(j>1)
\end{eqnarray*}
\begin{eqnarray*}
k_1( \tilde{z}_1,\cdots,\tilde{z}_\ell)
&=& \ln \cosh \Xi(\tilde{z}_1,\cdots,\tilde{z}_\ell) - \frac{1}{2}\tilde{z}_1 ^2
\end{eqnarray*}
giving for $j>1$:
\begin{eqnarray}
 k_j ( \tilde{z}_1^0,\ldots,
\tilde{z}_{j-1}^0, \tilde{z}_j,\ldots,\tilde{z}_\ell)
& =&
 \ln \cosh \Xi (\tilde{z}_1 ^0,\cdots,
\tilde{z}_{j-1}^0, \tilde{z}_j ,\cdots,\tilde{z}_\ell)
\nonumber \\
&&
 - \frac{1}{2} \tilde{z}_j ^2
- \frac{1}{2}\sum_{k=1}^{j-1} (\tilde{z}_k ^0) ^2
\label{eq:ks_found}
\end{eqnarray}
The identity \bd \frac{\partial}{ \partial \tilde{z}_j} k_{j}(
\tilde{z}_1^0,\cdots, \tilde{z}_{j-1}^0, \tilde{z}_j
,\cdots,\tilde{z}_\ell) =  \tilde{a}_j \tanh \Xi(\tilde{z}_1
^0,\cdots, \tilde{z}_{j-1}^0, \tilde{z}_j ,\cdots,\tilde{z}_\ell)-
\tilde{z}_j \ed shows that the values of all  $\{\tilde{z}_j^0\}$
are ultimately to be solved from the following $\ell$ coupled
saddle-point equations:
\begin{eqnarray}
\tilde{z}_j ^0 = \tilde{a}_j \tanh \Xi
 ( \tilde{z}_1^0,\ldots,\tilde{z}_\ell^0) ~~~~~~~~ ( j=1, \ldots, \ell)
 \label{eq:z0_SPE}
\end{eqnarray}
We
insert (\ref{eq:leading_K}) and (\ref{eq:ks_found}) into
(\ref{eq:f_iterative_L}), and find  that for $m_1\to\infty$ the free energy per spin
reduces in leading order simply to
\begin{eqnarray}
f& \simeq & \frac{1}{2}\pi_1 q_1-\frac{1}{\beta} \log 2
+\frac{1}{4}\sum_{k=2}^\ell q^2_k\pi_k m_1 m_2 \cdots
m_{k-1}(m_k-1) \nonumber \\ && +\frac{1}{4}q_1^2 \pi_1(m_1\!-\!1)
-\beta^{-1}\left[
 \ln \cosh \Xi (\tilde{z}_1 ^0,\cdots,\tilde{z}_\ell^0)
 - \frac{1}{2}\sum_{k=1}^{\ell} (\tilde{z}_k ^0) ^2\right]
\label{eq:f_large_m1}
\end{eqnarray}
with $\Xi(\tilde{z}_1^0,\ldots,\tilde{z}_\ell^0)=\sum_{k=1}^\ell
\tilde{a}_k \tilde{z}_k^0$. \vsp

 We can now derive the
saddle-point equations for $m_1\to\infty$ by variation of
(\ref{eq:f_large_m1}) with respect to the order parameters
$\{q_k\}$. Let us first derive the equation for $q_1$ in leading
order:
\begin{eqnarray*}
0& = &
\frac{1}{2}\pi_1
 +\frac{1}{2}q_1 \pi_1(m_1\!-\!1)
 -\frac{1}{\beta}\sum_{k=1}^\ell \frac{\partial
 \tilde{z}_k^0}{\partial q_1}
 \frac{\partial}{\partial \tilde{z}_k^0}\left[
 \ln \cosh \Xi - \frac{1}{2}\sum_{k=1}^{\ell} (\tilde{z}_k ^0)^2\right]
 \\
&&-\frac{1}{\beta} \sum_{k=1}^\ell \tanh\Xi \frac{\partial
\Xi}{\partial \tilde{a}_k}\frac{\partial\tilde{a}_k}{\partial
q_1}~+~\ldots~~~~~~~~(m_1\to\infty)
 \end{eqnarray*}
The term with the partial derivatives
$\partial/\partial\tilde{z}_k^0$ is identical zero due to the
saddle-point equations (\ref{eq:z0_SPE}), and hence, upon working
out the derivatives $\partial\tilde{a}_k/\partial
q_1=\zeta_k^{-1}\partial a_k/\partial q_1$ and using
(\ref{eq:z0_SPE}) to eliminate occurrences of $\tilde{z}_1^0$, we
just retain in leading order
\begin{eqnarray*}
0& = & \frac{1}{2}\pi_1 m_1[
 q_1
 -  \tanh^2\Xi]~+~\ldots ~~~~~~(m_1\to \infty)
\end{eqnarray*}
Hence
\begin{eqnarray*}
 q_1 &=& \tanh^2 \Xi ~~~~~~(m_1\to\infty)
\end{eqnarray*}
Similarly we can deal with the derivatives of $f$
(\ref{eq:f_large_m1}) with respect to the other $q_j$. With the
simple identity
\bd
 a_k\frac{\partial a_k}{\partial q_j}=
 \frac{1}{2}\beta
 \frac{\partial}{\partial q_j}[q_k\pi_k-
 q_{k+1}\pi_{k+1}]
 =
  \frac{1}{2}\beta \pi_j[\delta_{j,k}-\delta_{j,k+1}]
\ed the saddle-point equation for $q_j$ with $j>1$ is seen to
become in leading order
\begin{eqnarray*}
0& = &
 \frac{1}{2}q_j \pi_j m_1 \ldots m_{j-1}(m_j\!-\!1)
 -\frac{1}{\beta}\sum_{k=1}^\ell
\frac{\partial
 \tilde{z}_k^0}{\partial q_j}
 \frac{\partial}{\partial \tilde{z}_k^0}\left[
 \ln \cosh \Xi - \frac{1}{2}\sum_{k=1}^{\ell} (\tilde{z}_k ^0)^2\right]
 \\
&&\hspace*{20mm}-\frac{1}{\beta} \sum_{k=1}^\ell \tanh\Xi
\frac{\partial \Xi}{\partial
\tilde{a}_k}\frac{\partial\tilde{a}_k}{\partial q_j} ~+~\ldots
~~~~~~~~(m_1\to\infty)
\\
& = &
 \frac{1}{2}q_j \pi_j m_1 \ldots m_{j-1}(m_j\!-\!1)
-\frac{1}{\beta} \tanh^2\Xi \sum_{k=1}^\ell m_1\ldots m_k a_k
\frac{\partial a_k}{\partial q_j} ~+~\ldots
\\
& = & \frac{1}{2}\pi_j m_1 m_2 \ldots m_{j-1}(m_j\!-\!1)\left[ q_j
- \tanh^2\Xi \right]~+~\ldots
 \end{eqnarray*}
Hence, for all $j\geq 1$ one simply has
\begin{eqnarray}
 q_j &=& q^\star=\tanh^2 \Xi ~~~~~~(m_1\to\infty)
 \label{eq:qs_found}
\end{eqnarray}

\begin{figure}[t]
\vspace*{1mm} \setlength{\unitlength}{1.04mm}
\begin{picture}(150,85)
 \put(7,20){\epsfxsize=91\unitlength\epsfbox{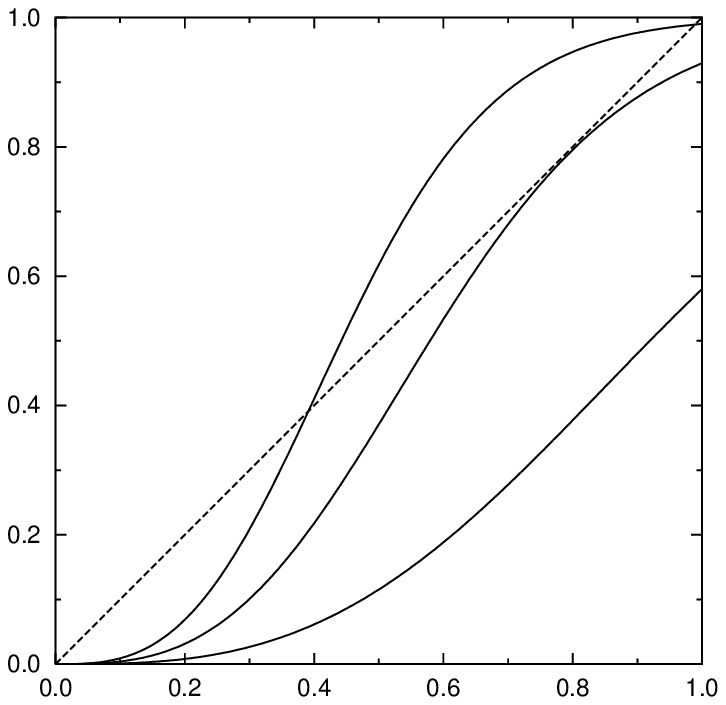}}
\put(84,20){\epsfxsize=91\unitlength\epsfbox{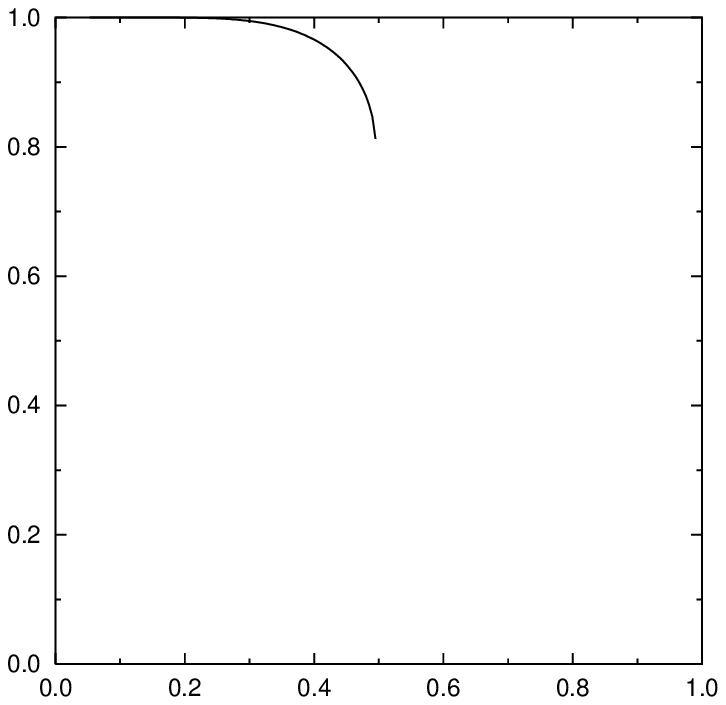}}
 \put(-2,52){ $F_x(q)$}   \put(39,13){$q$}
 \put(80,52){$q^\star$} \put(119,13){$x^{-1}$}
\end{picture}
\vspace*{-16mm} \caption{Properties of the equation $q=F_x(q)$,
where $F_x(q)=\tanh^2(x q^{\frac{3}{2}})$, from which to solve the
$m_1\to \infty$ order parameter amplitude $q^\star$ (see main
text). Left panel: shape of the function $F_x(q)$ for
$x\in\{1,2,3\}$ (solid curves, from bottom to top), together with
the diagonal (dashed) whose intersection with $F_x(q)$ defines
$q^\star$. Right panel: largest solution $q^\star$ of the equation
$q=F_x(q)$ as a function of $x^{-1}$. The jump occurs at
 $q^\star\approx 0.7911$ and
$x^{-1}\approx 0.4958$.} \label{fig:9}
\end{figure}

The final stage of our argument is to determine the amplitude
$q^\star$ in (\ref{eq:qs_found}).
 We note that, by virtue of (\ref{eq:z0_SPE}), we can write
$\Xi=\sum_{k=1}^\ell \tilde{a}_k \tilde{z}_k^0$ in leading order
as $\Xi \simeq Y \tanh \Xi$, where $Y = \sum _{k=1}^\ell
\tilde{a}_k ^2$. It now follows, upon inserting
(\ref{eq:qs_found}) into this relation, that $q^\star$ is the
solution of
\be
 q^\star=\tanh^2[Y\sqrt{q^\star}]~~~~~~(m_1\to\infty)
\label{eq:eqn_for_qstar} \ee In working out the quantity $Y$ in
leading order for $m_1\to\infty$ we have to take care not to
forget about the $m_1$ dependence of the factors $\pi_k$, defined
in (\ref{eq:pis}). In particular,
 \bd
 \pi_j\zeta_j^{-2}
 = \frac{\epsilon_{j}}{\mu_{j}}+
 \sum_{k=j+1}^{L}\frac{1}{m_{j+1}\ldots m_{k}}
 ~\frac{\epsilon_{k}}{\mu_{k}}
\ed
Using this relation we find
\begin{eqnarray*}
Y & = & \beta q^\star\left[ \sum _{j=1}^{\ell-1} (\pi_j
-\pi_{j+1})\zeta_j^{-2}+
  \pi_\ell \zeta_\ell^{-2}\right]~+~\ldots~~~~~~(m_1\to\infty)
  \\
& = & \beta q^\star\left[ \sum _{j=1}^{\ell}
\frac{\epsilon_{j}}{\mu_{j}}+
 \sum_{j=\ell+1}^{L}\frac{1}{m_{\ell+1}\ldots m_{j}}
 ~\frac{\epsilon_{j}}{\mu_{j}}
 \right]
~+~\ldots
\end{eqnarray*}
Our final result on the limit $m_1\to\infty$ is the following: in
the phase $SG_\ell$ one has $q_1=\ldots = q_\ell=q^\star >0$ and
$q_{\ell+1}=\ldots=q_L=0$, where $q^\star$ is the solution of
\begin{eqnarray}
q^\star & = & \tanh^2[ \beta \bar{\omega}_\ell (q^*) ^{3/2} ]
\label{eq:q_star_found}
\\
\bar{\omega}_\ell &=&
 \sum _{j=1}^{\ell}
\frac{\epsilon_{j}}{\mu_{j}}+
 \sum_{j=\ell+1}^{L}\frac{1}{m_{\ell+1}\ldots m_{j}}
 ~\frac{\epsilon_{j}}{\mu_{j}}
 \label{eq:bar_omega}
\end{eqnarray}
We note that, for any $\bar{\omega}_\ell>0$, equation
(\ref{eq:q_star_found}) will have a non-zero solution (and hence
phase $SG_\ell$ will indeed exist) when $T$ is sufficiently low,
and that $\lim_{T\to 0}q^\star=1$.  The behaviour of equation
(\ref{eq:q_star_found}) as a function of the the effective control
parameter $x=\beta\bar{\omega}_\ell$ is illustrated in figure
\ref{fig:9}. \vsp

From (\ref{eq:q_star_found},\ref{eq:bar_omega}) we can immediately
extract the critical temperature $T_{p,\ell}^{\rm 1st}$, signaling
the (first order) transition $P\to SG_\ell$. The condition for
nonzero solutions of (\ref{eq:q_star_found}) to bifurcate follows
from solving (\ref{eq:q_star_found}) simultaneously with the
equation $1 = \frac{d}{dq^\star} \tanh^2[ \beta \bar{\omega}_\ell
(q^*)^{3/2} ]$, which leads us to the explicit and simple result
\be
 T_{p,\ell}^{\rm 1st} =3\bar{\omega}_\ell~ q^\star(1-q^\star)
~~~~~~~~~~
 q^\star=\tanh^2[ \frac{\sqrt{q^\star}}{3(1-q^\star)}]
\label{eq:qstar_transition} \ee
Numerical solution of
(\ref{eq:qstar_transition}) shows that $q^\star\approx 0.7911$ and
$T_{p,\ell}^{\rm 1st}/\bar{\omega}_\ell\approx 0.4958$. In order
to assess the dependence on $\ell$ of the transition temperatures
$T_{p,\ell}^{\rm 1st}$, we subtract
\begin{eqnarray*}
T_{p, \ell+1} ^{{\rm 1st}} -T_{p,\ell} ^{{\rm 1st}} &=& 3
q^\star(1-q^\star)[\bar{\omega}_{\ell+1}-\bar{\omega}_\ell]
\\
&=& 3 q^\star(1-q^\star)[\frac{m_{\ell+1}- 1}{m_{\ell+1}}]\left[
\frac{\epsilon_{\ell+1}}{\mu_{\ell+1}} +
 \sum_{j=\ell+2}^{L}\frac{1}{m_{\ell+2}\ldots m_{j}}
 \frac{\epsilon_{j}}{\mu_{j}}
\right]
\end{eqnarray*}
Thus $T_{p,\ell+1}^{\rm 1st}>T_{p,\ell}$ for $m_{\ell+1}>1$ (i.e.
for  relatively low noise levels in the level-$(\ell+1)$
couplings), whereas $T_{p,\ell+1}^{\rm 1st}<T_{p,\ell}$ for
$m_{\ell+1}<1$ (for relatively high noise levels in the
level-$(\ell+1)$ couplings). Upon inserting the appropriate values
of the control parameters, one can easily verify that the general
results obtained in this sub-section regarding the values of the
order parameters in the phase  $SG_\ell$ and of the $P\to SG_\ell$
transition temperatures $T_{p,\ell}^{\rm 1st}$ for arbitrary $L$
are in perfect agreement with our solutions as calculated for the
special choices $L=2$ and $L=3$ in previous sections.

\subsection{The Limit $m_1\to 0$ in the $L$-level Hierarchy}

We next turn to the limit $m_1\to 0$, for fixed $T_1$ (i.e. $T\to
0$). We observed in the phase diagrams of figures \ref{fig:1},
\ref{fig:3}, \ref{fig:5} and \ref{fig:7} for $L=2$ and $L=3$ that
along the line $T=0$ one finds critical values
$T_{\ell-1,\ell}^{\rm 2nd}$ for $T_1$ which signal the creation of
phase $SG_{\ell}$ (where $q_1\geq \ldots \geq q_{\ell}>0$ and
$q_{\ell+1}=\ldots=q_L=0$) from $SG_{\ell-1}$ (where $q_1\geq
\ldots \geq q_{\ell-1}>0$ and $q_{\ell}=\ldots=q_L=0$). In this
subsection we calculate these critical values for the general
$L$-level system. Our starting point is again expression
(\ref{eq:f_iterative_L}) for the free energy per spin (obtained by
putting $q_{\ell+1}=\ldots=q_L=0$). In order to make all
occurrences of $m_1$ explicit, we substitute $T=T_1 m_1$, and we
write the factors $\pi_\ell$ of (\ref{eq:pis}) as \bd
\pi_\ell=\frac{\tilde{\pi}_\ell}{m_1}:~~~~~~
 \tilde{\pi}_1
 =\frac{\epsilon_{1}}{\mu_{1}} +
 \sum_{\ell^\prime=2}^{L}\frac{1}{m_{2}\ldots m_{\ell^\prime}}
 ~\frac{\epsilon_{\ell^\prime}}{\mu_{\ell^\prime}}
,~~~~~~
  \tilde{\pi}_{\ell>1}
 =\sum_{\ell^\prime=\ell}^{L}\frac{1}{m_{2}\ldots m_{\ell^\prime}}
 ~\frac{\epsilon_{\ell^\prime}}{\mu_{\ell^\prime}}
\ed so that $\tilde{\pi}_\ell=\order(m_1^0)$ for all $\ell$ as
$m_1\to 0$. This gives
\begin{eqnarray}
 m_1f(q_1,\ldots,q_\ell)& = &
\frac{1}{4}\tilde{\pi}_1 -\frac{1}{4}\tilde{\pi}_1(q_1\minus 1)^2
  +\order(m_1^2)
  -\frac{T_1
 m_1}{\prod_{k=2}^{\ell}m_{k}}\log K_\ell
 \nonumber
 \\
 && \hspace*{-25mm}
 +\frac{1}{4}m_1\left[
 q_1^2 \tilde{\pi}_1+
  q^2_2\tilde{\pi}_2(m_2\minus 1)
 +\sum_{k=3}^\ell q^2_k\tilde{\pi}_k m_2 \ldots
 m_{k-1}(m_k\minus 1)
\right] \label{eq:m1f}
\end{eqnarray}
with
\begin{eqnarray}
K_\ell & = &
 \int\!Dz_\ell \left\{ \int\!Dz_{\ell-1} \left\{ \ldots \int\!Dz_2
 \left\{ \int\!Dz_{1}
 \left\{ \cosh \Xi \right\}^{m_1} \right\}^{m_{2}} \!\!\! \ldots
 \right\}^{m_{\ell-1}} \right\}^{m_\ell}
\nonumber\\
 \Xi&=& m_1^{-1}\tilde{\Xi},~~~~~~ \tilde{\Xi}
=\sum_{k=1}^{\ell}z_k \tilde{a}_k \nonumber\\
\tilde{a}_\ell&=&\sqrt{T_1^{-1} q_\ell \tilde{\pi}_\ell},~~~~~~~
\tilde{a}_{k<\ell}=\sqrt{T_1^{-1} (q_k\tilde{\pi}_k\minus
q_{k+1}\tilde{\pi}_{k+1})} \nonumber
\end{eqnarray}
Since $\tilde{\Xi}=\order(m_1^0)$ as $m_1\to 0$, by virtue of our
definitions, we may conclude that $K_\ell=\order(m_1^0)$: \bd
 \left\{ \cosh \Xi
 \right\}^{m_1}=2^{-m_1}\left[e^{\tilde{\Xi}/m_1}+e^{-\tilde{\Xi}/m_1}\right]^{m_1}
 =e^{|\tilde{\Xi}|}(1+\order(m_1))~~~~~~~~(m_1\to 0)
\ed
\begin{eqnarray}
K_\ell & = &
 \int\!Dz_\ell \left\{ \int\!Dz_{\ell-1} \left\{ \ldots \int\!Dz_2
 \left\{ \int\!Dz_{1}
 e^{|\tilde{\Xi}|}
\right\}^{m_{2}} \!\!\! \ldots
 \right\}^{m_{\ell-1}} \right\}^{m_\ell}
 +\ldots
 \label{eq:Kl_m1zero}
\end{eqnarray}
Similarly one gets in leading order for $m_1\to 0$:
\begin{eqnarray}
\frac{1}{K_1}\int\!Dz_1~\left\{ \cosh \Xi
 \right\}^{m_1}\tanh\Xi&=&
\frac{\int\!Dz_1~e^{|\tilde{\Xi}|}~{\rm sgn}(\tilde{\Xi})}
{\int\!Dz_1~e^{|\tilde{\Xi}|}} ~+~\ldots
 \label{eq:tanh_av}
\end{eqnarray}
 According to (\ref{eq:m1f}) the saddle-point
value of $q_1$ is of the form $q_1=1-\order(\sqrt{m_1})$. If,
however, we substitute $q_1=1-\kappa \sqrt{m_1}+\order(m_1)$ into
(\ref{eq:m1f}), we find that the saddle-point obeys $\kappa=0$,
hence the true scaling of $q_1$ is
\be
 q_1=1-\kappa m_1+\order(m_1^2)
 \label{eq:scaling_q1}
\ee
 Insertion of (\ref{eq:scaling_q1}) into (\ref{eq:m1f}) gives
\begin{eqnarray}
 m_1f(q_2,\ldots,q_\ell)& = &
\frac{1}{4}(1+ m_1)\tilde{\pi}_1
  +\order(m_1^{2})
  -\frac{T_1 m_1}{\prod_{k=2}^{\ell}m_{k}}\log K_\ell|_{q_1=1}
 \nonumber
 \\
 && \hspace*{-25mm}
 +\frac{1}{4}m_1\left[
  q^2_2\tilde{\pi}_2(m_2- 1)
 +\sum_{k=3}^\ell q^2_k\tilde{\pi}_k m_2 \ldots
 m_{k-1}(m_k- 1)
\right]
\end{eqnarray}
and hence the $m_1\to 0$ saddle-point equations for the order
parameters $q_{\ell^\prime}$ with $1<\ell^\prime\leq \ell$ are
given by
\begin{eqnarray}
 q_2&=&
 \frac{2T_1}{\tilde{\pi}_2 (m_2\minus 1)
(\prod_{k=2}^{\ell}m_{k})}~\frac{\partial}{\partial q_2}\log
\tilde{K}_\ell \\
  q_{\ell^\prime}&=&
   \frac{2T_1}{\tilde{\pi}_{\ell^\prime}
(m_{\ell^\prime}\minus 1)(\prod_{k=2}^{\ell^\prime-1}m_k)
(\prod_{k=2}^{\ell}m_{k})}~\frac{\partial}{\partial
q_{\ell^\prime}}\log \tilde{K}_\ell
\end{eqnarray}
with $\tilde{K}_\ell(q_2,\ldots,q_\ell)=\lim_{q_1\to
1}\lim_{m_1\to 0} K_\ell$ (as given by (\ref{eq:Kl_m1zero})). At
this stage it is advantageous to use our earlier results
(\ref{eq:spRSfinal1},\ref{eq:spRSfinal2}), which in the present
context and in combination with (\ref{eq:tanh_av}) translate into
\begin{eqnarray}
q_{2} &=&\bra \ldots \bra ~\left[
\frac{\int\!Dz_1~e^{|\tilde{\Xi}|}~{\rm sgn}(\tilde{\Xi})}
{\int\!Dz_1~e^{|\tilde{\Xi}|}}\right]^2~
 \ket_{2} \ldots  \ket_{\ell}
\label{eq:q2}
\\
q_{k>2} &=&\bra \ldots \bra ~\left[\bra\ldots \bra \left[
\frac{\int\!Dz_1~e^{|\tilde{\Xi}|}~{\rm sgn}(\tilde{\Xi})}
{\int\!Dz_1~e^{|\tilde{\Xi}|}}\right]
\ket_2\ldots\ket_{k-1}\right]^2~
 \ket_{k} \ldots  \ket_{\ell}
\label{eq:qmore}
\end{eqnarray}
with \bd \tilde{K}_1 =\int\!\!Dz_1~e^{|\tilde{\Xi}|},~~~~~~
\tilde{K}_{\ell^\prime}= \int\!\!Dz_{\ell^\prime}
\tilde{K}_{\ell^\prime-1} ^{m_{\ell^\prime}},~~~~~~
 \bra f\ket_{\ell^\prime}=
\tilde{K}_{\ell^\prime}^{-1}\int\!Dz_{\ell^\prime}~\tilde{K}_{\ell^\prime-1}^{m_{\ell^\prime}}
f(z_{\ell^\prime}) \ed \bd \tilde{\Xi}= \sum_{k=1}^{\ell}z_k
\tilde{a}_k|_{q_1=1} =z_1\tilde{a}_1+\tilde{\Xi}_1
 \ed
  Working out the $z_1$ integrations in (\ref{eq:q2},\ref{eq:qmore})
  leads to \be
\frac{\int\!Dz_1~e^{|\tilde{\Xi}|}~{\rm sgn}(\tilde{\Xi})}
{\int\!Dz_1~e^{|\tilde{\Xi}|}}
=
 \frac{\sinh(\tilde{\Xi}_1)
 +\frac{1}{2}
e^{\tilde{\Xi}_1} {\rm Erf}
\left[\frac{\tilde{a}_1^2+\tilde{\Xi}_1}{\tilde{a}_1\sqrt{2}}\right]
-
\frac{1}{2} e^{-\tilde{\Xi}_1} {\rm Erf}
\left[\frac{\tilde{a}_1^2-\tilde{\Xi}_1}{\tilde{a}_1\sqrt{2}}\right]
} {\cosh(\tilde{\Xi}_1)
 +\frac{1}{2}
e^{\tilde{\Xi}_1} {\rm Erf}
\left[\frac{\tilde{a}_1^2+\tilde{\Xi}_1}{\tilde{a}_1\sqrt{2}}\right]
+ \frac{1}{2} e^{-\tilde{\Xi}_1} {\rm Erf}
\left[\frac{\tilde{a}_1^2-\tilde{\Xi}_1}{\tilde{a}_1\sqrt{2}}\right]
} \label{eq:integrals} \ee The transition value
$T_{\ell-1,\ell}^{\rm 2nd}$ for $T_1$ is the one which gives a
second order bifurcation of $q_\ell$ away from zero in
(\ref{eq:qmore}). \vsp

Let us work out these results first for $\ell=2$. Here one simply
has $\tilde{a}_1= (\sqrt{\tilde{\pi}_1-
q_2\tilde{\pi}_{2}})/\sqrt{T_1}$ and $\tilde{\Xi}_1=z_2 \sqrt{
q_2\tilde{\pi}_2}/\sqrt{T_1}$, with $q_2$ to be solved from
\begin{eqnarray}
q_{2} &=&\bra ~\left[
 \frac{\sinh(\tilde{\Xi}_1)
 +\frac{1}{2}
e^{\tilde{\Xi}_1} {\rm Erf}
\left[\frac{\tilde{a}_1^2+\tilde{\Xi}_1}{\tilde{a}_1\sqrt{2}}\right]
-
\frac{1}{2} e^{-\tilde{\Xi}_1} {\rm Erf}
\left[\frac{\tilde{a}_1^2-\tilde{\Xi}_1}{\tilde{a}_1\sqrt{2}}\right]}
{\cosh(\tilde{\Xi}_1)
 +\frac{1}{2}
e^{\tilde{\Xi}_1} {\rm Erf}
\left[\frac{\tilde{a}_1^2+\tilde{\Xi}_1}{\tilde{a}_1\sqrt{2}}\right]
+ \frac{1}{2} e^{-\tilde{\Xi}_1} {\rm Erf}
\left[\frac{\tilde{a}_1^2-\tilde{\Xi}_1}{\tilde{a}_1\sqrt{2}}\right]
} \right]^2~
 \ket_{2}
 \nonumber
 \\
 &=&
\bra ~\tilde{\Xi}_1^2\left[
 \frac{1+
{\rm Erf}\left[\frac{\tilde{a}_1}{\sqrt{2}}\right] +
\frac{\sqrt{2}}{\tilde{a}_1\sqrt{\pi}}e^{-\frac{1}{2}\tilde{a}^2_1}
 +\order(\tilde{\Xi}_1)
}
 {1 + {\rm Erf}
\left[\frac{\tilde{a}_1}{\sqrt{2}}\right] } \right]^2~
 \ket_{2}
  \\
 &=&
  \frac{q_2\tilde{\pi}_2}{T_1}
\left[ 1+
 \frac{
 \frac{\sqrt{2T_1}}{\sqrt{\tilde{\pi}_1}\sqrt{\pi}}
 e^{-\frac{1}{2}\tilde{\pi}_1/T_1}}
 {1 +{\rm Erf}
\left[\frac{ \sqrt{\tilde{\pi}_1}}{\sqrt{2T_1}}\right] } \right]^2
 +\order(q_2^{\frac{3}{2}})
\end{eqnarray}
Hence the condition for the second order  $SG_1\to SG_2$
transition,  $T_1=T_{1,2}^{\rm 2nd}$, is to be solved from
\begin{eqnarray}
1 &=&
 \sqrt{\frac{ \tilde{\pi}_2}{T_1}}
\left[ 1+
 \frac{
 (\frac{2T_1}{\tilde{\pi}_1\pi})^{\frac{1}{2}}~e^{-\frac{1}{2}\tilde{\pi}_1/T_1}}
 {1 +{\rm Erf}
\left[(\frac{\tilde{\pi}_1}{2T_1})^{\frac{1}{2}}\right] } \right]
\label{eq:case_l=2}
\end{eqnarray}
Equivalently, upon substituting $x^2=\tilde{\pi}_1/2T_1$:
\be
T_{1,2}^{\rm 2nd}=\frac{\tilde{\pi}_1}{2x^2},~~~~~~~~
 \sqrt{\frac{\tilde{\pi}_1}{2\tilde{\pi}_2}} =
 x+ \frac{1}{\sqrt{\pi}}\frac{e^{-x^2}}
 {1 +{\rm Erf}[x]}
\label{eq:case_l=2b}
\ee
Insertion of the parameter values appropriate to the $L=2$ and
$L=3$ examples, as studied in detail in a previous section, sheds interesting new light
on the nature of the jumps observed in the $SG_1\to SG_2$ transition
lines
of the $L=2$ and $L=3$ phase diagrams for small $T$:
\begin{eqnarray*}
m_2=0.5: &~~~~& T_1^{\rm jump}\approx 12.58,~~~~T_{1,2}^{\rm
2nd}\approx 9.884\\
m_2=1.5: &~~~~& T_1^{\rm jump}\approx 2.17,~~~~~~T_{1,2}^{\rm
2nd}\approx 2.095
\end{eqnarray*}
These results indicate that for low $T$ the $SG_1\to SG_2$
transition exhibits re-entrance, the extent of which decreases
with increasing $m_2$.
In fact, we found another second order transition line
from $SG_1$ to $SG_2$ which shows re-entrance for each $m_2$.
See Figure 11 and 12.
\begin{figure}[t]
\vspace*{0mm} \setlength{\unitlength}{1.04mm}
\begin{picture}(150,75)
 \put(7,20){\epsfxsize=71\unitlength\epsfbox{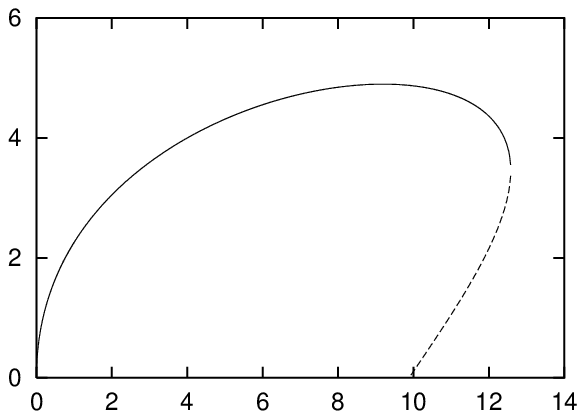}}
\put(84,20){\epsfxsize=71\unitlength\epsfbox{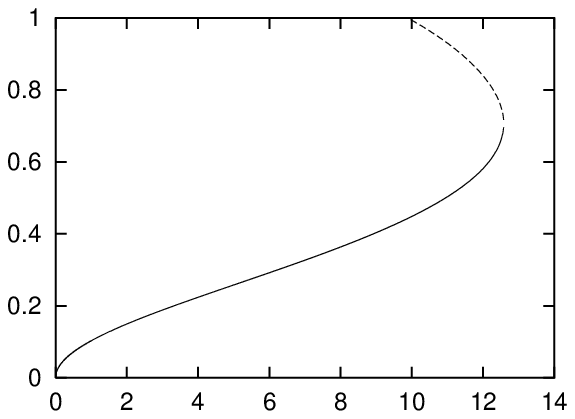}}
 \put(3,45){ $T$}   \put(42,15){$T_1$}
 \put(82,45){$q_1$} \put(121,15){$T_1$}
\end{picture}
\vspace*{-18mm} \caption{Re-entrance phenomena in the second order
$SG_1\to SG_2$ transition for $L=3$ and $m_2=0.5$. Left panel:
phase diagram in the $(T_1,T)$ plane. Right panel: value of the
order parameter $q_1$ at the transition line as a function of
$T_1$. In both cases the solid part of the curve indicates the
physical part of the transition line; the dashed part indicates
the non-physical (re-entrant) part. } \label{fig:11}
\end{figure}
\begin{figure}[t]
\vspace*{0mm} \setlength{\unitlength}{1.04mm}
\begin{picture}(150,75)
 \put(7,20){\epsfxsize=71\unitlength\epsfbox{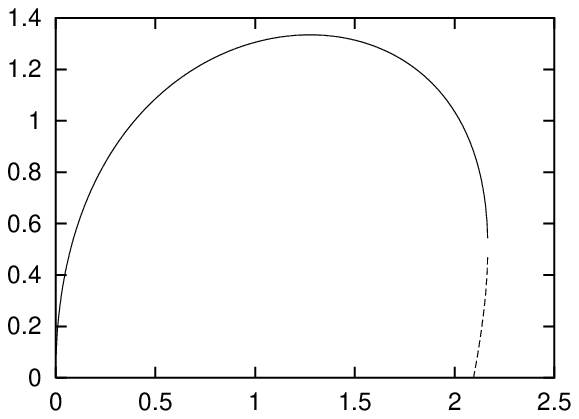}}
\put(84,20){\epsfxsize=71\unitlength\epsfbox{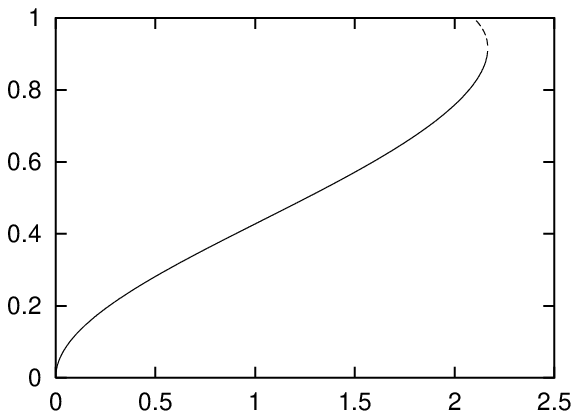}}
 \put(3,45){ $T$}   \put(42,15){$T_1$}
 \put(82,45){$q_1$} \put(121,15){$T_1$}
\end{picture}
\vspace*{-18mm} \caption{Re-entrance phenomena in the second order
$SG_1\to SG_2$ transition for $L=3$ and $m_2=1.5$. Left panel:
phase diagram in the $(T_1,T)$ plane. Right panel: value of the
order parameter $q_1$ at the transition line as a function of
$T_1$. In both cases the solid part of the curve indicates the
physical part of the transition line; the dashed part indicates
the non-physical (re-entrant) part. } \label{fig:12}
\end{figure}
On reflection, this is not entirely surprising, since for
$m_1\to 0$ one should expect replica symmetry to break.
It has been observed frequently in many disordered systems
(see e.g. \cite{Parisi})
that replica-symmetric
solutions show unphysical re-entrance phenomena in RSB regions
(which are removed by the proper RSB solution).
\vsp

For general values of $\ell$ we have to work out the following
continuous bifurcation condition of $q_\ell\neq 0$, derived from
equation (\ref{eq:qmore}): \bd \hspace*{-25mm} 1 = \lim_{q_\ell\to
0}\frac{\partial}{\partial q_\ell}
 \bra \ldots \bra
~\left[\bra\ldots \bra \left[
 \frac{\sinh(\tilde{\Xi}_1)
 +\frac{1}{2}
e^{\tilde{\Xi}_1} {\rm Erf}
\left[\frac{\tilde{a}_1^2+\tilde{\Xi}_1}{\tilde{a}_1\sqrt{2}}\right]
-
\frac{1}{2} e^{-\tilde{\Xi}_1} {\rm Erf}
\left[\frac{\tilde{a}_1^2-\tilde{\Xi}_1}{\tilde{a}_1\sqrt{2}}\right]
} {\cosh(\tilde{\Xi}_1)
 +\frac{1}{2}
e^{\tilde{\Xi}_1} {\rm Erf}
\left[\frac{\tilde{a}_1^2+\tilde{\Xi}_1}{\tilde{a}_1\sqrt{2}}\right]
+ \frac{1}{2} e^{-\tilde{\Xi}_1} {\rm Erf}
\left[\frac{\tilde{a}_1^2-\tilde{\Xi}_1}{\tilde{a}_1\sqrt{2}}\right]
} \right] \ket_2\ldots\ket_{k-1}\right]^2
 \ket_{k} \ldots  \ket_{\ell}
\ed
This is done in \ref{app:SGSG}, where we show that the value $T_{\ell-1,\ell}^{\rm 2nd}$
for $T_1$ which marks the continuous transition
from SG$_{\ell-1}$ to SG$_{\ell}$ is to be solved from
\begin{eqnarray}
1&=&\sqrt{\frac{\tilde{\pi}_{\ell}}{T_1}}
 \left\{ 1+\kappa + \sum_{k=2}^{\ell-1}m_2 \ldots m_{k-1}
(m_{k}-1)q_{k}\right \}
\label{eq:general_condition}
\end{eqnarray}
where $\kappa=\lim_{m_1\to 0}(1-q_1)/m_1$, where $q_\ell=0$, and where for the values of
the $\{q_k\}$ with $1<k<\ell$ one has to substitute the solution of
equations (\ref{eq:q2},\ref{eq:qmore}).
The quantity $\kappa$ is calculated along the lines of our
previous calculation of (\ref{eq:Kl_m1zero}) and
(\ref{eq:tanh_av}):
\begin{eqnarray}
\kappa&=&\lim_{m_1\to 0}\frac{1}{m_1}
\bra \ldots \bra~
\frac{\int\!Dz_1~\cosh^{m_1}(\Xi)[1-\tanh^2(\Xi)]}
{\int\!Dz_1~\cosh^{m_1}(\Xi)}~
 \ket_{2} \ldots  \ket_{\ell}
\nonumber\\
&=&
\lim_{m_1\to 0}\frac{1}{m_1}
\bra \ldots \bra~
\frac{\int\!Dz_1~\cosh^{m_1-2}(\tilde{\Xi}/m_1)}
{\int\!Dz_1~\cosh^{m_1}(\tilde{\Xi}/m_1)}~
 \ket_{2} \ldots  \ket_{\ell}
\end{eqnarray}
We know that $K_1=\int\!Dz_1~\cosh^{m_1}(\tilde{\Xi}/m_1)\to
\int\!Dz_1~e^{|\tilde{\Xi}|}$ for $m_1\to 0$. Let us calculate the
term in the numerator in leading order.
\begin{eqnarray*}
 \int\!Dz_1~\cosh^{m_1-2}(\tilde{\Xi}/m_1) &=&
I_+({\tilde{\Xi}_1}) + I_- ({\tilde{\Xi}_1})
\end{eqnarray*}
with
\begin{eqnarray*}
 I_+({\tilde{\Xi}_1})& =&\int_{\Xi >0} \!Dz_1~\cosh^{m_1 -2}(\tilde{\Xi}/m_1)
=\int_{-\frac{\tilde{\Xi}_1}{\tilde{a}_1}}^{\infty} \!Dz_1
\{ e^{\tilde{\Xi}/m_1}( 1 + e^{-2\tilde{\Xi}/m_1})/2\}^{m_1-2}
 \\
 I_-( {\tilde{\Xi}_1})& =&\int_{\Xi <0}\!Dz_1~\cosh^{m_1 -2}(\tilde{\Xi}/m_1)
=\int^{-\frac{\tilde{\Xi}_1}{\tilde{a}_1}}_{-\infty} \!Dz_1
\{ e^{-\tilde{\Xi}/m_1}( 1 + e^{2\tilde{\Xi}/m_1})/2\}^{m_1-2}\\
& =& I_+(-{\tilde{\Xi}_1}).
\end{eqnarray*}
Putting $b=2 \tilde{a}_1/m_1, b_n=nb$,
fixing $m_1$ to a small number, introducing a cut-off  $\varepsilon$
 and using
$(1+x)^{-2}=\sum_{n=0}^{\infty}( n+1) (-x)^n$ and
 the asymptotic
properties of the error function \cite{AbraSteg},
we obtain
\begin{eqnarray*}
\hspace*{-5mm}
 I_+({\tilde{\Xi}_1})& =&
\int_{-\frac{\tilde{\Xi}_1}{\tilde{a}_1}}^{\infty} \!Dz_1
 e^{-\tilde{\Xi}(2/m_1 -1)}\{( 1 + e^{-2\tilde{\Xi}/m_1})/2\}^{m_1-2}\\
&=&
\int_{-\frac{\tilde{\Xi}_1}{\tilde{a}_1}}^{\infty} \!Dz_1
 e^{-\tilde{\Xi}(2/m_1 -1)}\{( 1 + e^{-2\tilde{\Xi}/m_1})/2\}^{-2}
( 1 + {\cal O}(m_1))\\
&=&
4 \lim_{\varepsilon \rightarrow +0}
\int_{\varepsilon - \frac{\tilde{\Xi}_1}{\tilde{a}_1}}^{\infty} \!Dz_1
 e^{-\tilde{\Xi}(2/m_1 -1)}
\sum_{n=0}^{\infty}( n+1) (-)^n  e^{-2 n \tilde{\Xi}/m_1}
( 1 + {\cal O}(m_1))\\
&=&
4 \lim_{\varepsilon \rightarrow +0}
\sum_{n=0}^{\infty}( n+1) (-)^n
\int_{\varepsilon - \frac{\tilde{\Xi}_1}{\tilde{a}_1}}^{\infty} \!Dz_1
 e^{- (b z_1 + \frac{2\tilde{\Xi}_1}{m_1})(n+1 -\frac{m_1}{2})}
( 1 + {\cal O}(m_1))\\
&=&
4 \lim_{\varepsilon \rightarrow +0}
\sum_{n=0}^{\infty}( n+1) (-)^n
 e^{-  \frac{2\tilde{\Xi}_1}{m_1}(n+1 -\frac{m_1}{2})}
 e^{ b^2 (n+1 -\frac{m_1}{2})^2/2}\\
&& \times
\frac{1}{2} \{ 1- {\rm Erf}
\left[\frac{\varepsilon - \frac{\tilde{\Xi}_1}{\tilde{a}_1}+
b (n+1 -\frac{m_1}{2})}{\sqrt{2}}\right]\}
+\ldots \\
&=&
4 \lim_{\varepsilon \rightarrow +0}
\sum_{n=0}^{\infty}( n+1) (-)^n
 e^{-  \frac{2\tilde{\Xi}_1}{m_1}(n+1 -\frac{m_1}{2})}
 e^{ b^2 (n+1 -\frac{m_1}{2})^2/2
-(\varepsilon - \frac{\tilde{\Xi}_1}{\tilde{a}_1}
+ b (n+1 -\frac{m_1}{2}))^2/2}\\
&& \times \frac{1}{\sqrt{2\pi}}
\frac{1}
{\varepsilon - \frac{\tilde{\Xi}_1}{\tilde{a}_1}+ b (n+1 -\frac{m_1}{2})}
 +\ldots \\
&=&
4 \lim_{\varepsilon \rightarrow +0}
\sum_{n=0}^{\infty}( n+1) (-)^n
 e^{-  \frac{2\tilde{\Xi}_1}{m_1}(n+1 -\frac{m_1}{2})}
 e^{ -(\varepsilon - \frac{\tilde{\Xi}_1}{\tilde{a}_1}) b(n+1 -\frac{m_1}{2})
-(\varepsilon - \frac{\tilde{\Xi}_1}{\tilde{a}_1})^2/2}\\
&&\times  \frac{1}{\sqrt{2\pi}b(n+1)}
\frac{1}{ (\varepsilon - \frac{\tilde{\Xi}_1}{\tilde{a}_1})
\frac{1}{b(n+1)}
+ 1 -\frac{m_1}{2(n+1)}} +\ldots \\
&=&
  \frac{4}{\sqrt{2\pi}b}
\lim_{\varepsilon \rightarrow +0}
 e^{  \tilde{a}_1 \varepsilon
-(\varepsilon - \frac{\tilde{\Xi}_1}{\tilde{a}_1})^2/2}
\sum_{n=0}^{\infty} (-)^n
( e^{  - b \varepsilon })^{n+1}
( 1+ {\cal O}(m_1))+\ldots \\
&=&
  \frac{4}{\sqrt{2\pi}b}
\lim_{\varepsilon \rightarrow +0}
 e^{  \tilde{a}_1 \varepsilon
-(\varepsilon - \frac{\tilde{\Xi}_1}{\tilde{a}_1})^2/2}
\frac{ e^{  - b \varepsilon }}{1+ e^{  - b \varepsilon }}+\ldots \\
& = &
  \frac{2}{\sqrt{2\pi}b}
 e^{  -(\frac{\tilde{\Xi}_1}{\tilde{a}_1})^2/2}
+\ldots
  =  \frac{m_1}{\sqrt{2\pi}\tilde{a}_1}
 e^{  -(\frac{\tilde{\Xi}_1}{\tilde{a}_1})^2/2}
+\ldots.
\end{eqnarray*}
Therefore,
\begin{eqnarray*}
\int\!Dz_1~\cosh^{m_1-2}(\tilde{\Xi}/m_1) &=&
 I_+({\tilde{\Xi}_1}) + I_- ({\tilde{\Xi}_1})
 =  \frac{2m_1}{\sqrt{2\pi}\tilde{a}_1}
 e^{  -(\frac{\tilde{\Xi}_1}{\tilde{a}_1})^2/2}
+\ldots\\
& = &\frac{m_1}{\tilde{a}_1}\sqrt{\frac{2}{\pi}}
e^{-\tilde{\Xi}_1^2/2\tilde{a}_1^2}+\ldots
~~~~~~~~(m_1\to 0).
\end{eqnarray*}
Hence
\begin{eqnarray}
\kappa&=&
\frac{1}{\tilde{a}_1}\sqrt{\frac{2}{\pi}}
\bra \ldots \bra~
\frac{
e^{-\tilde{\Xi}_1^2/2\tilde{a}_1^2}}
{\int\!Dz_1~e^{|\tilde{\Xi}|}}~
 \ket_{2} \ldots  \ket_{\ell}
 \nonumber
 \\
 &=&
\bra \ldots \bra
\left[
\frac{
\tilde{a}_1^{-1}(2/\pi)^{\frac{1}{2}}~
e^{-\tilde{\Xi}_1^2/2\tilde{a}_1^2-\tilde{a}_1^2/2}}
{\cosh(\tilde{\Xi}_1)
 + \frac{1}{2}
e^{\tilde{\Xi}_1} {\rm Erf}
\left[\frac{\tilde{a}_1^2+\tilde{\Xi}_1}{\tilde{a}_1\sqrt{2}}\right]
+  \frac{1}{2}  e^{-\tilde{\Xi}_1} {\rm Erf}
\left[\frac{\tilde{a}_1^2-\tilde{\Xi}_1}{\tilde{a}_1\sqrt{2}}\right]}
\right]
 \ket_{2} \ldots  \ket_{\ell}
\label{eq:kappa_found}
\end{eqnarray}
We have now calculated explicitly all ingredients necessary for
determining $T_{\ell-1,\ell}^{\rm 2nd}$: one is to solve equation
(\ref{eq:general_condition}), upon insertion of the solution of
equations (\ref{eq:q2},\ref{eq:qmore}) for $\{q_{k>1}\}$ and
of expression (\ref{eq:kappa_found}) for $\kappa$.

As a simple test one can verify the outcome of (\ref{eq:general_condition})
for the case $\ell=2$ which we analysed before. This corresponds
to $\tilde{\Xi}_1=0$, insertion of which into
(\ref{eq:kappa_found}) and subsequently into
(\ref{eq:general_condition}) indeed brings us back to
(\ref{eq:case_l=2}), as it should.

\section{Discussion}

In this paper we have studied
self-programming in recurrent neural networks where both
neurons (the `processors') and synaptic interactions (`the
programme') evolve in time simultaneously.
 In contrast to previous models
involving coupled dynamics of fast neurons and slow interactions,
 the interactions in our model do not
evolve on a single time-scale; they are divided randomly into a
hierarchy of $L$ different groups of prescribed sizes, each with their own
characteristic time-scale $\tau_\ell$ and noise level $T_\ell$ ($\ell=1,\ldots,L$),
describing increasingly
non-volatile programming levels.

 We have solved our model in
equilibrium upon making the replica-symmetric (i.e.
ergodic) ansatz within each level of our hierarchy, leading to
 a theory which
resembles, but is not identical to Parisi's $L$-level replica
symmetry breaking (RSB) solution for spin systems with frozen disorder.
In addition to a paramagnetic phase, the phase diagram of our model involves a hierarchy of distinct spin
glass
phases $SG_\ell$, which differ in terms of the largest time-scale
on which the system (spins and couplings) will be found in a frozen state.
Our theory involves $L$ replica dimensions $m_\ell$, reminiscent of
the block sizes in Parisi's RSB scheme, which here represent ratios of the
temperatures of subsequent levels in the hierarchy of equilibrating
couplings, and which can take any value in the interval $[0,\infty\ket$.

We have solved our order parameter equations in full detail for
the choices $L=2$ and $L=3$, leading to extremely rich phase
diagrams, with an abundance of first-order transitions especially
when the level of stochasticity in the interaction dynamics is
chosen to be low, i.e. when one or more of the $m_\ell$ become
large (which can never happen in the Parisi scheme, where always
$0\leq m_\ell\leq 1$). We also studied the  asymptotic properties
of our model for arbitrary values of $L$ in the limits $m_1\to
\infty$ for fixed $T$ (deterministic dynamics of the level-1
interactions) and $m_1\to 0$ for fixed $T_1$ (deterministic
dynamics of the neuronal processors). This revealed further
non-trivial properties, such as the universal nature of the values
of the order parameters (i.e. $q_k=q^\star\approx 0.7911$ for all
$k\leq \ell$, independent of the control parameters) at the first
order $P\to SG_\ell$ transitions for $m_1=\infty$, and re-entrance
phenomena at the $SG_{\ell-1}\to SG_\ell$ borders in the phase
diagrams for $m_1=0$ (the latter are likely to be replaced by
discrete non-reentrant jumps when replica symmetry breaking is
taken into account).

In the present study we have not attempted to validate the
predictions of our theory by numerical simulations. With the
present CPU resources this would not have been possible, since our
model requires $L$ nested equilibrations of disordered
sub-systems. Even with fixed couplings it would not have been
possible to reach equilibrium with a system size sufficiently
large to suppress finite size effects; as a consequence, even for
a similar $L=1$ system it was already found to be impossible to
carry out simulations which achieve more than a very rough
qualitative agreement with the theory
 \cite{Jongenetal2}. Experimental verification for a recurrent
 self-programming network with $L>1$  will probably
 require hardware realizations.

In retrospect, we found ourselves pleasantly surprised by the
extent to which the present model allows for analytical solution,
in spite of the complications induced by the nested
equilibrations. Rather than being restricted by mathematical
intractability, our problem was rather how to select the control parameters
for which to show full phase diagrams.
This allowed us to gain qualitative and quantitative insight into the structural properties of
simple recurrent self-programming systems, in particular regarding the quesntions of when and how these systems switch from
random motion of processors and programme (the paramagnetic phase)
to states where processors and some (or all) of the levels of the hierarchy of
programme routines are `locked' into specific fixed-points (the different types spin-glass phases).

At the same time it will also be clear, however, that both in
terms of understanding the physics of realistic self-programming
systems and in terms of understanding the mathematical theory by
which they are described, this paper represents only a modest
step. In realistic self-programming systems one should obviously
expect the programming levels not to evolve only on the basis of
pair-correlations in processor states, a simple decay term, and
Gaussian noise, but one would as a minimum introduce external
symmetry-braking forces into both the processor dynamics and the
coupling dynamics, representing data to be processed and
programming objectives to be met, respectively. At a theoretical
level it would be interesting to investigate the form taken in the
present model of replica symmetry  breaking, for which we have
already found indirect evidence (in the form of re-entrance
phenomena) in studying the $m_1\to 0$ limit, by calculating AT
lines \cite{AT}. Since in our present model the replica symmetric
(RS) theory is already similar to an $L$-step RSB theory \'{a} la
Parisi, it is not immediately obvious what structure to expect
when replica symmetry is broken at one or more levels in our
hierarchy. These and other questions and extensions will hopefully
be addressed in future studies.

\section*{Acknowledgements}

One of the authors (T U) would like to acknowledge  the
Ministry of Education, Culture, Sports,
Science and Technology of the Japanese Goverment for their support of
the Overseas Research Fellowship $\#$12-kho-158.


\clearpage
\appendix
\section{Asymptotic form of $T_{1,2}^{\rm 2nd}$ when $m_1 \gg 1$}
\label{app:asym}

In this first appendix, we derive the asymptotic form of the
transition temperature
 $T_{1,2}^{\rm 2nd}$ for the case where $m_1 \gg 1$.
 The value of
 $T_{1,2}^{\rm 2nd}$ is determined by solving the coupled
 equations
\begin{eqnarray}
1& = &\sqrt{\beta \pi _2}[1+(m_1-1)q_1], \label{eq:app1}\\
q_1& = & \frac{1}{K_1}\int Dz_1 \cosh^{m_1}(a_1 z_1)\tanh ^2 (a_1 z_1),
\label{eq:app2}
\end{eqnarray}
where $a_1=\sqrt{\beta \pi _1 q_1}$. Upon assuming $q_1 \ll 1$ and
$m_1 \gg 1$, we can derive from (\ref{eq:app2}) the equation
for $\tilde{q}\equiv m_1 q_1$ to take the form
\begin{eqnarray}
1& \simeq &\sqrt{\beta \pi _1}[1+\beta \pi _1 \tilde{q}(1 +
\tilde{q})]. \label{eq:app3}
\end{eqnarray}
From (\ref{eq:app1}) and  (\ref{eq:app3}) we subsequently obtain
the following equations for the quantities $w\equiv 1/\sqrt{\beta
\pi _2}$ and $\tilde{q}$,
\begin{eqnarray}
\tilde{q}& = & w -1,  \label{eq:app4} \\
g(w)& \equiv & w ^3 - a w + b  =  0, \label{eq:app5}
\end{eqnarray}
where $a=\frac{\tilde{\pi} _1}{\tilde{\pi} _2} (1 +
\frac{\tilde{\pi} _1}{\tilde{\pi} _2})$, $b=(\frac{\tilde{\pi}
_1}{\tilde{\pi} _2})^2$ and $\tilde{\pi}_i=m_1 \pi _i$, both of
which are independent of $m_1$. From $\tilde{\pi} _1 = \tilde{\pi}
_2 + \frac{\epsilon _1}{\mu _1}$ it follows that
$g(1)=1-\frac{\tilde{\pi}_1}{\tilde{\pi}_2}<0$.  Thus $g(w)=0$ has
the unique solution $w^*$, which is greater than 1. Therefore we
obtain our desired result for $m_1\to \infty$:
\begin{eqnarray}
q_1 &\simeq & \frac{\tilde{q}}{m_1}=\frac{1}{m_1}(w^* -1)>0,\label{eq:app6}\\
T & \simeq & \tilde{\pi}_2 w^* \sqrt{T_1}. \label{eq:app7}
\end{eqnarray}

\section{Building Blocks of the General Saddle-Point Equations}
\label{app:saddle}

In this appendix we calculate the derivatives used in deriving the
saddle-point equations for general values of $L$ in section 4.4.
In this and subsequent appendices it will be helpful to define a
number of convenient short-hands (further notation is as in
section 4.4):
\begin{eqnarray*}
M_{\ell>0} &= \prod_{k=1}^{\ell}m_{k},~~~~~~~~& M_0=1
\\
\psi_{\ell>0}& = \bra \ldots \bra\tanh \Xi\ket_1\ldots
\ket_{\ell},~~~~~~& \psi_{0}= \tanh \Xi
\\
J_{\ell} &= K_{\ell}^{-1} \frac{\partial}{\partial \Xi}(K_{\ell}
\psi_{\ell})
\end{eqnarray*}

\subsection{Calculation of $ K_{\ell}^{-1}\partial
K_{\ell}/\partial \Xi$ and $J_{\ell} = K_{\ell}^{-1}\partial
(K_{\ell} \psi_{\ell})/\partial \Xi$ }

We first calculate $K_\ell^{-1}\partial K_{\ell}/\partial \Xi$ for
$\ell \ge 1$:
\begin{eqnarray*}
K_\ell^{-1} \frac{ \partial K_{\ell}}{\partial \Xi} &=&
 \frac{m_{\ell}}{K_\ell} \int\! Dz_{\ell} K_{\ell-1}^{m_{\ell}-1}
 \frac{ \partial K_{\ell-1}}{\partial \Xi}
\end{eqnarray*}
\begin{eqnarray}
~~~~~~ &=& \frac{ \prod_{k=1}^{\ell}  m_{k}}{K_\ell} \int\!
Dz_{\ell} K_{\ell-1} ^ {m_{\ell}-1} \int\! Dz_{\ell-1} K_\ell ^
{m_{\ell-1}-1} \cdots \int\! Dz_{1} \cosh^ {m_1-1} \Xi \sinh \Xi
\nonumber \\ &=& M_{\ell}~ \bra \cdots \bra\tanh \Xi\ket_1\cdots
\ket_{\ell} = M_{\ell}~\psi_{\ell} \label{eqn:dkdxi}
\end{eqnarray}
Next we calculate  $J_{\ell} = K_{\ell}^{-1} \partial(K_{\ell}
\psi_{\ell})/\partial \Xi$ for $\ell \ge 1$. We commence with
$J_1$:
\begin{eqnarray}
J_{1} &=& K_1^{-1} \int\! Dz_{1}~\frac{\partial}{\partial
\Xi}(K_{0}^{m_{1}}\psi_{0}) \nonumber \\ &=& K_1^{-1}\int\!
Dz_{1}~ K_{0}^{m_{1}}\left[1+(m_1 -1) \tanh^2 \Xi \right]
\nonumber \\ &=& 1+(m_1 -1) \bra \tanh ^2 \Xi \ket_1 \label{eq:J1}
\end{eqnarray}
We move on to $\ell >1$, abbreviating $L_\ell= K_\ell^{-1}
\partial K_{\ell}/\partial \Xi$:
\begin{eqnarray*}
J_{\ell}  &=&K_\ell^{-1}\int\! Dz_{\ell}~\frac{\partial}{\partial
\Xi}(K_{\ell-1}^{m_{\ell}-1}K_{\ell-1}\psi_{\ell-1})
\\ &=&
K_\ell^{-1}\!\int\! Dz_{\ell} \left\{
 (m_{\ell}-1) K_{\ell-1}^{m_{\ell}-2}
 \frac{\partial K_{\ell-1}}{\partial \Xi} K_{\ell-1}\psi_{\ell-1}+
K_{\ell-1}^{m_{\ell}-1} \frac{\partial} {\partial \Xi}(
K_{\ell-1}\psi_{\ell-1}) \right\}
\\ &=&
K_\ell^{-1} \int\! Dz_{\ell}  \left\{ ( m_{\ell}-1)
K_{\ell-1}^{m_{\ell}}L_{\ell-1}\psi_{\ell-1}+
K_{\ell-1}^{m_{\ell}} J_{\ell-1}\right\}
\\ &=&
(m_{\ell}-1)  \bra L_{\ell-1} \psi_{\ell-1}\ket_{\ell} +\bra
J_{\ell-1}\ket_{\ell}
\\
&=& M_{\ell-1}(m_{\ell}-1) \bra\psi_{\ell-1}^2\ket_{\ell}
 + \bra J_{\ell-1}\ket_{\ell}
\end{eqnarray*}
Iteration of this relation gives:
\begin{eqnarray}
J_\ell &=& \sum_{k=2}^{\ell}M_{k-1}(m_{k}-1)
 \bra \ldots \bra\psi_{k-1}^2\ket_{k}\ldots\ket_{\ell}
+\bra\ldots \bra J_1\ket_2 \ldots \ket_{\ell} \nonumber
\\ &=&
\sum_{k=2}^{\ell}M_{k-1}(m_{k}-1)
 \bra \ldots \bra\psi_{k-1}^2\ket_{k}\ldots\ket_{\ell}
+1+(m_1 -1)\bra\cdots \bra \tanh^2 \Xi \ket_1 \cdots \ket_{\ell}
\nonumber
\\
 &=& 1 +   \sum_{k=1}^{\ell}M_{k-1}(m_{k}-1)
 \bra \ldots \bra\psi_{k-1}^2\ket_{k}\ldots\ket_{\ell}
 \label{eq:Jell}
\end{eqnarray}
Finally we prove that $J_{\ell}>0$  for all $\ell \ge 1$. We put
$q_k = \bra \ldots \bra\psi_{k-1}^2\ket_{k}\ldots\ket_{\ell}$ and
use the inequalities $q_1\geq q_2\geq \ldots\geq q_\ell\geq 0$:
\begin{eqnarray*}
 J_{\ell}&= & \sum_{k=2}^{\ell}(M_{k} - M_{k-1}) q_k +1+(m_1 -1)q_1 \\
& = &  \sum_{k=2}^{\ell-1} M_{k} ( q_{k}-  q_{k+1}) + M_{\ell}
q_{\ell} -m_1 q_2 +1+(m_1 -1)q_1 \\ &= &
\sum_{k=2}^{\ell-1}M_{k}(q_{k}-q_{k+1})+M_{\ell}q_{l}+m_1(q_1-
q_2)+1 -q_1 >0
\end{eqnarray*}

\subsection{Calculation of $\partial K_\ell/\partial q_{\ell^\prime}$}

In this second part of \ref{app:saddle} we will repeatedly need
the following simple  relations:
\be
 \frac{\partial\Xi}{\partial q_1}=
 \frac{1}{2}\beta \pi_1 z_1 ~(\frac{\partial\Xi}{\partial z_1})^{-1},
 ~~~~~~~~ \frac{\partial\Xi}{\partial q_{\ell>1}}=\frac{1}{2}\beta\pi_\ell
 \left\{ z_{\ell}(\frac{\partial\Xi}{\partial z_{\ell}})^{-1}\!
 - z_{\ell-1}( \frac{\partial\Xi}{\partial z_{\ell-1}})^{-1}\right\}
\label{eq:partial_ql} \ee We first calculate $\partial
K_\ell/\partial q_{\ell^\prime}$  for $1 < \ell^\prime \le \ell$:
\begin{eqnarray*}
\frac{\partial K_\ell}{\partial q_{\ell^\prime}}&=&
 m_\ell \int\! Dz_\ell  K_{\ell-1} ^{m_\ell -1 }
 \frac{\partial K_{\ell-1}}{\partial q_{\ell^\prime}}\\
&=&  [\prod_{k=\ell^\prime-1}^\ell m_k]
 \int\! Dz_\ell  K_{\ell-1} ^{m_\ell -1 }\int\! Dz_{\ell-1} K_{\ell-2}^{m_{\ell-1} -1 ~}\\
&& \hspace*{20mm} \ldots~ \int\! Dz_{\ell^\prime}
K_{\ell^\prime-1} ^{m_{\ell^\prime} -1 } \int\! Dz_{\ell^\prime-1}
K_{\ell^\prime-2} ^{m_{\ell^\prime-1} -1 }
 \frac{\partial K _{\ell^\prime-2}}{\partial q_{\ell^\prime}}\\
&=&  [\prod_{k=\ell^\prime-1}^\ell m_k]
 \int\! Dz_\ell  K_{\ell-1}^{m_\ell -1 }
 \ldots \int\! Dz_{\ell^\prime} K_{\ell^\prime-1}^{m_{\ell^\prime} -1 }
\int\! Dz_{\ell^\prime-1} K_{\ell^\prime-2}^{m_{\ell^\prime-1} -1
}\\ && \hspace*{20mm} \times \frac{\beta \pi _{\ell^\prime}}{2}
 \left\{ z_{\ell^\prime}(\frac{\partial\Xi}{\partial z_{\ell^\prime}})^{-1}\!
-z_{\ell^\prime-1}(\frac{\partial\Xi}{\partial
z_{\ell^\prime-1}})^{-1}\right\}
 \frac{\partial K_{\ell^\prime-2} }{\partial \Xi} \\
&=& \frac{\beta \pi _{\ell^\prime}}{2}
  [\prod_{k=\ell^\prime-1}^\ell m_k]
 \int\! Dz_\ell  K_{\ell-1}^{m_\ell -1 }\int\! Dz_{\ell-1} K_{\ell-2} ^{m_{\ell-1} -1 }\\
&& \hspace*{20mm} \ldots \int\! Dz_{\ell^\prime}
 \frac{\partial  K_{\ell^\prime-1}^{m_{\ell^\prime} -1 } }{\partial \Xi}
\int\! Dz_{\ell^\prime-1} K_{\ell^\prime-2} ^{m_{\ell^\prime-1} -1
}
 \frac{\partial K_{\ell^\prime-2} }{\partial \Xi} \\
&=&  \frac{\beta \pi _{\ell^\prime}}{2}
  [\prod_{k=\ell^\prime-1}^\ell m_k]
(m_{\ell^\prime}-1) K_\ell \\ && \hspace*{20mm} \times \bra\bra
\ldots \bra \frac{1}{ K_{\ell^\prime-1} } \frac{\partial
K_{\ell^\prime-1} }{\partial \Xi} \bra \frac{1}{
K_{\ell^\prime-2} } \frac{\partial K_{\ell^\prime-2} }{\partial
\Xi} \ket_{\ell^\prime -1}
\ket_{\ell^\prime}\cdots\ket_{\ell-1}\ket_\ell
\end{eqnarray*}
Hence, upon using the relation $K_{\ell}^{-1}\partial  K_{\ell}
/\partial \Xi  =   M_\ell \psi_\ell$, we obtain for $1 <
\ell^\prime \le \ell$ the simple result:
\begin{eqnarray}
K_\ell^{-1}\frac{\partial K_\ell}{\partial q_{\ell^\prime}} &=&
  \frac{\beta \pi _{\ell^\prime}}{2}
  [\prod_{k=1} ^\ell m_k]
 [\prod_{j=1} ^{\ell^\prime-1}m_j](m_{\ell^\prime}-1)
\bra\bra \ldots \bra \psi_{\ell^\prime-1} \bra
\psi_{\ell^\prime-2} \ket_{\ell^\prime-1}
\ket_{\ell^\prime}\ldots\ket_{\ell-1}\ket_\ell
\nonumber
\\ &=&
\frac{\beta \pi _{\ell^\prime}}{2} M_\ell
M_{\ell^\prime-1}(m_{\ell^\prime}-1) \bra\bra \ldots \bra
\psi_{\ell^\prime-1} ^2
\ket_{\ell^\prime}\ldots\ket_{\ell-1}\ket_\ell
\label{eq:lprime>1}
\end{eqnarray}
For $\ell^\prime=1 \le \ell$ we obtain, similarly:
\begin{eqnarray}
K_\ell^{-1} \frac{\partial K_\ell}{\partial q_1}&=&
\frac{ m_\ell}{K_\ell} \int\!
Dz_\ell K_{\ell-1} ^{m_\ell -1 } \frac{\partial K
_{\ell-1}}{\partial q_1} \nonumber
\\ &=& \frac{\prod_{k=1}^\ell m_k}{K_\ell}
 \int\! Dz_\ell  K_{\ell-1} ^{m_\ell -1 }\int\! Dz_{\ell-1}
 K_{\ell-2}^{m_{\ell-1} -1 }
 \ldots \int\! Dz_1 K_0^{m_{1} -1 }
 \frac{\partial K _{0}}{\partial q_1}
\nonumber \\ &=& \frac{M_\ell}{K_\ell}
 \int\! Dz_\ell  K_{\ell-1} ^{m_\ell -1 }
 \ldots \int\! Dz_1 K_0 ^{m_{1} -1 }
\frac{\beta \pi _{1}}{2}
 z_{1}(\frac{\partial\Xi}{\partial z_1})^{-1}
 \frac{\partial K_0 }{\partial \Xi}
\nonumber \\ &=& \frac{\beta \pi _{1}}{2} \frac{M_\ell}{K_\ell}
 \int\! Dz_\ell  K_{\ell-1} ^{m_\ell -1 }
\ldots \int\! Dz_1
 \frac{\partial }{\partial \Xi} \left\{
K_{0} ^{m_{1} -1 } \frac{\partial K_0 }{\partial \Xi} \right\}
\nonumber \\ &=& \frac{\beta \pi _{1}}{2} \frac{M_\ell}{K_\ell}
 \int\! Dz_\ell  K_{\ell-1}^{m_\ell -1 }
 \ldots \int\! Dz_1
\left\{1 +(m_{1} -1 )\tanh^2 \Xi \right\} K_{0} ^{m_{1}} \nonumber
\\ &=& \frac{\beta \pi _{1}}{2} M_\ell \left\{ 1 + (m_{1} -1 )
\bra\ldots \bra\tanh^2 \Xi\ket_1 \ldots \ket_\ell \right\}
\label{eq:lprime=1}
\end{eqnarray}
Next we turn to the case where  $\ell^\prime > \ell$, noting that
$K_\ell$ is a function of $\{z_{\ell+1},\cdots z_{L}\}$. If
$\ell^\prime >\ell+1$ one finds
\begin{eqnarray*}
K _\ell^{-1} \frac{\partial K _\ell}{\partial q_{\ell^\prime}} &=&
\frac{\beta \pi_{\ell^\prime}}{2} \left\{
z_{\ell^\prime}(\partial\Xi/\partial z_{\ell^\prime})^{-1}\!
-z_{\ell^\prime-1}(\partial\Xi/\partial
z_{\ell^\prime-1})^{-1}\right\} \frac{1}{ K_\ell} \frac{\partial
K_\ell}{\partial \Xi}\\ &=& \frac{\beta \pi_{\ell^\prime}}{2}
\left\{ z_{\ell^\prime}(\partial\Xi/\partial
z_{\ell^\prime})^{-1}\! -z_{\ell^\prime-1}(\partial\Xi/\partial
z_{\ell^\prime-1})^{-1}\right\} M_\ell \psi_\ell
\end{eqnarray*}
For $\ell^\prime=\ell+1$, on the other hand, we obtain
\begin{eqnarray*}
\frac{\partial K_\ell}{\partial q_{\ell+1}}&=&
 m_\ell \int\! Dz_\ell  K_{\ell-1}^{m_\ell -1 } \frac{\partial K
_{\ell-1}} {\partial q_{\ell+1}}
\\ &=& \frac{m_\ell\beta \pi
_{\ell+1}}{2} \int\! Dz_\ell K_{\ell-1} ^{m_\ell -1 }
 \left\{ z_{\ell+1}(\partial\Xi/\partial z_{\ell+1})^{-1}\!
-z_{\ell}(\partial\Xi/\partial z_{\ell})^{-1} \right\}
 \frac{\partial K_{\ell-1} }{\partial \Xi} \\
&=&
\frac{m_\ell \beta \pi _{\ell+1}}{2}\left\{
 z_{\ell+1} (\frac{\partial\Xi}{\partial z_{\ell+1}})^{-1} \int\! Dz_\ell  K_{\ell-1}^{m_\ell -1 }
 \frac{\partial K_{\ell-1} }{\partial \Xi}
 \right.
 \\
 &&\left.\hspace*{20mm}
 -  \int\! Dz_\ell \frac{\partial}
{\partial \Xi} \left[
 K_{\ell-1} ^ {m_\ell -1 } \frac{\partial K_{\ell-1} } {\partial \Xi}\right]
  \right\}
\end{eqnarray*}
This latter term will appear inside an integral over $z_{\ell+1}$
whenever  $q_{\ell+1}> 0$. Thus, we will at some point have to
evaluate integrals of the following form, with a function $
g(\Xi(z_{l+1},\cdots, z_L))$:
\begin{eqnarray}
\int\! Dz_{\ell+1} g(\Xi)\frac{\partial K _\ell}{\partial
q_{\ell+1}} &=&
 \frac{m_\ell \beta \pi_{\ell+1}}{2} \int\! Dz_{\ell+1}  \frac{\partial  g(\Xi)}{\partial \Xi}
 \int\! Dz_\ell  K_{\ell-1}^{m_\ell -1 }
 \frac{\partial K_{\ell-1} }{\partial \Xi}
 \label{eq:integ_zell+1}
\end{eqnarray}
We calculate the derivative in the second term of $\partial
K_\ell/\partial q_{\ell+1}$ as follows: \bd \frac{\partial
}{\partial \Xi}
 \left\{
 K_{\ell-1}^{m_\ell -1 } \frac{\partial K_{\ell-1} } {\partial \Xi} \right\}
 =  M_{\ell-1} \frac{\partial }{\partial \Xi} \left\{
 K_{\ell-1} ^ {m_\ell -1 }  K_{\ell-1} \psi _{\ell-1} \right\}
\ed
\begin{eqnarray*}
~~~~ & =& M_{\ell-1} \left\{ (m_\ell -1) K_{\ell-1} ^ {m_\ell -2 }
\frac{\partial K_{\ell-1}}{\partial \Xi} K_{\ell-1} \psi _{\ell-1}
+
 K_{\ell-1} ^ {m_\ell -1 }   \frac{\partial }{\partial \Xi}
( K_{\ell-1} \psi _{\ell-1}) \right\}\\ & =& M_{\ell-1} \left\{
(m_\ell -1) K_{\ell-1} ^ {m_\ell} M_{\ell-1}\psi_{\ell-1}^2+
 K_{\ell-1} ^ {m_\ell -1 }   \frac{\partial }{\partial \Xi}
( K_{\ell-1} \psi _{\ell-1}) \right\}
\end{eqnarray*}
Upon using formula (\ref{eq:Jell}) we arrive at
\bd
M_{\ell-1}^{-1}\frac{\partial }{\partial \Xi}
 \left\{
 K_{\ell-1} ^ {m_\ell -1 } \frac{\partial K_{\ell-1} } {\partial \Xi} \right\}
= (m_\ell -1) K_{\ell-1} ^ {m_\ell } M_{\ell-1}\psi_{\ell-1}^2 \ed
\bd \hspace*{30mm}
 +   K_{\ell-1}^{m_\ell} \left[
 1+ \sum_{j=1}^{\ell-1}M_{j-1}(m_j-1)\bra\cdots\bra\psi_{j-1}^2\ket_j \cdots \ket_{\ell-1}
\right] \ed
Hence
\begin{eqnarray*}
K_{\ell}^{-1} \int\! Dz_\ell~ \frac{\partial }{\partial \Xi}
 \left\{
 K_{\ell-1} ^ {m_\ell -1 } \frac{\partial K_{\ell-1} } {\partial \Xi} \right\}
& =& M_{\ell-1} \left\{
 1+ \sum_{j=1}^{\ell}M_{j-1}(m_j-1)\bra\ldots\bra\psi_{j-1}^2\ket_j \ldots \ket_{\ell}
 \right\}\\
& =& M_{\ell-1} J_{\ell}
\end{eqnarray*}
and therefore
\begin{eqnarray}
K_\ell^{-1}\frac{\partial K _\ell}{\partial q_{\ell+1}} &=&
\frac{m_\ell}{K_\ell} \frac{\beta \pi _{\ell+1}}{2} \left\{
 z_{\ell+1} (\partial\Xi/\partial z_{\ell+1})^{-1}K_\ell
M_{\ell-1}\bra\psi_{\ell-1}\ket_\ell - K_\ell M_{\ell-1} J_{\ell}
\right\}\nonumber \\ &=& \frac{\beta \pi _{\ell+1}}{2} M_{\ell-1}
m_\ell \left\{
 z_{\ell+1} (\partial\Xi/\partial z_{\ell+1})^{-1}\psi_{\ell} -  J_{\ell} \right\},
\label{eq:lprime=l+1}
\end{eqnarray}

\section{Condition for Second-Order $SG_{\ell-1}\to SG_{\ell}$ Transitions}
\label{app:SGSG}

In this Appendix we derive the condition for the second order
transition $SG_{\ell-1}$ to $SG_{\ell}$, for arbitrary values of
$1 \le \ell \le L$. We generalize our previous notational
conventions, and write the general saddle-point equations as
\begin{eqnarray*}
q_\ell& = &\varphi_\ell(q_1, \ldots,
q_L),~~~~~~~~(\ell=1,\ldots,L)
\end{eqnarray*}
Similarly we define, for $ 1 \le k \le \ell$:
\begin{eqnarray*}
\varphi_k^{(\ell)} (q_1, \ldots, q_L)& = &
 \bra\ldots \bra \psi_{k-1}^2 \ket_k \ldots  \ket_\ell
\end{eqnarray*}
Note that, with these conventions, $\varphi_k(q_1, \ldots, q_L)=
\bra \ldots \bra \varphi_k^{(\ell)} \ket_{\ell+1} \ldots \ket_L$.
The second order $SG_{\ell-1}\to SG_\ell$ transition temperature
is defined by
\begin{eqnarray}
&&\frac{\partial \varphi_\ell ^{(\ell)}}{\partial
q_\ell}|_{(q_{\ell-1}>0;~q_\ell=q_{\ell+1}= \ldots= q_L=0)}=1
\label{eq:thing_to_find}
\end{eqnarray}
We can write $\varphi_\ell^{(\ell)}(q_1, \ldots, q_L)$ as $\varphi
_\ell^{(\ell)}=K_\ell^{-1} \int\! Dz_\ell K_{\ell-1}^{m_\ell}
\psi_{\ell-1}^2 = \bra \psi_{\ell-1} ^2 \ket _\ell$, with
$\psi_{\ell-1}=\bra\ldots  \bra  \tanh \Xi\ket_1 \ldots
\ket_{\ell-1}$. From this expression  we obtain
\begin{eqnarray*}
\frac{\partial \varphi_\ell^{(\ell)}}{\partial q_\ell} & =&
 -\frac{1}{K_\ell}
\frac{\partial K _\ell}{\partial q_\ell}  \varphi _\ell^{(\ell)}
+\frac{1}{K_\ell} \int\! Dz_\ell \frac{\partial}{\partial q_\ell}
[K_{\ell-1} ^{m_\ell} \psi_{\ell-1}^2]
\\
&=& -\frac{1}{K_\ell} \frac{\partial K _\ell}{\partial q_\ell}
\varphi _\ell^{(\ell)} +\frac{m_\ell-2}{K_\ell} \int\! Dz_\ell
 K_{\ell-1} ^{m_\ell -3 } \frac{\partial K _{\ell-1}}{\partial q_\ell}
(K_{\ell-1}\psi_{\ell-1})^2
\\&&
\hspace*{20mm}
 +\frac{2}{K_\ell} \int\! Dz_\ell K_{\ell-1} ^{m_\ell-1}
\psi_{\ell-1} \frac{\partial}{\partial
q_\ell}[K_{\ell-1}\psi_{\ell-1}]
\\
&=& -\frac{1}{K_\ell} \frac{\partial K _\ell}{\partial q_\ell}
\varphi _\ell^{(\ell)} +\frac{m_\ell-2}{K_\ell} \int\! Dz_\ell
 K_{\ell-1} ^{m_\ell -3 } \frac{\partial K _{\ell-1}}{\partial q_\ell}
[K_{\ell-1}\psi_{\ell-1}]^2
\\ &&
\hspace*{20mm} +\frac{2}{K_\ell} \int\! Dz_\ell K_{\ell-1}
^{m_\ell-1} \psi_{\ell-1} \int\! Dz_{\ell-1}
\frac{\partial\Xi}{\partial q_\ell}
 \frac{\partial }{\partial \Xi}[K_{\ell-2}
 ^{m_{\ell-1}}\psi_{\ell-2}]
 \\
&=& -\frac{1}{K_\ell} \frac{\partial K _\ell}{\partial q_\ell}
\varphi _\ell^{(\ell)} +\frac{m_\ell-2}{K_\ell} \int\! Dz_\ell
 K_{\ell-1} ^{m_\ell -3 } \frac{\partial K _{\ell-1}}{\partial q_\ell}
[K_{\ell-1}\psi_{\ell-1}]^2
\\ &&
 +\frac{\beta \pi_\ell}{K_\ell} \int\! Dz_\ell \left\{
(\partial\Xi/\partial z_\ell)^{-1}\! \frac{\partial}{\partial
z_\ell} (K_{\ell-1} ^{m_\ell-1} \psi_{\ell-1} ) \right\} \int\!
Dz_{\ell-1}
 \frac{\partial }{\partial \Xi}[K_{\ell-2}
 ^{m_{\ell-1}}\psi_{\ell-2}]
 \\
&=& -\frac{1}{K_\ell} \frac{\partial K _\ell}{\partial q_\ell}
\varphi _\ell^{(\ell)} +\frac{m_\ell-2}{K_\ell} \int\! Dz_\ell
 K_{\ell-1} ^{m_\ell -3 } \frac{\partial K _{\ell-1}}{\partial q_\ell}
[K_{\ell-1}\psi_{\ell-1}]^2
\\ &&
+\frac{\beta \pi _\ell}{K_\ell} \int\! Dz_\ell \left \{ (m_\ell-2)
K_{\ell-1} ^{m_\ell-2} \frac{\partial K_{\ell-1}}{\partial
\Xi}\psi_{\ell-1} + K_{\ell-1}^{m_\ell-2}\frac{\partial}{\partial
\Xi}( K_{\ell-1} \psi_{\ell-1} ) \right\}
\\&&
\hspace*{60mm} \times
 \frac{\partial }{\partial \Xi}[ K_{\ell-1}\psi_{\ell-1}]
\\
&=& -\frac{1}{K_\ell} \frac{\partial K _\ell}{\partial q_\ell}
\varphi_\ell^{(\ell)} +(m_\ell-2)\bra
 K_{\ell-1} ^{-1 } \frac{\partial K _{\ell-1}}{\partial
 q_\ell}\psi_{\ell-1}^2\ket_\ell
 \\&&
\hspace*{20mm} +\beta \pi _\ell (m_\ell-2)\bra K_{\ell-1} ^{-2}
\frac{\partial K_{\ell-1}}{\partial \Xi}\psi_{\ell-1}
 \frac{\partial }{\partial \Xi} (
 K_{\ell-1}\psi_{\ell-1})\ket_\ell
 \\&&
 \hspace*{20mm}
+ \beta \pi _\ell \bra K_{\ell-1} ^{-2} \left\{
\frac{\partial}{\partial \Xi}[ K_{\ell-1}  \psi_{\ell-1} ]
\right\}^2\ket_\ell
\end{eqnarray*}
For $\ell>1$ this reduces to
\begin{eqnarray}
\frac{\partial \varphi _\ell^{(\ell)}}{\partial q_\ell} &=&
-\frac{1}{K_\ell} \frac{\partial K _\ell}{\partial q_\ell}
\varphi_\ell^{(\ell)}
 +\beta \pi_\ell
(m_\ell-2)\bra \frac{1}{K_{\ell-1}} \frac{\partial
K_{\ell-1}}{\partial \Xi}\psi_{\ell-1} J_{\ell-1}\ket_\ell
 + \beta \pi_\ell \bra J_{\ell-1} ^{2}\ket_\ell
\nonumber \\ && +(m_\ell-2)\frac{\beta \pi_\ell}{2
K_\ell}m_{\ell-1} \int\! Dz_\ell \left\{ \frac{\partial}{\partial
\Xi} [ K_{\ell-1} ^{m_\ell -1 }\psi_{\ell-1}^2] \right \} \int\!
Dz_{\ell-1} K_{\ell-2} ^{m_{\ell-1} -1 } \frac{\partial K
_{\ell-2}}{\partial \Xi}
\nonumber\\ &=&
-\frac{1}{K_\ell}
\frac{\partial K _\ell}{\partial q_\ell} \varphi _\ell^{(\ell)}
+\beta \pi _\ell (m_\ell-2)\bra \frac{1}{K_{\ell-1}}
\frac{\partial K_{\ell-1}}{\partial \Xi}\psi_{\ell-1}
J_{\ell-1}\ket_\ell
 + \beta \pi_\ell \bra J_{\ell-1} ^{2}\ket_\ell
\nonumber\\&& +(m_\ell-2)\frac{\beta \pi_\ell}{2}m_{\ell-1} \bra
\left\{ (m_\ell -3) \frac{1}{K_{\ell-1}}\frac{\partial
K_{\ell-1}}{\partial \Xi}
 \psi_{\ell-1}^2 \right.\nonumber\\
&& \left. \hspace*{20mm}
 +~ 2 \psi_{\ell-1}
\frac{1}{K_{\ell-1}}\frac{\partial }{\partial \Xi} (K_{\ell-1}
\psi_{\ell-1})\right\}\bra \frac{1}{K_{\ell-2}} \frac{\partial K
_{\ell-2}}{\partial \Xi}\ket_{\ell-1}\ket_{\ell} \nonumber \\ &=&
-\frac{1}{K_\ell} \frac{\partial K _\ell}{\partial q_\ell} \varphi
_\ell^{(\ell)} +\beta \pi _\ell (m_\ell-2)
M_{\ell-1}\bra\psi_{\ell-1}^2 J_{\ell-1}\ket_\ell
 + \beta \pi _\ell \bra J_{\ell-1} ^{2}\ket_\ell
\nonumber\\&& +(m_\ell-2)\frac{\beta \pi_\ell}{2}m_{\ell-1} \bra
\left\{ (m_\ell -3) M_{\ell-1}\psi_{\ell-1}^3 + 2 \psi_{\ell-1}
J_{\ell-1} \right\}\bra M
_{\ell-2}\psi_{\ell-2}\ket_{\ell-1}\ket_{\ell} \nonumber \\ &=&
\frac{\beta \pi_\ell}{2} \left\{ -M_\ell M_{\ell-1}(m_\ell- 1)
(\varphi _\ell ^{(\ell)})^2
 + 2 (m_\ell- 2)m_{\ell-1}M _{\ell-2}\bra
\psi_{\ell-1}^2 J_{\ell-1} \ket_{\ell} \right. \nonumber\\
&&\left. \hspace*{20mm}
 +(m_\ell- 2)m_{\ell-1}M _{\ell-2} (m_\ell -3)
 M_{\ell-1}\bra \psi_{\ell-1}^4\ket_\ell
  \right.\nonumber\\
&& \left. \hspace*{20mm} +2(m_\ell- 2)
M_{\ell-1}\bra\psi_{\ell-1}^2 J_{\ell-1}\ket_\ell + 2 \bra
J_{\ell-1} ^{2}\ket_\ell
 \right\}\nonumber\\
&=& \frac{\beta \pi_\ell}{2} \left\{ -M_\ell M_{\ell-1}(m_\ell-1)
(\varphi _\ell ^{(\ell)})^2 +4(m_\ell-2)
 M_{\ell-1}\bra\psi_{\ell-1}^2 J_{\ell-1}\ket_\ell
\right. \nonumber\\ &&\left.
 +(m_\ell-2)m_{\ell-1}M _{\ell-2} (m_\ell -3) M_{\ell-1}\bra
 \psi_{\ell-1}^4\ket_\ell
    + 2 \bra J_{\ell-1} ^{2}\ket_\ell\right\}
\label{eq:nearly_l>1}
\end{eqnarray}
whereas for $\ell=1$ one finds
\begin{eqnarray*}
\frac{\partial \varphi_1 ^{(1)} }{ \partial q_1} &=&
 -\frac{1}{K_1} \frac{\partial K_1}{\partial
q_1} \varphi_1 ^{(1)} + \frac{1}{K_1}\int\! Dz_1
\frac{\partial}{\partial q_1}[K_0 ^{m_1}\tanh^2 \Xi] \nonumber\\
&=& -\frac{1}{K_1} \frac{\partial K_1}{\partial q_1} \varphi_1
^{(1)} + \frac{1}{K_1}\int\! Dz_1 \frac{\beta \pi_1}{2}
z_{1}(\partial\Xi/\partial z_{1})^{-1}
 \frac{\partial}{\partial \Xi}[K_0 ^{m_1}\tanh^2 \Xi]
 \nonumber\\
&=& -\frac{1}{K_1} \frac{\partial K_1}{\partial q_1} \varphi_1
^{(1)} +  \frac{\beta \pi_1}{2}\frac{1}{K_1}\int\! Dz_1
 \frac{\partial ^2 }{\partial \Xi ^2}[K_0 ^{m_1}\tanh^2 \Xi]
\end{eqnarray*}
The second derivative of $K_0 ^{m_1}\tanh^2 \Xi$ is given by
\begin{eqnarray*}
 \frac{\partial^2  }{\partial \Xi^2}[K_0 ^{m_1}\tanh^2 \Xi]
& =&
 \frac{\partial}{\partial\Xi}\left[ (m_1 -2) \cosh{m_1-3}\Xi
\sinh^3 \Xi +2 \cosh^{m_1-1} \Xi \sinh \Xi \right]
\end{eqnarray*}
\begin{eqnarray*}
~~~~~~~~~~~&=& \cosh^{m_1} \Xi  \left[ (m_1 -2) (m_1 -3) \tanh^4
\Xi + (5m_1 -8) \tanh^2 \Xi  +2 \right]
\end{eqnarray*}
Insertion of this expression, and using (\ref{eq:lprime=1}), we
now arrive at
\begin{eqnarray}
 \frac{\partial \varphi_1 ^{(1)}}{ \partial q_1}
&=&\frac{\beta \pi_1}{2} \left\{ - m_1(m_1-1)( \varphi_1 ^{(1)})^2
+(m_1 -2) (m_1 -3)\bra \tanh^4 \Xi \ket_1 \right. \nonumber\\
&&\left.\hspace*{60mm}
 + 4(m_1 -2)\varphi_1
^{(1)} +2 \right\} \label{eq:nearly_l=1}
\end{eqnarray}
Finally we note that the property $q_{\ell}=0$ induces the
simplifying relations $\varphi_\ell^{(\ell)} = \psi_{\ell-1} = 0$,
with which expressions (\ref{eq:nearly_l>1}) and
(\ref{eq:nearly_l=1}) lead to \bd \frac{\partial \varphi_\ell
^{(\ell)}}{\partial q_\ell}|_{(q_{\ell-1}>0;~q_\ell=q_{\ell+1}=
\ldots= q_L=0)} = \beta \pi _\ell J_{\ell-1}^2 \ed Together with
(\ref{eq:Jell}) we can now work out the condition
(\ref{eq:thing_to_find}) for the second order $SG_{\ell-1}\to
SG_\ell$ transition explicitly, giving the final result:
\be
\sqrt{\beta \pi _\ell} \left\{ 1+
\sum_{k=1}^{\ell-1}M_{k-1}(m_{k}-1)q_{k} \right \}=1
\label{eq:final_result}
 \ee

\section{Derivation of Partial Derivatives for $L=2$ and $L=3$}
\label{app:L3}

In this appendix we will find it convenient to use the following
abbreviations:
\begin{eqnarray*}
\varphi_1(q_1,q_2,q_2) &=&  \bra\bra \bra \tanh^2 \Xi\ket_1\ket_2
 \ket_3
\\
 \varphi_2(q_1,q_2, q_3)&=&
\bra \bra \bra \tanh \Xi\ket_1^2 \ket_2\ket_{3}
\\
 \varphi_3(q_1, q_1, q_3)& =& \bra  \bra  \bra  \tanh \Xi\ket_1
\ket_{2}^2\ket_{3}
\end{eqnarray*}
(these are the right-hand sides of the $L=3$ saddle-point
equations for $\{q_1,q_2,q_3\}$),
\begin{eqnarray*}
\varphi_1^{(1)}(q_1,q_2,q_3) &= \bra\tanh^2 \Xi \ket_1,~~~~~~ &
\varphi_1^{(2)}(q_1,q_2,q_3) = \bra\bra\tanh^2 \Xi\ket_1\ket_2\\
\varphi_2^{(1)}(q_1,q_2,q_3) &= \bra\tanh \Xi\ket_1^2,~~~~~~ &
\varphi_2^{(2)}(q_1,q_2,q_3) = \bra\bra\tanh \Xi\ket_1^2\ket_2\\
\varphi_3^{(2)}(q_1,q_2,q_3) &= \bra\bra\tanh\Xi\ket_1\ket_2^2
\end{eqnarray*}
and also $\psi_\ell= \bra \ldots\bra
\tanh\Xi\ket_1\ldots\ket_\ell$.
 Note that $\bra \varphi_k^{(1)}\ket_2=\varphi_k^{(2)}$,
$\varphi_k=\bra \varphi_k^{(2)}\ket_3$ ($k=1,2,3$),
$\psi^2_1=\varphi_2^{(1)}$ and $\psi_2^2=\varphi_3^{(2)}$. Further
notation in this appendix is as in sections 4 and 5 of the main
text and as in the previous appendix.

\subsection{Building Blocks of the Partial Derivatives}

First, we calculate the derivative of $\varphi_1^{(1)}(q_1, q_2,
q_3)$ with respect to $q_1$.
From the general formula (\ref{eq:nearly_l=1}) we obtain
\begin{eqnarray}
 \frac{\partial \varphi_1
^{(1)}}{ \partial q_1} &=&\frac{\beta \pi_1}{2} \left\{2 -
m_1(m_1-1)( \varphi_1 ^{(1)})^2 \right. \nonumber
\\ &&\left. +(m_1 -2) (m_1
-3)\bra \tanh^4 \Xi \ket_1 + 4(m_1 -2)\varphi_1 ^{(1)} \right\}
\label{eq:dphi11dq1}
\end{eqnarray}
Next we turn to
 the derivative of
$\varphi_2 ^{(2)}(q_1,q_2,q_3)$ with respect to $q_2$.
The general formula (\ref{eq:nearly_l>1}) with $k=2$ can be simplifid
by using (\ref{eq:Jell}):
\begin{eqnarray}
\frac{\partial \varphi_2 ^{(2)}}{ \partial q_2}
&=&  \frac{\beta \pi_2}{2}
\left\{ -M_2 m_{1}(m_2-1) (\varphi _2 ^{(2)})^2
 +m_{1}^2 (m_2-2)(m_2 -3)\bra \psi_{1}^4\ket_2 \right.
 \nonumber \\
&& \left.  \hspace*{20mm} +4m_{1}(m_2-2) \bra\psi_{1}^2
J_{1}\ket_2 + 2 \bra J_{1} ^{2}\ket_2 \right\} \nonumber \\ &=&
\frac{\beta \pi_2}{2} \left\{ - m_{1}^2 m_2 (m_2-1) (\varphi
_2^{(2)})^2
 +m_{1}^2 (m_2-2)(m_2 -3)\bra \psi_{1}^4\ket_2 \right.
\nonumber \\ &&    +4m_{1}(m_2-2)\varphi _2 ^{(2)} +4m_{1}(m_1
-1)(m_2-2) \bra \psi_{1}^2 \bra\tanh ^2 \Xi\ket_1 \ket_2 \nonumber
\\ && \left. +2 + 4(m_1 -1)\varphi _1 ^{(2)} +2(m_1
-1)^2\bra\bra\tanh ^2 \Xi\ket_1^2\ket_2 \right\}
\label{eq:dphi22dq2}
\end{eqnarray}
The third partial derivative which we will need is
\begin{eqnarray*}
 \frac{\partial \varphi_1 ^{(2)}}{ \partial q_1} &=&
-\frac{1}{K_2}\frac{\partial K_2}{ \partial q_1}\varphi_1 ^{(2)} +
\frac{1}{K_2}\int\! Dz_2   \frac{\partial }{ \partial q_1} ( K_1
^{m_2} \bra\tanh^2 \Xi\ket_1)\\ &=& -\frac{1}{K_2}\frac{\partial
K_2}{ \partial q_1}\varphi_1 ^{(2)} + \frac{m_2}{K_2}\int\! Dz_2
K_1 ^{m_2-1} \frac{\partial K_1}{ \partial q_1}
 \bra\tanh^2 \Xi\ket_1
+\bra \frac{\partial }{ \partial q_1} \bra\tanh^2 \Xi\ket_1\ket
\\
&=& -\frac{1}{K_2}\frac{\partial K_2}{ \partial q_1}\varphi_1
^{(2)} + m_2\bra\frac{1}{K_1} \frac{\partial K_1}{ \partial q_1}
\bra\tanh^2 \Xi\ket_1 \ket_2 + \bra \frac{\partial \varphi_1
^{(1)} }{ \partial q_1}\ket_2
\end{eqnarray*}
The last term in this expression is found to be
\begin{eqnarray*}
 \bra \frac{\partial \varphi_1 ^{(1)} }{ \partial q_1}\ket_2
&=&\frac{\beta \pi_1}{2} \left\{ -m_1(m_1-1)\bra\bra\tanh^2
\Xi\ket_1^2\ket_2 \right.\\ && \left.  +(m_1 -2) (m_1 -3)\bra\bra
\tanh^4 \Xi \ket_1\ket_2 + 4(m_1 -2) \varphi_1 ^{(2)}+2 \right\}
\end{eqnarray*}
Thus
\begin{eqnarray}
 \frac{\partial \varphi_1 ^{(2)}}{ \partial q_1}
&=& \frac{\beta \pi_1}{2} \left\{  4(m_1 -2) \varphi_1 ^{(2)}+2
 -
M_2\varphi_1 ^{(2)} - (m_1 -1)M_2( \varphi ^{(2)} _1) ^2 \right.
\nonumber \\ && +m_1 m_2\bra[ 1+ (m_1 -1)\bra\tanh ^2 \Xi\ket_1]
\bra\tanh^2 \Xi\ket_1 \ket_2 \nonumber \\ &&
 -m_1(m_1-1)\bra\bra\tanh^2 \Xi\ket_1^2\ket_2
+(m_1 -2) (m_1 -3)\bra\bra \tanh^4 \Xi \ket_1\ket_2 \nonumber \\
&& \left. +(m_1 -2) (m_1 -3)\bra\bra \tanh^4 \Xi \ket_1\ket_2
 + 4(m_1 -2) \varphi_1 ^{(2)}+2 \right\}
 \label{eq:dphi12dq1}
\end{eqnarray}
Similarly we calculate the fourth relevant partial derivative:
\begin{eqnarray*}
 \frac{\partial \varphi_1 ^{(2)}}{ \partial q_2} &=&
-\frac{1}{K_2}\frac{\partial K_2}{ \partial q_2}\varphi_1 ^{(2)} +
\frac{1}{K_2}\int\! Dz_2   \frac{\partial }{ \partial q_2} [
K_1^{m_2-1}\int\! Dz_1 \cosh ^{m_1} \Xi \tanh^2 \Xi]
\\
&=&
-\frac{1}{K_2}\frac{\partial K_2}{ \partial q_2}\varphi_1 ^{(2)} +
(m_2-1) \frac{1}{K_2}\int\! Dz_2  K_1 ^{m_2-2}
  \frac{\partial  K_1}{ \partial q_2} K_1 \bra\tanh^2 \Xi\ket_1
  \\
&& \hspace*{20mm} +  \frac{1}{K_2}\int\! Dz_2  K_1 ^{m_2-1} \int\!
Dz_1 \frac{\partial}{ \partial q_2}[  \cosh ^{m_1} \Xi \tanh^2
\Xi]
\\ &=&
-\frac{\beta \pi_2}{2} m_1^2 m_2(m_2 -1)\bra\psi _1
^2\ket_2\varphi_1 ^{(2)} + (m_2-1) \frac{\beta m_1
\pi_2}{2K_2}\int\! Dz_2
\\ &&
\hspace*{20mm} \times ~ \frac{\partial }{\partial \Xi}[ K_1
^{m_2-1} \bra\tanh^2 \Xi\ket_1]. \int\! Dz_1 \cosh ^{m_1-1} \Xi
\frac{\partial K_0}{\partial \Xi}
\\
&&
+ \frac{1}{K_2}\int\! Dz_2
K_1 ^{m_2-1}\frac{\beta \pi_2}{2} \int\! Dz_1
(\frac{z_2}{a_2}-\frac{z_1}{a_1})
 \frac{\partial}{ \partial \Xi}(  \cosh ^{m_1} \Xi \tanh^2 \Xi)
 \\
&=& \frac{\beta \pi_2}{2}\left\{ \frac{1}{K_2}\int\! Dz_2
\frac{\partial  K_1 ^{m_2-1}}{\partial \Xi}
\frac{\partial}{\partial \Xi}[K_1 \bra\tanh ^2\Xi\ket_1] - m_1^2
m_2(m_2 -1)\varphi_2 ^{(2)} \varphi_1 ^{(2)} \right.\\ && \left.
\hspace*{20mm} + (m_2-1) \frac{m_1}{K_2} \int\! Dz_2 K_1 \psi_1
 \frac{\partial }{\partial \Xi}[K_1 ^{m_2-1} \bra\tanh^2
 \Xi\ket_1]
 \right\}
\end{eqnarray*}
In this expression we need to calculate the following two objects:
\begin{eqnarray*}
 \frac{\partial }{\partial \Xi}[K_1 ^{m_2-1} \bra\tanh^2
\Xi\ket_1] &=& \frac{\partial }{\partial \Xi}[K_1 ^{m_2-2}
K_1\bra\tanh^2 \Xi\ket_1]
\end{eqnarray*}
\begin{eqnarray*}
~~~~~~~ &=&(m_2 -2) \frac{\partial K_1 }{\partial \Xi}K_1 ^{m_2-3}
K_1\bra\tanh^2 \Xi\ket_1 +K_1 ^{m_2-2} \frac{\partial }{\partial
\Xi}[ K_1\bra\tanh^2 \Xi\ket_1] \\
&=&
K_2 ^{m_1}\left\{ (m_2 -2) m_1 \psi_1 K_1 ^{-1}\bra\tanh^2
\Xi\ket_1 +K_1 ^{-2} \frac{\partial }{\partial \Xi}[
K_1\bra\tanh^2 \Xi\ket_1]\right\}
\end{eqnarray*}
and
\begin{eqnarray*}
 \frac{\partial }{\partial \Xi}[ K_1\bra\tanh^2 \Xi\ket_1] &=&
K_1 \left[ (m_1-2)\bra\tanh^3 \Xi\ket_1 +2\bra\tanh \Xi\ket_1
\right]
\end{eqnarray*}
 Insertion of these intermediate results
 into our previous expression for $\partial \varphi_1
^{(2)}/ \partial q_2$ gives
\begin{eqnarray}
 \frac{\partial \varphi_1 ^{(2)}}{ \partial q_2} &=& \frac{\beta
\pi_2}{2} m_1(m_2 -1)\left\{- m_1 m_2 \varphi_2 ^{(2)}\varphi_1
^{(2)}
 + m_1 (m_2 -2)\bra \psi_1 ^2 \bra\tanh^2 \Xi\ket_1\ket_2 \right.
 \nonumber
 \\
&& \left. \hspace*{20mm} + 2\bra\psi_1 \left( (m_1-2)\bra\tanh^3
\Xi\ket_1 +2\bra\tanh \Xi\ket_1 \right)\ket_2
 \right\}
 \nonumber \\
&=& \frac{\beta \pi_2}{2} m_1(m_2 -1)\left\{- m_1 m_2
\varphi_2 ^{(2)}\varphi_1 ^{(2)}
 + m_1 (m_2 -2)\bra \psi_1 ^2 \bra\tanh^2 \Xi\ket_1\ket_2 \right. \nonumber \\
&& \left. + 2 (m_1-2)\bra\psi_1\bra\tanh^3 \Xi\ket_1 \ket_2 +4\varphi_2 ^{(2)}
 \right\}
 \label{eq:dphi12dq2}
\end{eqnarray}
From this last result one immediately obtains also the remaining
partial derivative
\begin{eqnarray}
\frac{\partial \varphi_2 ^{(2)}}{ \partial q_1}
&=&\frac{\alpha_1}{\alpha_2} \frac{\partial \varphi_1 ^{(2)}}{ \partial q_2}
=\frac{\pi_1 (m_1 -1)}{\pi_2 m_1(m_2-1)}
 \frac{\partial \varphi_1 ^{(2)}}{ \partial q_2}
 \nonumber \\
&=& \frac{\beta \pi_1}{2} (m_1 -1)\left\{- m_1 m_2
\varphi_1 ^{(2)}\varphi_2 ^{(2)}
 + m_1 (m_2 -2)\bra \psi_1 ^2 \bra\tanh^2 \Xi\ket_1\ket_2 \right. \nonumber \\
&& \left. \hspace*{20mm} + 2 (m_1-2)\bra\psi_1\bra\tanh^3
\Xi\ket_1 \ket_2 +4\varphi_2 ^{(2)}
 \right\},
\label{eq:dphi22dq1}
\end{eqnarray}
where $\alpha_l=\frac{\beta \pi _\ell}{2}M_{\ell -1}(m_\ell-1).$

\subsection{Partial Derivatives of the Saddle-Point Equations}

Here we calculate all partial derivatives of the form $\partial
\varphi_\ell(q_1,q_2,q_3)/\partial q_{\ell^\prime}$, with
$\ell,\ell^\prime\in\{1,2,3\}$, which play a role in our
bifurcation analysis of the $L=2$ and $L=3$ saddle-point
equations.

\subsubsection*{Calculation of $\partial \varphi_3/\partial q_3$:}
$~$\\[3mm] Starting from the general formula (\ref{eq:nearly_l>1})
\begin{eqnarray*}
\frac{\partial \varphi_3}{\partial q_3} &=& \frac{\beta \pi_3}{2}
\left\{ -M_3 M_{2}(m_3-1) \varphi^2 _3
 +(m_3-2)m_{2}m _1 (m_3 -3) M_{2}\bra \psi_{2}^4\ket_3 \right.\\
&& \left.   +4(m_3-2) M_{2}\bra\psi_{2}^2 J_{2}\ket_3 + 2 \bra
J_{2} ^{2}\ket_3 \right\}
\end{eqnarray*}
we derive
\begin{eqnarray}
 \frac{\partial \varphi _3}{\partial q_3} &=&
 \frac{\beta
\pi_3}{2} \left\{ -M_3 M_{2}(m_3-1) \varphi^2 _3
 +(m_3-2)m_{2}m _1 (m_3 -3) M_{2}\bra \psi_{2}^4\ket_3 \right.
 \nonumber
 \\
&&  +4(m_3-2) M_{2}( \bra\psi_{2}^2\ket_3 + (m_1-1)\bra \psi_{2}^2
\bra\bra\tanh^2 \Xi\ket_1 \ket_2\ket_3 \nonumber
\\ && +
m_1(m_2-1)\bra\psi_{2}^2 \bra\psi_{1}^2\ket_2\ket_3) \nonumber
\\ && \left.
+ 2 \bra( 1+  (m_1-1)\bra\bra\psi_0^2\ket_1 \ket_2 +
m_1(m_2-1)\bra\psi_{1}^2\ket_2)^2\ket_3 \right\} \nonumber
\\ &=& \frac{\beta
\pi_3}{2} \left\{ -M_3 M_{2}(m_3-1) \varphi^2 _3
 +(m_3-2)m_{2}m _1 (m_3 -3) M_{2}\bra \psi_{2}^4\ket_3 \right.
 \nonumber
 \\
&&  +4(m_3-2) M_{2}( \varphi_3 + (m_1-1)\bra \psi_{2}^2
\bra\bra\tanh^2 \Xi\ket_1 \ket_2\ket_3 \nonumber
\\ && +
m_1(m_2-1)\bra\psi_{2}^2 \bra\psi_{1}^2\ket_2\ket_3)
 +  2+
4(m_1-1)\varphi_1 + 4m_1(m_2-1)\varphi_2 \nonumber
\\ &&
\left. + 2(m_1-1)^2 \bra(\bra\bra\tanh ^2 \Xi\ket_1 \ket_2)
^2\ket_3 + 2m_1^2(m_2-1)^2\bra(\bra\psi_{1}^2\ket_2) ^2 \ket_3
 \nonumber
 \right.\\
 && \left. + 4m_1 (m_1-1)(m_2-1)\bra\bra\bra\tanh^2 \Xi \ket_1
\ket_2 \bra\psi_{1}^2\ket_2\ket_3 \right\}
\label{eq:dphi3dq3}
\end{eqnarray}

\subsubsection*{Calculation of $\partial \varphi_3/\partial q_1$:}
$~$\\[1mm]
\begin{eqnarray*}
 \frac{\partial \varphi _3}{\partial q_1}& =&
  -\frac{1}{K_3}
\frac{\partial K _3}{\partial q_1}  \varphi _3 +\frac{1}{K_3}
\int\! Dz_3  \frac{\partial}{\partial q_1} (K_{2} ^{m_3}
\psi_{2}^2)
\\ &=&
 -\frac{1}{K_3} \frac{\partial K _3}{\partial
q_1}  \varphi _3 +\frac{m_3-2}{K_3} \int\! Dz_3
 K_{2} ^{m_3 -3 } \frac{\partial K _{2}}{\partial q_1}
(K_{2}\psi_{2})^2
\\ &&
\hspace*{20mm} +\frac{2}{K_3} \int\! Dz_3 K_{2} ^{m_3-1} \psi_{2}
\frac{\partial}{\partial q_1}(K_{2}\psi_{2})
\\
 &=&
 -\frac{1}{K_3} \frac{\partial K _3}{\partial
q_1}  \varphi _3 +\frac{m_3-2}{K_3} \int\! Dz_3
 K_{2} ^{m_3 -3 } \frac{\partial K _{2}}{\partial q_1}
(K_{2}\psi_{2})^2
\\ &&
\hspace*{20mm}
+\frac{2(m_2-1)}{K_3} \int\! Dz_3 K_{2} ^{m_3-1}
\psi_{2} \int\! Dz_2 K_{1}^{m_2-2}\frac{\partial K_1}{\partial
q_1}
 K_{1}\psi_{1} \\
&& \hspace*{20mm}
 +\frac{2}{K_3} \int\! Dz_3 K_{2} ^{m_3-1}
\psi_{2} \int\! Dz_2 K_{1}^{m_2-1}\frac{\partial }{\partial q_1} (
K_{1}\psi_{1})
 \\
&=& -\frac{1}{K_3} \frac{\partial K _3}{\partial q_1}  \varphi _3
+\frac{m_3-2}{K_3} \int\! Dz_3
 K_{2} ^{m_3 -3 } \frac{\partial K _{2}}{\partial q_1}
(K_{2}\psi_{2})^2\\ && +\frac{2(m_2-1)}{K_3} \int\! Dz_3 K_{2}
^{m_3-1} \psi_{2} \int\! Dz_2 K_{1}^{m_2-2}\frac{\partial
K_1}{\partial q_1}
 K_{1}\psi_{1} \\
&&+\frac{\beta \pi_2}{2} \frac{2}{K_3} \int\! Dz_3 K_{2} ^{m_3-1}
\psi_{2}\int\! Dz_2 K_{1}^{m_2-1}\int\! Dz_1
 z_1(\partial\Xi/\partial z_1)^{-1}\!
 \frac{\partial }{\partial \Xi} (K_{0}^{m_1} \psi_0)
 \\
&=& -\frac{1}{K_3} \frac{\partial K _3}{\partial q_1}  \varphi _3
+(m_3-2)\bra \frac{1}{K_2} \frac{\partial K _{2}}{\partial
q_1}\psi_{2}^2 \ket_3 +2(m_2-1) \bra\psi_{2}
\bra\frac{1}{K_1}\frac{\partial K_1}{\partial q_1}\psi_{1}\ket_2
\ket_3 \\ &&+\frac{\beta \pi_1}{2} \frac{2}{K_3} \int\! Dz_3
K_{2}^{m_3-1} \psi_{2}
 \int\! Dz_2   K_{1}^{m_2-1}\int\! Dz_1
 \frac{\partial ^2 }{\partial \Xi ^2 } (K_{0}^{m_1} \psi_0)
 \\
&=& -\frac{1}{K_3} \frac{\partial K _3}{\partial q_1}  \varphi _3
+(m_3-2)\bra \frac{1}{K_2} \frac{\partial K _{2}}{\partial
q_1}\psi_{2}^2 \ket_3 +2(m_2-1) \bra\psi_{2}
\bra\frac{1}{K_1}\frac{\partial K_1}{\partial q_1}\psi_{1}\ket_2
\ket_3
\\
&& \hspace*{20mm}
 +\frac{\beta \pi_1}{2} 2\bra\psi_{2} \bra
\frac{1}{K_{1}}
 \frac{\partial ^2 }{\partial \Xi ^2 } (K_{1} \bra\psi_0\ket_1)\ket_2 \ket_3.
\end{eqnarray*}
The second derivative  $\partial ^2(K_{1} \bra\psi_0\ket_1)/
\partial \Xi ^2$ in this expression is calculated as
\begin{eqnarray*}
 \frac{\partial ^2 }{\partial \Xi ^2 } (K_{1} \bra\psi_0\ket_1)
&=&  \int\! Dz_1 \frac{\partial^2  }{\partial \Xi^2  }
[\cosh^{m_1-1} \Xi  \sinh \Xi]
\end{eqnarray*}
\begin{eqnarray*}
&=& \int\! Dz_1 \left[(m_1 \minus 1)(m_1 \minus 2) \cosh^{m_1-3}
\Xi \sinh^3 \Xi +(3 m_1 \minus 2) \cosh^{m_1-1} \Xi  \sinh \Xi
\right]
\\
 & =&  K_1 \left( (m_1 -1)(m_1 -2)
\bra \tanh^3 \Xi\ket_1 + (3 m_1 -2) \psi_1 \right)
\end{eqnarray*}
Insertion into our previous expression for $\partial
\varphi_3/\partial q_1$ gives:
\begin{eqnarray*}
 \frac{\partial \varphi _3}{\partial q_1}
& =&  -\frac{1}{K_3} \frac{\partial K _3}{\partial q_1}  \varphi _3
+(m_3-2)\bra \frac{1}{K_2} \frac{\partial K _{2}}{\partial q_1}\psi_{2}^2 \ket_3
 +2(m_2-1)
\bra\psi_{2} \bra\frac{1}{K_1}\frac{\partial K_1}{\partial
q_1}\psi_{1}\ket_2 \ket_3 \\ &&+\frac{\beta \pi_1}{2} \left\{
2(m_1 -1)(m_1 -2) \bra\psi_{2} \bra   \bra \tanh^3 \Xi\ket_1
\ket_2 \ket_3 +2 (3 m_1 -2) \bra\psi_{2} \bra \psi_1  \ket_2
\ket_3 \right\}
\end{eqnarray*}
Upon using our previous result (\ref{eq:lprime=1}) of \ref{app:saddle}, this
subsequently translates into
\begin{eqnarray}
 \frac{\partial \varphi _3}{\partial q_1}
& =& \frac{\beta \pi_1}{2} \left\{ -M_3
 \varphi _3 -M_3(m_1 -1)\varphi_1 \varphi_3
+(m_3-2)M_2\bra \psi_{2}^2 \ket_3 \right. \nonumber
\\ && + (m_1
-1)(m_3-2)M_2 \bra\varphi^{(2)}_1 \psi_{2}^2 \ket_3 \nonumber
\\ &&  +2m_1
(m_2-1) \bra\psi_{2} \bra\psi_{1}\ket_2 \ket_3 +2 m_1(m_1 -1)
(m_2-1) \bra\psi_{2}\bra\bra\tanh ^2 \Xi\ket_1 \psi_{1}\ket_2
\ket_3 \nonumber
\\ && \left. + 2(m_1 -1)(m_1 -2) \bra\psi_{2} \bra   \bra
\tanh^3 \Xi\ket_1 \ket_2 \ket_3 +2 (3 m_1 -2) \bra\psi_{2} \bra
\psi_1  \ket_2 \ket_3 \right\} \nonumber
\\ & =&  \frac{\beta
\pi_1}{2}(m_1-1) \left\{ 4 \varphi _3 -M_3\varphi_1 \varphi_3
 + (m_3-2)M_2 \bra\varphi^{(2)}_1 \psi_{2}^2 \ket_3 \right.
\nonumber
 \\
&&  +2 m_1 (m_2-1) \bra\psi_{2}\bra\bra\tanh ^2 \Xi\ket_1
\psi_{1}\ket_2 \ket_3 \nonumber \\ &&\left. +  2(m_1 -2)
\bra\psi_{2} \bra \bra \tanh^3 \Xi\ket_1 \ket_2 \ket_3 \right\}
\label{eq:dphi3dq1}
\end{eqnarray}

\subsubsection*{Calculation of $\partial \varphi_3/\partial q_2$:}
$~$\\[1mm]
\begin{eqnarray*}
\frac{\partial \varphi _3}{\partial q_2}
& =&
 -\frac{1}{K_3}
\frac{\partial K _3}{\partial q_2}  \varphi _3 +\frac{1}{K_3}
\int\! Dz_3  \frac{\partial}{\partial q_2} (K_{2} ^{m_3}
\psi_{2}^2)
\\
&=&
-\frac{1}{K_3} \frac{\partial K _3}{\partial q_2}  \varphi _3
+\frac{m_3-2}{K_3} \int\! Dz_3
 K_{2} ^{m_3 -3 } \frac{\partial K _{2}}{\partial q_2}
(K_{2}\psi_{2})^2\\ && +\frac{2}{K_3} \int\! Dz_3 K_{2} ^{m_3-1}
\psi_{2} \int\! Dz_2 \frac{\partial}{\partial q_2}(K_{1}^{m_2-1}
K_{1}\psi_{1})
\\
&=&
-\frac{1}{K_3} \frac{\partial K _3}{\partial q_2}  \varphi _3
+\frac{m_3-2}{K_3} \int\! Dz_3
 K_{2} ^{m_3 -3 } \frac{\partial K _{2}}{\partial q_2}
(K_{2}\psi_{2})^2\\ && +\frac{2(m_2-1)}{K_3} \int\! Dz_3 K_{2}
^{m_3-1} \psi_{2} \int\! Dz_2 K_{1}^{m_2-2}\frac{\partial
K_1}{\partial q_2}
 K_{1}\psi_{1}
 \\
&&+\frac{2}{K_3} \int\! Dz_3 K_{2} ^{m_3-1} \psi_{2} \int\! Dz_2
K_{1}^{m_2-1} \int\! Dz_1 \frac{\partial \Xi}{\partial q_2}
 \frac{\partial }{\partial \Xi} (K_{0}^{m_1} \psi_0)
 \\
&=&
 -\frac{1}{K_3} \frac{\partial K _3}{\partial q_2}  \varphi _3
+\frac{m_3-2}{K_3} \int\! Dz_3
 K_{2} ^{m_3 -3 } \frac{\partial K _{2}}{\partial q_2}
(K_{2}\psi_{2})^2\\ && +\frac{2(m_2-1)}{K_3} \int\! Dz_3 K_{2}
^{m_3-1} \psi_{2} \int\! Dz_2 K_{1}^{m_2-2}\frac{\partial
K_1}{\partial q_2}
 K_{1}\psi_{1}
 \\
&&+\frac{\beta \pi_2}{2} \frac{2}{K_3} \int\! Dz_3 K_{2} ^{m_3-1}
\psi_{2}
 \int\! Dz_2 \frac{\partial  K_{1}^{m_2-1}}
{\partial \Xi}\int\! Dz_1
 \frac{\partial }{\partial \Xi} (K_{0}^{m_1} \psi_0)\\
&=&
-\frac{1}{K_3} \frac{\partial K _3}{\partial q_2}  \varphi _3
+\frac{m_3-2}{K_3} \int\! Dz_3
 K_{2} ^{m_3 -3 } \frac{\partial K _{2}}{\partial q_2}
(K_{2}\psi_{2})^2\\ && +\frac{2(m_2-1)}{K_3} \int\! Dz_3 K_{2}
^{m_3-1} \psi_{2} \int\! Dz_2 K_{1}^{m_2-2}\frac{\partial
K_1}{\partial q_2}
 K_{1}\psi_{1}
 \\
&&+\frac{\beta \pi_2}{2} \frac{2}{K_3} \int\! Dz_3 K_{2} ^{m_3-1}
\psi_{2}
 \int\! Dz_2 (m_2-1) K_{1}^{m_2-2}  \frac{\partial  K_{1}}{\partial \Xi}
K_{1} J_1
\\
&=&
 -\frac{1}{K_3} \frac{\partial K _3}{\partial q_2}  \varphi _3
+(m_3-2)\bra \frac{1}{K_2} \frac{\partial K _{2}}{\partial
q_2}\psi_{2}^2 \ket_3\\ && +2(m_2-1) \bra\psi_{2}
\bra\frac{1}{K_1}\frac{\partial K_1}{\partial q_2}\psi_{1}\ket_2
\ket_3 +\frac{\beta \pi_2}{2} 2 m_1 (m_2-1)  \bra\psi_{2} \bra
\psi _1  J_1\ket_2  \ket_3.
\end{eqnarray*}
Here we require the following object:
\begin{eqnarray*}
 \bra\frac{1}{K_1}\frac{\partial K_1}{\partial q_2}\psi_{1}\ket_2
&=& \frac{1}{K_2} \int\! Dz_2 K_{1}^{m_2-1}\psi_1\frac{\partial
K_1}{\partial q_2}
\end{eqnarray*}
\begin{eqnarray*}
~~~~~~  &=& \frac{1}{K_2} m_1 \frac{\beta \pi_2}{2}\int\!
Dz_2\left \{ \frac{\partial}{\partial \Xi} [ K_1
^{m_2-1}\psi_1]\right\} \int\! Dz_1 K_{0}^{m_1-1} \frac{\partial
K_0}{\partial \Xi}
\\
&=& \frac{1}{K_2} m_1 \frac{\beta \pi_2}{2}\int\! Dz_2 K_1 ^{m_2}
\left[ (m_2 -2) \frac{1}{K_1} \frac{\partial K_1}{\partial \Xi}
\frac{1}{K_1} \psi_1 + \frac{1}{K_1} \frac{\partial}{\partial \Xi}
( K_1 \psi_1) \frac{1}{K_1} \right]
\\
&&\times \int\! Dz_1 K_{0}^{m_1} \tanh \Xi\\
 &=&
 m_1 \frac{\beta \pi_2}{2}\bra \left[
 (m_2 -2) \frac{1}{K_1} \frac{\partial K_1}{\partial \Xi}
 \psi_1 + \frac{1}{K_1}  \frac{\partial}{\partial \Xi} (
K_1 \psi_1)  \right]
 \psi _1 \ket_2
 \\
&=&
 m_1 \frac{\beta \pi_2}{2} \left[
m_1 (m_2 -2) \bra\psi_1 ^3\ket_2 + \bra J_1 \psi_1\ket_2  \right]
\end{eqnarray*}
Insertion into our earlier expression for $\partial \varphi
_3/\partial q_2$, followed by usage of
(\ref{eq:lprime>1}-\ref{eq:lprime=l+1}) whenever appropriate,
allows us to write
\begin{eqnarray}
\frac{\partial \varphi _3}{\partial q_2} &=& -\frac{1}{K_3}
\frac{\partial K _3}{\partial q_2}  \varphi _3 +(m_3-2)\bra
\frac{1}{K_2} \frac{\partial K _{2}}{\partial q_2}\psi_{2}^2
\ket_3 \nonumber \\ && +2m_1 (m_2-1)
 \frac{\beta \pi_2}{2} \bra\psi_{2}
\left \{ m_1 (m_2 -2) \bra\psi_1 ^3\ket_2 + \bra J_1 \psi_1\ket_2
\right\}\ket_3 \nonumber \\ && +\frac{\beta \pi_2}{2} 2 m_1
(m_2-1) \bra\psi_{2} \bra  \psi _1  J_1\ket_2  \ket_3 \nonumber \\
&=& -\frac{1}{K_3} \frac{\partial K _3}{\partial q_2}  \varphi _3
+(m_3-2)\bra \frac{1}{K_2} \frac{\partial K _{2}}{\partial
q_2}\psi_{2}^2 \ket_3 \nonumber \\ &&
 +2m_1^2 (m_2-1) (m_2 -2) \frac{\beta
\pi_2}{2} \bra\psi_{2} \bra\psi_1 ^2\ket_2\ket_3 \nonumber
\\ &&
+\frac{\beta \pi_2}{2} 4 m_1 (m_2-1) \left\{ \bra\psi_{2} \bra
\psi _1 \ket_3  \ket_3 + (m_1 -1) \bra\psi_{2} \bra  \psi _1
\bra\tanh ^2 \Xi\ket_1 \ket_2  \ket_3. \right\} \nonumber
\\ &=&
\frac{\beta \pi_2}{2}(m_2-1)\left\{ -M_3 m_1 \varphi_2 \varphi _3
+M_2 m_1(m_3-2)\bra\bra\psi_1 ^2\ket_2\psi_{2}^2 \ket_3 \right.
\nonumber
\\ && +2m_1^2 (m_2 -2)  \bra\psi_{2} \bra\psi_1
^3\ket_2\ket_3 \nonumber
\\ && \left. + 4 m_1  \varphi_3 + 4
m_1(m_1 -1)\bra\psi_{2} \bra  \psi _1  \bra\tanh ^2 \Xi\ket_1
\ket_2 \ket_3 \right\} \label{eq:dphi3dq2}
\end{eqnarray}

\subsubsection*{Calculation of
 $\partial \varphi_1/\partial q_3$ and $\partial
\varphi_2/\partial q_3$:} $~$\\[3mm] These two partial
derivatives are found to be relatively easy to calculate, and come
out as follows:
\begin{eqnarray}
 \frac{\partial \varphi_1}{\partial q_3}&=& \frac{\alpha _3}{\alpha _1}\frac{\partial
\varphi _3}{\partial q_1} =\frac{\pi_3 M_2(m_3-1)}{\pi_1
(m_1-1)}\frac{\partial \varphi _3}{\partial q_1} \nonumber
\\
& =&
\frac{\beta \pi_3}{2} M_2 (m_3 -1)
 \left\{
4 \varphi _3 -M_3 \varphi_1 \varphi_3
 + (m_3-2)M_2 \bra\varphi^{(2)}_1 \psi_{2}^2 \ket_3 \right. \nonumber
 \\
&& \left. +2 m_1 (m_2\minus 1) \bra\psi_{2}\bra\bra\tanh ^2
\Xi\ket_1 \psi_{1}\ket_2 \ket_3 +  2(m_1- 2) \bra\psi_{2} \bra
\bra \tanh^3 \Xi\ket_1 \ket_2 \ket_3 \right\}\nonumber
\\
\label{eq:dphi1dq3}
\\
 \frac{\partial
\varphi _2}{\partial q_3}&=& \frac{\alpha _3}{\alpha
_2}\frac{\partial \varphi _3}{\partial q_2} =\frac{\pi_3 M_2 (m_3
-1)}{\pi_2 M_1 (m_2 -1)} \frac{\partial \varphi _3}{\partial q_2}
\nonumber
\\
& =& \frac{\beta \pi_3}{2}m_2(m_3\minus 1) \left\{ - M_3
m_1\varphi_2 \varphi _3 +M_2 m_1(m_3- 2)\bra\bra\psi_1
^2\ket_2\psi_{2}^2 \ket_3
 +4 m_1 \varphi_3
\right. \nonumber
\\ &&
\left.
+2m_1^2(m_2-2) \bra\psi_{2} \bra\psi_1 ^3 \ket_2 \ket_3
   +4 m_1(m_1 -1)  \bra\psi_{2} \bra  \psi _1 \bra\tanh^2
\Xi\ket_1 \ket_2  \ket_3 \right\}\nonumber \\
 \label{eq:dphi2dq3}
\end{eqnarray}

\subsubsection*{Calculation of $\partial \varphi_2/\partial q_1$ and
$\partial \varphi_1/\partial q_2$:} $~$\\[1mm]
\begin{eqnarray}
 \frac{\partial \varphi _1}{\partial q_2}
 &=&
-\frac{1}{K_3}\frac{\partial K_3}{ \partial q_2}\varphi_1 +
\frac{1}{K_3}\int\! Dz_3   \frac{\partial }{ \partial q_2} ( K_2
^{m_3} \varphi_1 ^{(2)}) \nonumber
\\
&=&
 -\frac{1}{K_3}\frac{\partial K_3}{
\partial q_2}\varphi_1 + m_3  \frac{1}{K_3}\int\! Dz_3   \frac{1}{
K_ 2} \frac{\partial K_2}{ \partial q_2} K_2 ^{m_3} \varphi_1
^{(2)} + \frac{1}{K_3}\int\! Dz_3 K_2 ^{m_3} \frac{\partial
\varphi_1 ^{(2)}}{ \partial q_2} \nonumber \\
&=& -\frac{\beta
\pi_2}{2} M_3 m_1(m_2 -1)\varphi_1 \varphi_2 + m_3 \bra \frac{1}{
K_ 2} \frac{\partial K_2}{ \partial q_2}
  \varphi_1 ^{(2)}\ket_3
+ \bra \frac{\partial \varphi_1 ^{(2)}}{ \partial q_2} \ket _3
\nonumber \\
&=&
 \frac{\beta \pi_2}{2} M_3 m_1(m_2 -1)
( - \varphi_1 \varphi_2 + \bra \varphi_1 ^{(2)} \varphi_2
^{(2)}\ket_3) \nonumber
\\ &&
+ \bra
 \frac{\beta \pi_2}{2} m_1(m_2 -1)\left\{- m_1 m_2
\varphi_1 ^{(2)}\varphi_2 ^{(2)}
 + m_1 (m_2 -2)\bra \psi_1 ^2 \bra\tanh^2 \Xi\ket_1\ket_2 \right.
\nonumber
 \\ && \left.
 \hspace*{20mm}+ 2 (m_1-2)\bra\psi_1\bra\tanh^3 \Xi\ket_1
\ket_2 +4\varphi_2 ^{(2)}
 \right\}
 \ket _3
\nonumber
\end{eqnarray}
\begin{eqnarray}
~~~~~~ &=& \frac{\beta \pi_2}{2}  m_1(m_2 -1) \left\{-M_3\varphi_1
\varphi_2
 + m_1 m_2(m_3-1) \bra \varphi_1 ^{(2)} \varphi_2 ^{(2)}\ket_3
+4 \varphi_2
 \right.
\nonumber \\ && \left. + m_1 (m_2 -2) \bra \bra \psi_1 ^2
\bra\tanh^2 \Xi \ket_1\ket_2 \ket_3 + 2 (m_1-2)\bra \bra
\psi_1\bra\tanh^3 \Xi\ket_1 \ket_2 \ket_3   \right\} \nonumber
\\ \label{eq:dphi1dq2}
\end{eqnarray}
From this result, in turn, immediately follows
\begin{eqnarray}
\frac{\partial \varphi_2 }{ \partial q_1}
&=&\frac{\alpha_1}{\alpha_2} \frac{\partial \varphi_1}{ \partial q_2}
=\frac{\pi_1 (m_1 -1)}{\pi_2 m_1(m_2-1)}
 \frac{\partial \varphi_1 }{ \partial q_2}
 \nonumber
 \\
&=& \frac{\beta \pi_1}{2} (m_1 -1)
\left\{-m_1 m_2 m_3 \varphi_1 \varphi_2
 + m_1 m_2(m_3-1) \bra \varphi_1 ^{(2)} \varphi_2 ^{(2)}\ket_3
  +4 \varphi_2  \right.
 \nonumber
 \\
&& \left. + m_1 (m_2 -2) \bra \bra \psi_1 ^2 \bra\tanh^2 \Xi
\ket_1\ket_2 \ket_3 + 2 (m_1-2)\bra \bra \psi_1\bra\tanh^3
\Xi\ket_1 \ket_2 \ket_3 \right\} \nonumber\\ \label{eq:dphi2dq1}
\end{eqnarray}

\subsubsection*{Calculation of $\partial \varphi_1/\partial q_1$
and $\partial\varphi_2/\partial q_2$:} $~$\\[3mm] Finally, let us
turn to the remaining two  partial derivatives:
 $\partial \varphi _1/\partial q_1$ and
$\partial \varphi _2/\partial q_2$:
\begin{eqnarray}
 \frac{\partial \varphi _1}{\partial q_1}
 & =&
 -\frac{1}{K_3}
\frac{\partial K _3}{\partial q_1}  \varphi _1 +\frac{1}{K_3}
\int\! Dz_3  \frac{\partial}{\partial q_1} [K_{2} ^{m_3}
\varphi_{1}^{(2)}] \nonumber
\\ &=&
 -\frac{1}{K_3} \frac{\partial K
_3}{\partial q_1}  \varphi _1 +\frac{m_3}{K_3} \int\! Dz_3
 K_{2} ^{m_3 -1 } \frac{\partial K _{2}}{\partial q_1}
\varphi_{1}^{(2)} +\bra \frac{\partial \varphi _1 ^{(2)}}{\partial
q_1}\ket _3 \nonumber
\\ & =& \frac{\beta \pi_1}{2} \left\{- M_3
\left\{ 1+ (m_1 -1)\varphi_1\right\}  \varphi _1 +M_3 \bra \left\{
1+ (m_1 -1)\varphi^{(2)}_1 \right\}\varphi_{1}^{(2)} \ket_3
\right. \nonumber
\\&&
 - (m_1 -1)M_2\bra( \varphi ^{(2)} _1) ^2 \ket_3
 + m_1(m_1-1)(m_2 -1)\bra\bra\bra\tanh^2 \Xi\ket_1^2\ket_2 \ket_3
\nonumber
 \\&& \left. +(m_1 -2) (m_1 -3)\bra \bra\bra \tanh^4 \Xi
\ket_1\ket_2 \ket_3
 + 4(m_1 -2)\bra \varphi_1 ^{(2)}\ket _3+2 \right\}
\nonumber
 \\ & =& \frac{\beta \pi_1}{2} \left\{ - (m_1 -1)M_3
(\varphi_1)^2
 + (m_1 -1)M_2(m_3 -1)\bra (\varphi^{(2)}_1)^2 \ket_3 \right.
\nonumber \\&&
 + m_1(m_1-1)(m_2 -1)\bra\bra\bra\tanh^2 \Xi\ket_1^2\ket_2 \ket_3
\nonumber \\&&
 \left. +(m_1 -2) (m_1 -3)\bra \bra\bra \tanh^4 \Xi \ket_1\ket_2 \ket_3
 + 4(m_1 -2)\varphi_1 +2 \right\}
\label{eq:dphi1dq1}
\end{eqnarray}
where we use (\ref{eq:dphi12dq1}),
and, similarly, using (\ref{eq:dphi22dq2})
\begin{eqnarray}
\frac{\partial \varphi _2}{\partial q_2} & =& -\frac{1}{K_3}
\frac{\partial K _3}{\partial q_2}  \varphi _2 +\frac{1}{K_3}
\int\! Dz_3  \frac{\partial}{\partial q_2} (K_{2} ^{m_3}
\varphi_{2}^{(2)}) \nonumber
\\ & =&
 -\frac{\beta \pi_2}{2}M_3 m_1(m_2 -1)(\varphi_2 )^2
+m_3\bra\frac{\beta \pi_2}{2}M_2 m_1(m_2 -1)\varphi^{(2)}_2
\varphi_{2}^{(2)} \ket_3 \nonumber
\\&& +\bra
 \frac{\beta \pi_1}{2}
\left\{ - m_{1}^2 m_2 (m_2-1) (\varphi _2 ^{(2)})^2
 +m_{1}^2 (m_2-2)(m_2 -3)\bra \psi_{1}^4\ket_2 \right.
\nonumber
 \\&&
   +4m_{1}(m_2-2)\varphi _2 ^{(2)}
+4m_{1}(m_1 -1)(m_2-2)  \bra \psi_{1}^2 \bra\tanh ^2 \Xi\ket_1
\ket_2 \nonumber
\\&& \left. +2 + 4(m_1 -1)\varphi _1 ^{(2)} +2(m_1
-1)^2\bra\bra\tanh ^2 \Xi\ket_1^2\ket_2 \right\}\ket_3 \nonumber
\\
& =&
 \frac{\beta \pi_2}{2} \left\{
- M_3 m_1(m_2 -1)(\varphi_2 )^2 +m_1^2 m_2 (m_2
-1)(m_3-1)\bra(\varphi^{(2)}_2)^2 \ket_3 \right. \nonumber
\\&&
 +m_{1}^2 (m_2-2)(m_2 -3)\bra \bra \psi_{1}^4 \ket_2 \ket_3
\nonumber
 \\&&
   +4m_{1}(m_2-2)\varphi _2
+4m_{1}(m_1 -1)(m_2-2) \bra \bra \psi_{1}^2 \bra\tanh ^2 \Xi
\ket_1 \ket_2 \ket_3 \nonumber
\\&& \left. +2 + 4(m_1 -1)\varphi _1
+2(m_1 -1)^2 \bra \bra \bra\tanh ^2 \Xi \ket_1^2 \ket_2 \ket_3
\right\} \label{eq:dphi2dq2}
\end{eqnarray}

\end{document}